\documentclass[12pt]{article}
\usepackage[utf8]{inputenc}
\usepackage{amsmath, amsthm, bm, amssymb, hyperref, xcolor, caption, subcaption, graphicx, dsfont, multirow, booktabs, array,rotating,enumitem, mathtools, placeins}

\hypersetup{
	hidelinks,
    colorlinks = true,
    linkbordercolor = {white},
    citecolor = blue,
    urlcolor = blue,
    linkcolor = blue,
    linktoc = all 
}

\usepackage[
  top=2cm,
  bottom=2cm,
  left=1in,
  right=1in,
  headheight=17pt, 
  includehead,includefoot,
  heightrounded, 
]{geometry} 

\usepackage{natbib}
\setcitestyle{authoryear}
\usepackage{grffile}

\author{Max Welz \\
  \href{mailto:max.welz@uzh.ch}{\texttt{max.welz@uzh.ch}} \\ University of Zurich}
  
\date{\today\\[1ex]}
    
\title{\textsc{Robust estimation of polyserial correlation coefficients: A density power divergence approach}\thanks{
This paper has been published in \emph{Psychometrika} at \url{https://doi.org/10.1017/psy.2026.10091}. This ArXiv version differs from the published version in layout, pagination, and typographic detail. Please cite the published version. I thank Andreas Alfons for valuable feedback. I further thank the editor as well as three anonymous reviewers, whose comments led to a substantially improved paper. Replication files are publicly available on GitHub at \url{https://github.com/mwelz/robust-polyserial-replication}.}}

\newcommand{\Eoperator}{\mathbb{E}}
\newcommand{\Poperator}{\mathbb{P}}
\newcommand{\E}[1]{\Eoperator \left[ #1 \right]}

\newcommand{\var}[1]{\mathbb{V}\mathrm{ar} \left[ #1 \right]}
\newcommand{\cor}[2]{\mathrm{Cor} \left[#1,\ #2\right]}
\newcommand{\cov}[2]{\mathrm{Cov} \left[#1,\ #2\right]}
\newcommand{\corF}[3]{\mathrm{Cor}_{#1} \left[#2,\ #3\right]}

\renewcommand{\Pr}[2]{\Poperator_{#1} \left[ #2 \right]}
\newcommand{\CDF}[1]{\Phi\left( #1 \right)}
\newcommand{\PDF}[1]{\phi\left( #1 \right)}
\newcommand{\PDFprime}[1]{\phi'\left( #1 \right)}
\renewcommand{\emph}[1]{\textit{#1}}
\renewcommand{\vec}[1]{\boldsymbol{#1}}
\newcommand{\mat}[1]{\boldsymbol{#1}}

\newcommand{\Y}{\mathcal{Y}}
\newcommand{\z}{\vec{z}}
\newcommand{\Z}{\vec{Z}}
\newcommand{\Zset}{\bm{\mathcal{Z}}}
\newcommand{\Btheta}{\bm{\theta}}
\newcommand{\Btau}{\vec{\tau}}
\newcommand{\ssigma}{\sigma^2}
\newcommand{\taustarfun}[2]{\tau^*_{#1}\left({#2}\right)}
\newcommand{\taustar}[1]{\taustarfun{#1}{x; \Btheta}}

\newcommand{\funmain}[5]{{#1}_{#2}^{#3}\left(#4; #5\right)}
\newcommand{\pxyfun}[3]{\funmain{p}{XY}{#3}{#1}{#2}}
\newcommand{\pxy}[1]{\pxyfun{x,y}{#1}{}}
\newcommand{\pxyemptyfun}[1]{\pxyfun{\cdot,\cdot}{#1}{}}
\newcommand{\pxyempty}{\pxyemptyfun{\Btheta}}
\newcommand{\pxfun}[3]{\funmain{p}{X}{#3}{#1}{#2}}
\newcommand{\px}[1]{\pxfun{x}{#1}{}}
\newcommand{\pyfun}[4]{\funmain{p}{Y|X}{#4}{#1\ |\ #2}{#3}}
\newcommand{\py}[1]{\pyfun{y}{x}{#1}{}}
\newcommand{\pymarg}[2]{\funmain{p}{Y}{}{#1}{#2}}
\newcommand{\pzfun}[3]{\funmain{p}{\Z}{#3}{#1}{#2}}
\newcommand{\pz}[1]{\pzfun{\z}{#1}{}}
\newcommand{\pzemptyfun}[1]{\pzfun{\cdot,\cdot}{#1}{}}
\newcommand{\pzempty}{\pzemptyfun{\Btheta}}
\newcommand{\Pxysymb}{P_{XY}}
\newcommand{\Pxetasymb}{P_{X\eta}}
\newcommand{\Pxyfun}[2]{\funmain{P}{XY}{}{#1}{#2}}
\newcommand{\Pxy}[1]{\Pxyfun{x,y}{#1}}

\newcommand{\pxetafun}[3]{\funmain{p}{X\eta}{#3}{#1}{#2}}
\newcommand{\Pxetafun}[2]{\funmain{P}{X\eta}{}{#1}{#2}}

\newcommand{\Hxy}{H_{XY}}
\newcommand{\Hxyfun}[1]{\Hxy\left(#1\right)}
\newcommand{\Hxeta}{H_{X\eta}}
\newcommand{\Hxetafun}[1]{\Hxeta\left(#1\right)}

\newcommand{\fhat}{\hat{f}_{N}}
\newcommand{\fhatfun}[2]{\fhat^{#2}\left(#1\right)}
\newcommand{\fhatxy}[1]{\fhatfun{#1}{}}
\newcommand{\fxy}{f_{\varepsilon,XY}}
\newcommand{\feps}{f_{\varepsilon,\vec{Z}}}
\newcommand{\fepsz}{\feps(\z)}
\newcommand{\Fxy}{F_{\varepsilon,XY}}
\newcommand{\Fxeta}{F_{\varepsilon,X\eta}}
\newcommand{\Fxyfun}[1]{\Fxy \left( #1\right)}
\newcommand{\Fxetafun}[1]{\Fxeta \left( #1\right)}

\newcommand{\divsymb}{\textnormal{D}}
\newcommand{\KLsymb}{\textnormal{KL}}
\newcommand{\divergencefun}[3]{{#1}\left(#2 \ \big|\big|\ #3\right)}
\newcommand{\KL}[2]{\divergencefun{\KLsymb}{#1}{#2}}
\newcommand{\divergence}[2]{\divergencefun{\divsymb_\alpha}{#1}{#2}}
\newcommand{\divergencenull}[2]{\divergencefun{\divsymb_0}{#1}{#2}}

\newcommand{\scorefun}[2]{\vec{u}_{#1}\left(#2\right)}
\newcommand{\scorexy}[1]{\scorefun{#1}{x,y}}
\newcommand{\scorez}[1]{\scorefun{#1}{\z}}
\newcommand{\fisher}[1]{\mat{I}_{#1}}
\newcommand{\Qfun}[2]{\mat{Q}_{#2}\left(#1\right)}
\newcommand{\Qz}[1]{\Qfun{\z}{#1}}
\newcommand{\Qxy}[1]{\Qfun{x,y}{#1}}
\newcommand{\gradient}{\nabla_{\Btheta}}
\newcommand{\hessian}{\nabla^2_{\Btheta}}

\newcommand{\matfun}[2]{\mat{#1}\left( #2 \right)}
\newcommand{\matinv}[2]{\matfun{#1}{#2}^{-1}}

\newcommand{\hatmatfun}[2]{\hat{\mat{#1}}_N\left( #2 \right)}
\newcommand{\hatmatinv}[2]{\hatmatfun{#1}{#2}^{-1}}

\newcommand{\hatN}[1]{\hat{#1}_N}
\newcommand{\hatNMLE}[1]{\hatN{#1}^{\mathrm{\ MLE}}}
\newcommand{\thetahat}{\hatN{\Btheta}}
\newcommand{\thetahatMLE}{\hatNMLE{\Btheta}}
\newcommand{\rhohat}{\hatN{\rho}}
\newcommand{\rhohatt}{\rhohat^{(t)}}
\newcommand{\BTheta}{\bm{\Theta}}

\newcommand{\gauss}{\textnormal{N}}
\newcommand{\I}[1]{\mathds{1}\left\{ #1 \right\}}
\renewcommand{\d}{\textnormal{d}}

\renewcommand{\hat}[1]{\widehat{#1}}

\renewcommand{\tilde}[1]{\widetilde{#1}}
\newcommand{\R}{\mathbb{R}}

\newcommand{\convP}{\stackrel{\Poperator}{\longrightarrow}}
\newcommand{\convweak}{\stackrel{\mathrm{d}}{\longrightarrow}}

\newcommand{\vecfun}[2]{\vec{#1}\left({#2}\right)}

\newcommand{\partialderivative}[2]{\frac{\partial #1 }{\partial #2}}
\newcommand{\partialderivativetwice}[2]{\frac{\partial^2 #1 }{\partial {#2}\partial {#2}^\top}}
\newcommand{\partialderivativetwiced}[3]{\frac{\partial^2 #1 }{\partial {#2}\partial{#3}}}

\newcommand{\QED}{\hfill$\blacksquare$}
\newcommand{\proglang}[1]{\textsf{#1}}
\newcommand{\pkg}[1]{\texttt{#1}}
\newcommand{\rhohatmean}{\hatN{\rho}^{\ \text{ave}}}
\newcommand{\SE}[1]{\textnormal{SE}\left(#1\right)}
\newcommand{\SEapprox}[1]{\textnormal{SE}^{\textnormal{approx}}\left(#1\right)}
\newcommand{\SEhat}[1]{\hat{\textnormal{SE}}\left(#1\right)}
\newcommand{\SEbarhat}[1]{\hat{\textnormal{SE}}^{\textnormal{ave}}\left(#1\right)}

\newcommand{\edit}[1]{\textcolor{black}{#1}}

\newcommand{\matnorm}[1]{{\left\vert\kern-0.25ex\left\vert\kern-0.25ex\left\vert #1 
    \right\vert\kern-0.25ex\right\vert\kern-0.25ex\right\vert}}

\theoremstyle{plain}
\newtheorem{theorem}{\color{blue}Theorem}

\newtheorem{assumption}{\color{blue}Assumption}

\theoremstyle{definition}

\usepackage{apptools}
\AtAppendix{\counterwithin{lemma}{section}}
\AtAppendix{\counterwithin{proposition}{section}}
\AtAppendix{\counterwithin{corollary}{section}}

\makeatletter
\newcommand\notsotiny{\@setfontsize\notsotiny{8}{8.8}}
\makeatother

\begin{document}

\maketitle

\begin{abstract}
\noindent
The association between a continuous and an ordinal variable is commonly modeled through the polyserial correlation model. However, this model, which is based on a partially-latent normality assumption, may be misspecified in practice, due to, for example (but not limited to), outliers or careless responses. The typically used maximum likelihood (ML) estimator is highly susceptible to such misspecification: One single observation not generated by partially-latent normality can suffice to produce arbitrarily poor estimates. As a remedy, we propose a novel estimator of the polyserial correlation model designed to be robust against the adverse effects of observations discrepant to that model. The estimator leverages \emph{density power divergence estimation} to achieve robustness by implicitly downweighting such observations; the ensuing weights constitute a useful tool for pinpointing potential sources of model misspecification. The proposed estimator generalizes ML and is consistent as well as asymptotically Gaussian. As price for robustness, some efficiency must be sacrificed, but substantial robustness can be gained while maintaining more than~98\% of ML efficiency. We demonstrate our estimator's robustness and practical usefulness in simulation experiments and an empirical application in personality psychology where our estimator helps identify outliers. Finally, the proposed methodology is implemented in free open-source software.
\end{abstract}

\textsc{Keywords:} Polyserial correlation, model misspecification, robust estimation, mixed data, careless responding

\newpage


\section{Introduction}
Empirical research in the psychological, social, and health sciences often features data that contain both continuous and ordered categorical (ordinal) random variables. Examples of continuous variables are household income, blood pressure, or time spent doing an activity of interest, while examples of ordinal variables are responses to rating scales (measuring, e.g., personality traits, wellbeing, or overall health), relationship status (with categories such as \emph{single}, \emph{in a relationship}, and \emph{married}), or grouped measurements of continuous variables like one's income group. The association between a continuous and an ordinal variable is typically modeled by means of \emph{polyserial correlation} \citep{pearson1909}. Polyserial correlation is a key building block in the analysis of mixed data, particularly structural equation models (SEMs). For instance, the popular \proglang{R} package \pkg{lavaan} \citep{lavaan} for SEM analyses by default uses polyserial correlation for SEMs involving both continuous and ordinal variables. 

The polyserial correlation model postulates the existence of a latent continuous variable that underlies and governs the observed ordinal variable through an unobserved discretization process. The correlation between the observed continuous variable and the latent variable is called \emph{polyserial correlation}, whereas the correlation between the observed continuous and the observed ordinal variable is called \emph{point} polyserial correlation, where the latter can be computed from the former. For identification, the polyserial correlation model  assumes that the observed continuous variable and the latent continuous variable are jointly normally distributed, thereby making it a partially-latent normality model. Estimation is typically conducted by means of maximum likelihood \citep{cox1974,olsson1982}. However, the validity of the partially-latent normality assumption is often questionable in practice \citep[e.g.,][]{barbiero2025,demirtas2016,bedrick1995}.
\edit{Disquietingly, though, central statistical properties of polyserial correlation crucially depend on this very assumption to hold true: Violations of partially-latent normality have been shown to have potentially devastating effects on identification \citep[e.g.,][]{moss2023} and maximum likelihood estimation \citep[e.g.,][]{bedrick1995} of polyserial correlation. Nonnormality of (partially) latent variables can also introduce serious biases in SEM analyses conducted with software that assumes such normality \citep[e.g.,][]{foldnes2022,foldnes2020polycor}, such as \pkg{lavaan} \citep{lavaan}, \pkg{LISREL} \citep{lisrel}, and \pkg{Mplus} \citep{Mplus}.}

\edit{Motivated by the susceptibility of polyserial correlation to partially-latent nonnormality, t}he contributions of this paper are twofold. First, we study estimation of the polyserial correlation model when a (possibly empty) subset of the observed data are of low-quality and therefore may have \emph{not} been generated by a normal distribution, such as (but not limited to) outliers in the continuous variable and/or careless responses in the ordinal variable. Consequently, such observations are uninformative for estimating the polyserial correlation model. This situation is called \emph{partial model misspecification} because the assumption of partially-normality might be violated for parts of the data, but is satisfied for the remaining data. We demonstrate that already one single uninformative observation can suffice for maximum likelihood (ML) estimation of polyserial correlation to yield arbitrary results and, furthermore, that the accuracy of ML estimation is highly susceptible to even minor misspecification of partially-latent normality. Partial misspecification stems from classic literature on robust statistics \citep[e.g.,][]{huber2009} where it is known as \emph{Huber contamination model}, owing to \citet{huber1964}.

Second, in wake of the non-robustness of ML estimation to uninformative observations generated by partially-latent nonnormality, we propose an alternative estimator that is designed to be robust against such partial misspecification. The proposed \edit{methodology applies} \emph{density power divergence} estimat\edit{ion} \citep{basu1998} \edit{to the polyserial correlation model} and achieves robustness by implicitly downweighting observations that cannot be sufficiently well fitted by \edit{that} model. \edit{The ensuing weights are a useful tool for pinpointing potential sources of (partial) model misspecification. As additional methodological contribution, we devise a simple rescaling of the weights to ensure that they are contained in the unit interval. Overall, to the best of our knowledge, the proposed methodology is the first contamination-robust approach to polyserial correlation.}

In line with the partial misspecification framework, the \edit{robust} estimator allows the model to be misspecified for an unknown fraction of uninformative observations, but makes \emph{no assumption} on \emph{how} and \emph{where} partial misspecification occurs (which may be absent altogether). Consequently, partial misspecification can manifest through an unlimited and unrestricted variety of ways, such as (but not limited to) outliers or careless responses. Conversely, if the polyserial correlation model is correctly specified for all observations in a sample, then \edit{the robust estimator is, just like ML, consistent for the true parameter vector, and, therefore}, generaliz\edit{es} ML estimation. 

\edit{Studying their respective behavior under model misspecification, we show that both ML and the robust estimator still converge in probability, but to different limits. Crucially, the robust estimator converges to a parameter vector that is closer to the true parameter vector than ML, thereby gaining its robustness. Moreover, the robust estimator and ML remain asymptotically normally distributed}, allowing for statistical inference \edit{both under correct and incorrect specification of the polyserial correlation model}.

In robust statistics, there is a well-established fundamental tension between robustness and efficiency for estimation procedures for data involving continuous variables \citep[e.g.,][]{huber2009,hampel1986}. Our estimator is no exception: The price to pay for its enhanced robustness manifests in the form of diminished efficiency compared to ML. However, we show that sacrificing as little as~2\% of efficiency suffices to obtain a substantial gain in robustness, so the efficiency loss is only comparatively minor. An additional price of our robust estimator is increased computational intensity. Nevertheless, using our implementation, the robust estimator usually executes in less than \edit{two} seconds on a regular laptop, so the overall computational burden should remain small in practical applications. This implementation of our proposed methodology is publicly and freely available as part of the \proglang{R} package \pkg{robcat} \citep[for ``ROBust CATegorical data analysis'';][]{robcat} on CRAN (the Comprehensive \proglang{R} Archive Network) at \url{https://CRAN.R-project.org/package=robcat}.

This paper is organized as follows. Section~\ref{sec:literature} reviews related literature, while Section~\ref{sec:polyserial-model} summarizes the polyserial correlation model and its ML estimation. Section~\ref{sec:misspecification} describes the partial misspecification framework adopted in this paper. Section~\ref{sec:estimator} introduces our robust estimator, and Section~\ref{sec:properties} derives its theoretical and computational properties. Section~\ref{sec:simulation} carries out simulation studies to compare the performance of the robust estimator to ML in a variety of settings. Section~\ref{sec:application} provides an empirical application on data from personality psychology. 
Section~\ref{sec:conclusion} discusses and concludes.

\section{Literature} \label{sec:literature}
Potential violations of the normality assumption that underlies the polyserial correlation model have been studied in previous literature, though the misspecification framework used therein is fundamentally different to the partial misspecification framework adopted in this paper. Specifically, previous literature focuses on \emph{distributional misspecification}, where the polyserial correlation model is misspecified (usually through non-normality) for the \emph{entire} observed sample. In contrast, in partial misspecification, the model is only misspecified for a (possibly empty) subset of the sample.  We explain the differences between partial and distributional misspecification in more detail in Section~\ref{sec:distributional-subsection}.

Focusing on distributional misspecification. \citet{bedrick1995} shows that the accuracy of normality-based ML estimation of polyserial correlation crucially depends on whether or not the marginal distribution of the latent variable that underlies the observed ordinal variable is normal. If that distribution is not normal, but, for instance, an exponential or $t$-distribution, ML estimates of polyserial correlation may be attenuated \citep{kraemer1981,brogden1949,lord1963}. For a dichotomous ordinal variable, \citet{demirtas2017} and \citet{demirtas2016} devise an algorithm that relates  polyserial correlation  to  point polyserial correlation when the underlying joint distribution (which they assume to be known) is not bivariate normal. For a given potentially nonnormal joint distribution and a dichotomous ordinal variable, \citet{cheng2016} derive a general expression for the maximum point polyserial coefficient (in population). \citet{barbiero2025} generalizes the results of \citet{cheng2016} to a polytomous ordinal variable. 

\edit{
\citet[][Section~2]{moss2023} and \citet[][Section~2.5]{gronneberg2020} use partial identification analyses to study distributional misspecification of the polyserial model from a theoretical perspective. Specifically, given known marginal distributions of the observed continuous and latent continuous variables but keeping their joint distribution unspecified, they derive partial identification sets for the partially-latent correlation. They show that the partial identification sets tend to be uninformatively wide. Consequently, polyserial correlation is very sensitive to partially-latent joint nonnormality because different nonnormal distributions can yield widely different correlations. To reduce the width of the partial identification sets, one must make restrictive assumptions on the partially-latent joint distribution, such as Gaussian-like characteristics or equipping it with a parametric structure. 
} 

In a broader context, violations of normality assumptions have been studied extensively in the psychometric literature, especially with respect to the distribution of latent variables 
(e.g., \citealp{asparouhov2016}; \citealp{lyhagen2023}; \citealp{monroe2018}; \edit{\citealp{moss2023}}; \citealp{roscino2006}; \citealp{yuan2004}, and references therein). 
A particular focus of recent literature has been the \emph{polychoric} correlation model of \citet{pearson1922}, which models through a latent bivariate normality distribution the association of two latent variables that govern two observed ordinal variables \citep[see][for a modern exposition]{olsson1979poly}. \citet{foldnes2020polycor} and \citet{jin2017} show that ML estimation of polychoric correlation is highly susceptible to distributional misspecification, leading to possibly large biases in SEM analyses based on polychoric correlation \citep{foldnes2022,gronneberg2022}.\footnote{It should be noted that ML estimation of polychoric correlation was originally believed to be fairly robust against distributional misspecification of latent normality \citep[e.g.,][]{li2016,coenders1997,flora2004}, an assessment based on simulation experiments using the method of \citet{vale1983} for generating nonnormal bivariate data that are then discretized to be ordinal. However, \citet{gronneberg2019} show that ordinal data generated in this way are indistinguishable from ordinal data generated by latent normality. Using a method that ensures proper violation of latent normality \citep[the VITA method of][]{gronneberg2017vita} reveals strong susceptibility of ML estimation of polychoric correlation to such violations \citep[e.g.,][]{foldnes2020polycor,foldnes2022}.}
\citet{welz2025polycor} are concerned with partial misspecification of the polychoric correlation model, where latent normality is violated due to a (possibly empty) subset of data points that were generated by an unspecified and unknown nonnormal process. We use the same (partial) misspecification framework in this paper. \citet{welz2025polycor} show that having in a sample already about~5\% of such observations, commonly referred to as \emph{contamination}, can suffice for a substantial estimation bias. Contamination might arise due to, for instance but not limited to, careless responding to polytomous items, which has been identified as a major threat to the validity of psychometric analyses \citep[e.g.,][and references therein]{alfons2024spc,ward2023,arias2020,bowling2016,huang2015ier,meade2012}. 
As a remedy, \citet{welz2025polycor} propose a fully efficient contamination-robust estimator of the polychoric correlation model that makes no assumptions on the prevalence and type of contamination (which is possibly absent altogether). Their estimator exploits the theory of~$C$-estimation \citep{welz2024robcat}, being a general framework of robust estimation with categorical data. 

In the context of item response theory (IRT), \citet{itaya2025} focus on the seminal model by \citet{rasch1960}, whose estimation may be compromised by contamination due to, for instance, careless responding or random guessing. They therefore robustify marginal ML estimation thereof by using a minimum \emph{density power divergence} (DPD) estimator \citep{basu1998} and design a majorization–minimization algorithm for this purpose. In this paper, we also utilize DPD estimation theory to robustify the (joint) ML estimation of the polyserial correlation model.

\section{Polyserial correlation} \label{sec:polyserial-model}
%

This section defines (point) polyserial correlation and then reviews ML estimation thereof.

\subsection{The polyserial correlation model} \label{sec:model}
Suppose we observe a continuous real-valued random variable~$X$ with unknown population mean~$\mu = \mu_X \in\R$ and unknown population variance $\ssigma = \ssigma_X > 0$. In addition, suppose that we also observe a polytomous ordinal random variable~$Y$ that takes values in some finite set~$\Y$ of known cardinality~$r$. Without loss of generality, we assume throughout this paper that $\Y = \{1,2,\dots, r\}$. Further assume that there exists a latent continuous random variable~$\eta$ governing the ordinal variable~$Y$ through the unobserved discretization process
\begin{equation}
Y =
\begin{cases}
1 & \textnormal{ if } \eta < \tau_1, \\
2 & \textnormal{ if } \tau_1 \leq \eta < \tau_2,\\ 
3 & \textnormal{ if } \tau_2 \leq \eta < \tau_3,\\
\vdots & \\ 
r & \textnormal{ if } \tau_{r-1} \leq \eta,
\end{cases}
\label{eq:discretization}
\end{equation}
where $-\infty < \tau_1 < \tau_2 < \cdots < \tau_{r-1} < +\infty$ are fixed but unknown \emph{threshold} parameters. In practice,~$Y$ often denotes the responses to a Likert-type rating item with~$r$  response categories.

The primary object of interest is the fixed but unknown population correlation between the observed~$X$ and the latent~$\eta$, denoted by
\[
	\rho = \cor{X}{\eta}.
\]
To identify the correlation coefficient~$\rho$, one usually assumes that~$X$ and~$\eta$ are jointly normally distributed according to
\begin{equation}
	\begin{pmatrix}
	X \\ \eta
	\end{pmatrix}
	\sim
	\gauss_2
	\left(
	\begin{pmatrix}
	\mu \\ 0
	\end{pmatrix}
	,
	\begin{pmatrix}
	\ssigma & \rho\sigma \\ \rho\sigma & 1
	\end{pmatrix}
	\right).
\label{eq:normality}
\end{equation}
The bivariate normality model~\eqref{eq:normality} implies the marginal normality properties $X\sim \gauss(\mu, \sigma^2)$ and $\eta\sim\gauss(0,1)$. It also identifies the correlation coefficient $\rho\in (-1,1)$ through the familiar identity $\cor{X}{\eta} = \cov{X}{\eta}\big/\sqrt{\var{X}\var{\eta}} = \rho$. Note that since the variable~$\eta$ is unobserved, its population mean and variance are not jointly identifiable, which is why they are fixed to~0 and~1, respectively.

Combining the discretization model~\eqref{eq:discretization} with the bivariate normality model~\eqref{eq:normality} yields the \emph{polyserial correlation model} \citep{pearson1913}, or, in short, polyserial model. In this model, the correlation parameter $\rho = \cor{X}{\eta}$ is referred to as the \emph{polyserial correlation coefficient}. If~$Y$ is dichotomous, the polyserial model reduces to the \emph{biserial correlation model} of \citet{pearson1909}. The theoretical properties of biserial correlation are studied by \citet{tate1955jasa,tate1955biometrika} as well as \citet{jaspen1946}, and those of polyserial correlation by \citet{olsson1982}. 

The polyserial model is subject to $d = r + 2$ parameters, namely the mean and variance parameters $(\mu, \ssigma)$ of the observed~$X$, the polyserial correlation coefficient~$\rho$ from the normality model~\eqref{eq:normality}, as well as the $r-1$ thresholds from the discretization process~\eqref{eq:discretization} of the latent~$\eta$. These parameters are jointly collected in a~$d$-dimensional parameter vector 
\[
	\Btheta = \left(\rho, \mu, \ssigma, \Btau^\top \right)^\top,
\]
where the vector $\Btau = (\tau_1,\dots, \tau_{r-1})^\top$ contains the $r-1$ thresholds.

Under the polyserial model evaluated at a parameter vector~$\Btheta\in\BTheta$, the density of the observed-latent pair $(X,\eta)$ of continuous variables is given by
\[
	\pxetafun{x,v}{\Btheta}{} = 
	\phi_2
	\left(
	\begin{pmatrix}
	x \\ v
	\end{pmatrix}
	;
	\begin{pmatrix}
	\mu \\ 0
	\end{pmatrix}
	,
	\begin{pmatrix}
	\ssigma & \rho\sigma \\ \rho\sigma & 1
	\end{pmatrix}
	\right),
	\qquad x,v\in\R,
\]
where $\phi_2(\cdot; \vec{m}, \mat{S})$ is the density of the bivariate normal distribution with population mean~$\vec{m}$ and covariance matrix~$\mat{S}$. Further denote by $\Pxetafun{\cdot,\cdot}{\Btheta}$ the distribution function corresponding to the normal density $\pxetafun{\cdot,\cdot}{\Btheta}{}$.

For the observed variables $(X,Y)$, the joint density at a realization $x\in\R$ and a response $y\in\Y = \{1,\dots, r\}$ of the ordinal~$Y$ under the polyserial model at parameter~$\Btheta$ reads
\begin{equation}\label{eq:jointpdf}
	\pxy{\Btheta} = \int_{\tau_{y-1}}^{\tau_y} \pxetafun{x,v}{\Btheta}{} \d v,
\end{equation}
where we adopt the conventions $\tau_0 = -\infty$ and $\tau_r = +\infty$.\footnote{\edit{Being a random vector comprising a continuous and discrete variable, $(X,Y)$ technically does not have a joint density in a regular sense, i.e., a Randon-Nikod{\'y}m derivative with respect to the two-dimensional Lebesgue measure. The formally correct way to refer to~$p_{XY}$ in~\eqref{eq:jointpdf}  would be that it is the Randon-Nikod{\'y}m derivative of the distribution~$P_{XY}$ in~\eqref{eq:jointcdf} with respect to the product of the one-dimensional Lebesgue measure and the counting measure. For the sake of brevity, however, we refer to~$p_{XY}$ simply as joint density in this paper, acknowledging an abuse of terminology.}}
The joint distribution function of the observed~$(X,Y)$ under the polyserial model can now be expressed as
\begin{equation}\label{eq:jointcdf}
	\Pxy{\Btheta} = \Pr{\Btheta}{X \leq x, Y \leq y} 
	= \int_{-\infty}^x \sum_{w\leq y} \pxyfun{u,w}{\Btheta}{} \d u
	=
	\int_{-\infty}^{x}\int_{-\infty}^{\tau_y} \pxetafun{u,v}{\Btheta}{} \d v\d u ,
\end{equation}
where the third equality follows from~\eqref{eq:jointpdf} in conjunction with the linearity of the integral operator. As such,~$\Pxy{\Btheta}$ is the distribution function associated with  density~$\pxy{\Btheta}$, so we refer to it as the \emph{polyserial model distribution}.

Our expressions for the polyserial model distribution and density are different but equivalent to the more commonly used expressions in \citet{olsson1982}, which are provided in Appendix~\ref{app:olsson1982notation}.

\subsection{Point polyserial correlation}\label{sec:pointpolyserial}
In addition to the correlation between the observed~$X$ and the latent~$\eta$, one might also be interested in the correlation between~$X$ and the observed ordinal~$Y$. The correlation between the observed~$X$ and~$Y$ is known as \emph{point polyserial correlation}. In order to identify the desired point polyserial correlation $\cor{X}{Y}$, one needs to assign a numerical interpretation to the~$r$ answer categories of~$Y$, that is, introduce a \emph{scoring system}. \edit{Given a scoring system, the point polyserial correlation coefficient $\tilde{\rho} = \cor{X}{Y}$ is identified by the polyserial model and can be estimated by using estimates of the polyserial model parameters~$\Btheta$. We provide details in Appendix~\ref{app:pointpolyserial}.}

\subsection{Maximum likelihood estimation}\label{sec:MLE}
Suppose we observe a sample $\{(X_i, Y_i)\}_{i=1}^N$ of~$N$ independent copies of $(X,Y)$ generated by the polyserial model at some \emph{true} parameter vector~$\Btheta_* = \left(\rho_*, \mu_*, \ssigma_*, \Btau_*^\top\right)^\top$. The statistical problem is to estimate the true~$\Btheta_*$ from the observed sample, which is traditionally achieved by the maximum likelihood (ML) estimator proposed by \citet{cox1974} and \citet{olsson1982}. 

The ML estimator (MLE) of $\Btheta_*$ is defined as the log-likelihood maximizer
\begin{equation}\label{eq:MLE}
	\thetahatMLE 
	= 
	\arg\max_{\Btheta\in\BTheta} \left\{\sum_{i=1}^N \log\big( \pxyfun{X_i, Y_i}{\Btheta}{} \big)\right\},
\end{equation}
where the parameter space 
\begin{equation}\label{eq:Theta}
	\BTheta = 
	\left\{
	 \left(\rho, \mu, \ssigma, \Btau^\top\right)^\top\ \Big|\ 
	 \rho\in (-1,1),\ \mu\in\R,\ \sigma > 0,\ -\infty < \tau_1 < \cdots < \tau_{r-1} < +\infty 
	\right\}
\end{equation}
is the set of legal parameters~$\Btheta$ the MLE maximizes over. As such,~$\BTheta$ rules out degenerate cases such as $\rho = \pm 1$, non-positive standard deviations, or non-monotonic thresholds.  
Assuming that the polyserial model is correctly specified for the data at hand, \citet{cox1974} and \citet{olsson1982} show that the MLE is consistent for the true~$\Btheta_*$, asymptotically normally distributed, and fully efficient.

As a computationally attractive alternative to ML estimation, \citet{olsson1982} propose a two-step (TS) estimation procedure. In the first step, one computes as estimators of~$\mu$,~$\sigma^2$, and $\tau_k,k=1,\dots,r-1,$ the sample statistics
\begin{equation}\label{eq:twostep}
	\hat{\mu}_{\textrm{TS}} = \frac{1}{N}\sum_{i=1}^N X_i, \quad 
	\hat{\sigma}^2_{\textrm{TS}} = \frac{1}{N-1} \sum_{i=1}^N \left(X_i - \hat{\mu}_{\textrm{TS}}\right)^2, \quad 
	\hat{\tau}_{k,\textrm{TS}} = \Phi^{-1}\left(\frac{1}{N}\sum_{j=1}^k N_{Y,j}  \right),
\end{equation}
respectively, where $N_{Y,j} = \sum_{i=1}^N \I{Y_i = j}$ denotes the empirical marginal frequency of the~$j$-th response option of~$Y$. In the second step, one substitutes for these sample statistics in the log-likelihood in~\eqref{eq:MLE} and maximizes the ensuing log-likelihood with respect to the remaining parameter,~$\rho$, via conditional maximum likelihood (conditional on the sample statistics). The main advantage of the two-step approach is reduced computing time because one needs to numerically solve the maximization problem~\eqref{eq:MLE} only with respect to one parameter,~$\rho$, rather than all~$d$ parameters in~$\Btheta$. A drawback is diminished efficiency as compared to (joint) ML where all~$d$ parameters are estimated simultaneously. By means of simulation experiments, \citet{olsson1982} find that if the polyserial model is correctly specified, ML and the two-step estimator produce similar results, with differences decreasing with an increasing number of response options for~$Y$. The two-step approach is the default estimator for polyserial correlation in the SEM software package \pkg{lavaan} \citep{lavaan}.

\edit{While it is in principle possible to robustify the two-step approach approach against contamination by using the same estimation theory as in this paper, doing so would result in  substantial theoretical and computational drawbacks. We discuss this in detail in Appendix~\ref{app:twostep}.}

Alternative estimators \edit{of polyserial correlation} have been proposed by \citet{bedrick1996}, \citet{lord1963}, and \citet{brogden1949}, with \citet{bedrick1992,bedrick1990}, \citet{koopman1983} and \citet{kraemer1981} studying the theoretical properties of the latter two. We discuss these approaches in more detail in Section~\ref{sec:distributional-subsection}, after conceptualizing misspecification of the polyserial model.

\section{Model misspecification}\label{sec:misspecification}

In order to study the effects of partial model misspecification, we first define this concept and then explain how it differs from distributional misspecification.

\subsection{Partial misspecification of the polyserial model}

The polyserial model is misspecified if at least one observation for the continuous-ordinal variable pair~$(X,Y)$ has not been generated by the bivariate partially-latent normality model in~\eqref{eq:normality}. Akin to \citet{welz2025polycor}, we consider a \emph{partial misspecification} framework where only a fraction $(1-\varepsilon)$ of observations in a given sample are generated by an underlying normal distribution~$\Pxetasymb$ with true parameter~$\Btheta_*$, whereas a fixed but unknown fraction~$\varepsilon$ of observations are generated from some different but unspecified underlying distribution~$\Hxeta$.\footnote{Being a population quantity and not a sample quantity, it is strictly speaking imprecise to call~$\varepsilon$ a fraction. It would be more accurate to call~$\varepsilon$ a (contamination) population probability with which one samples from the contamination distribution~$\Hxeta$ instead of the model distribution~$\Pxetasymb$. However, to be consistent with robust statistics literature, we refer to~$\varepsilon$ as a fraction throughout this paper.} Since~$\Hxeta$ is unspecified, its correlation structure may differ from~$\Pxetasymb$ so that observations~$(X,Y)$ generated by the underlying~$\Hxeta$ (after discretization) may be uninformative for the true polyserial correlation coefficient. 

Formally, the polyserial model is said to be \emph{partially} misspecified if the unknown sampling distribution of the observed-latent variable pair $(X,\eta)$ is given by
\begin{equation}\label{eq:contamdist}
	\Fxetafun{x,v} = (1-\varepsilon) \Pxetafun{x,v}{\Btheta_*} + \varepsilon \Hxetafun{x,v},
\end{equation}
for $x,v\in\R$. Correspondingly, the implied unknown sampling distribution of the observed variables~$(X,Y)$ reads
\begin{equation}\label{eq:contamdistXY}
	\Fxyfun{x,y} = (1-\varepsilon) \Pxyfun{x,y}{\Btheta_*} + \varepsilon \Hxyfun{x,y},
\end{equation}
for $x\in\R, y\in\Y$, where
\[
	\Hxyfun{x,y} = \int_{-\infty}^{\tau_{\varepsilon,y}} \Hxetafun{x,\d v}
\]
is the distribution of the observations for which the polyserial model is (partially) misspecified. The unknown and unspecified discretization thresholds $-\infty = \tau_{\varepsilon,0} <  \tau_{\varepsilon,1}<\dots < \tau_{\varepsilon,r-1}<  \tau_{\varepsilon,r} = +\infty$ need not equal the true discretization thresholds   $\tau_{*,1} < \dots < \tau_{*,r-1}$ of the polyserial model. Furthermore, denote by~$\fxy$ the unknown density corresponding to the sampling distribution~$\Fxy$. 

Misspecification models of the type in~\eqref{eq:contamdist} are standard in the robust statistics literature, where they are known as \emph{Huber contamination models}, owing to pioneering work of \citet{huber1964}. Following \citet{welz2025polycor}, we therefore adopt terminology from robust statistics and call~$\varepsilon$ the \emph{contamination fraction}, the uninformative~$\Hxeta$ the \emph{contamination distribution} (or simply \emph{contamination}), and~$\Fxeta$ the \emph{contaminated} distribution.  In the case of a zero-valued contamination fraction ($\varepsilon = 0$), there is no misspecification so that the polyserial model is correctly specified. Indeed, if $\varepsilon = 0$, then $F_{0,X\eta}(\cdot,\cdot) = \Pxetafun{\cdot,\cdot}{\Btheta_*}$ and  $F_{0,XY}(\cdot,\cdot) = \Pxyfun{\cdot,\cdot}{\Btheta_*}$, so the partial misspecification framework in~\eqref{eq:contamdist} nests the correctly specified polyserial model.

Neither the contamination fraction~$\varepsilon$ nor the contamination distribution~$\Hxeta$ in~\eqref{eq:contamdist} are assumed to be known. Thus, both quantities are left completely unspecified in practice, which \emph{``means that we are not making any assumptions on the degree, magnitude, or type of contamination (which is possibly absent altogether)''} \citep{welz2025polycor}. Hence,  the polyserial model may be misspecified due to an unlimited variety of reasons, for instance but not limited to outliers or nonnormality in the continuous~$X$ , and/or careless responding or item misunderstanding in the ordinal~$Y$. 
Because~$\varepsilon$ and~$\Hxeta$ are unspecified in practice, the polyserial model distribution~$\Pxyfun{\cdot,\cdot}{\Btheta_*}$ remains the distribution of interest: Our only aim is to estimate the parameter~$\Btheta_*$ of the polyserial model while reducing the impact of potential contamination in the observed data. Consequently, the contaminated distributions~$\Fxy$ and~$\Fxeta$ are never estimated. They are purely theoretical objects to study the  properties of estimators of the polyserial model when that model is partially misspecified due to data contamination. 

While no assumption is made on the specific value of the contamination fraction~$\varepsilon$, we impose the restriction $\varepsilon\in [0,0.5)$, which is common in robust statistics \citep[e.g.,][p.~67]{hampel1986}. This restriction ensures proper identification: If half or more of the data points were contaminated, it would no longer be possible to distinguish between contamination and observations generated by the polyserial model, at least not without further assumptions. Under such additional assumptions, also values of~$\varepsilon \geq 0.5$ could be considered. We refer to \citet{welz2025polycor} for a more detailed discussion.

\begin{figure}[t]
	\centering
	\includegraphics[width = 0.8\textwidth]{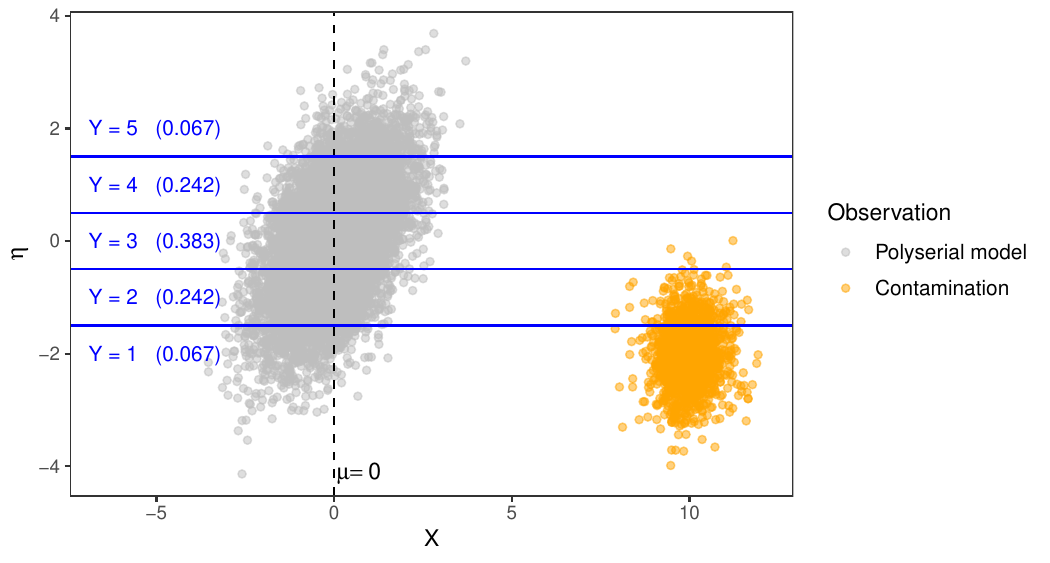}
\caption{Simulated data where the polyserial model is misspecified for a fraction $\varepsilon = 0.15$ of the $N=10,000$ points. The gray dots are draws of $(X,\eta)$ from the polyserial model with true parameters $\rho = 0.5, \mu = 0, \sigma^2 = 1$, while the orange dots are draws from a contamination distribution~$\Hxeta$,  being a bivariate $t$-distribution here with noncentrality parameter $(10,-2)^\top$, scale matrix $\textnormal{diag}(0.25, 0.25)$, and~10 degrees of freedom. The horizontal lines mark the thresholds that discretize the latent~$\eta$ to the observed ordinal~$Y$ with five response options. The numbers in parentheses indicate the population marginal probability of the respective response option under the true polyserial model.}
\label{fig:main-simdesign}
\end{figure}

Figure~\ref{fig:main-simdesign} visualizes a simulated example data set generated by the contaminated distribution~$\Fxeta$ in~\eqref{eq:contamdist} with contamination fraction $\varepsilon = 0.15$. In this example, the contamination distribution~$\Hxeta$, whose random draws are orange dots, is a shifted bivariate $t$-distribution with noncentrality parameter set to $(10,-2)^\top$, scale matrix $\textnormal{diag}(0.25, 0.25)$, and~10 degrees of freedom. The remaining realizations (gray dots) are generated by the bivariate normal distribution~$\Pxetasymb$ with true parameters $\mu_* = 0$, $\sigma_*^2 = 1$, and $\rho_* = 0.5$. Here, contamination manifests through somewhat larger values in the~$X$-dimension and inflation of the first two response options in the~$Y$-dimension (after discretization), resulting in a (partially) misspecified polyserial model. Applying the ML estimator in~\eqref{eq:MLE} to the observed data for~$(X,Y)$ in Figure~\ref{fig:main-simdesign} yields a sign-flipped polyserial correlation estimate of $-0.522$, representing a substantial bias with respect to the true value of $\rho_* = 0.5$. 
In contrast, the robust estimator---which will be introduced in Section~\ref{sec:estimator} and uses the exact same information as ML---estimates a correlation of~$0.498$, which is accurate for the true value of $\rho_* = 0.5$.

\edit{
	We stress that there exist nonnormal joint distributions of $(X,\eta)$ that, after discretizing the latent~$\eta$ variable with fixed thresholds, result in the same density for the observed $(X,Y)$ as if the partially-latent $(X,\eta)$ were jointly normal. This phenomenon is known as \emph{discretize equivalence} of latent variables \citep{foldnes2019identification}. It follows that there exist contamination distributions~$\Hxeta$ and contamination fractions $\varepsilon > 0$ in~\eqref{eq:contamdist} under which the contaminated density~$\fxy$ of the observed $(X,Y)$ is exactly equal to the polyserial model density in~\eqref{eq:jointpdf} at the true parameter, that is, $\fxy (x,y) = \pxy{\Btheta_*}$ for all $(x,y)\in\R\times\Y$. Hence, in this scenario, the polyserial model is misspecified, but the misspecification does not have adverse effects because the true polyserial model density of $(X,Y)$ remains unaffected. To avoid cumbersome notation in our analysis, we follow \citet{welz2025polycor} and assume \emph{consequential} misspecification throughout this article, that is, $\fxy (x,y) \neq \pxy{\Btheta_*}$ for at least one response $y\in\Y$ whenever $\varepsilon > 0$, given $x\in\R$. Nevertheless, \emph{``it is silently understood that misspecification need not be consequential''} \citep{welz2025polycor}, in which case there is no practical problem and both ML and robust estimator are consistent for the true~$\Btheta_*$.
}

Before we define our robust estimator, we briefly juxtapose the partial misspecification framework to distributional misspecification.

\subsection{Distributional misspecification}\label{sec:distributional-subsection}
The polyserial model is said to be \emph{distributionally} misspecified if it is misspecified for \emph{all} observations in a sample. This contrasts the partial misspecification framework in~\eqref{eq:contamdist} where the model is only misspecified for a (possibly zero-valued) fraction of the observations. Let~$G = G_{X,\eta}$ denote the unknown joint distribution of~$X$ and the latent~$\eta$, which, under distributional misspecification, is nonnormal for all data points. In distributional misspecification, the object of interest is the correlation coefficient between~$X$ and~$\eta$ under the distribution~$G$, denoted $\rho_G = \corF{G}{X}{\eta}$, instead of the correlation coefficient under bivariate normality (which would be the polyserial correlation coefficient). Although our robust estimator is designed for partial misspecification rather than distributional misspecification, it can in certain situations also offer a robustness gain under distributional misspecification. We discuss this in more detail in Section~\ref{sec:additional-simulations}.

\section{Robust estimation of polyserial correlation}\label{sec:estimator}
It is well-known that ML estimation is highly susceptible to partial model misspecification due to data contamination \citep[e.g.,][]{huber2009,hampel1986,huber1964}, so it might be desirable to construct alternative estimators that are more robust against contamination. 
However, attempting to robustify estimation of the polyserial model against contamination bears the inherent challenge of how to handle the mixed-type structure of the data with one continuous and one ordinal variable. 

While there exist general frameworks for robust estimation with exclusively continuous variables and exclusively categorical variables, none of them are applicable to mixed variables. On the one hand, robust~$M$-estimation \edit{as proposed by \citet{huber1964}}  and related methods \citep[e.g.,][]{huber2009} are intended for continuous random variables, and, broadly speaking, achieve robustness by downweighting observations with extreme values. For instance, \citet{alfons2022} use a variation of~$M$-estimation,~$MM$-estimation, to robustify mediation analyses against outliers, heavy-tailedness, or skewness of the observed distribution. Such estimators are not applicable to mixed data because the ordinal variable cannot take extreme values due to being categorical, nor does it admit a direct numerical interpretation to begin with. On the other hand, the theory of robust~$C$-estimation \citep{welz2024robcat} is designed for exclusively  categorical variables. $C$-estimation downweighs categories whose empirical frequency disagrees with their corresponding theoretical frequency under a postulated model. Since one variable in the polyserial model is continuous, it does not have discrete categories, so~$C$-estimators cannot be applied. However, it turns out that the fundamental idea of downweighting data points that cannot be modeled sufficiently well by a postulated model can be exploited to achieve robustness through minimum \emph{density power divergence} (DPD) estimation \citep{basu1998}. We explain in this section how minimum DPD estimation can be applied to the polyserial model.

For further reference, define the Kullback-Leibler (KL) divergence \citep{kullback1951} between two bivariate densities~$g_1$ and~$g_2$, each defined on~$\R^2$ (or a common subset thereof), by
\begin{equation}\label{eq:KLgeneral}
	\KL{g_1}{g_2}
	=
	\int\int 
	g_1(s,t) \log\left( \frac{g_1(s,t)}{g_2(s,t)} \right) \d t \d s,
\end{equation}
where the integrals are taken over the densities' domain. 
If the second dimension of the densities corresponds to a discrete random variable, that is, the first variable is continuous, but the second one is discrete, replace the inner integral in~\eqref{eq:KLgeneral} by a summation over the second variable's domain.\footnote{To explicitly accommodate discrete distributions in the definition of the KL divergence, we could have adapted the Riemann-integral-based definition in~\eqref{eq:KLgeneral} to a definition based on Lebesgue integration with respect to a general measure that can either be the Lebesgue measure (for continuous distributions) or the counting measure (for discrete distributions). However, we believe that is more instructive to refrain from introducing measure theoretic concepts in this paper.}

Throughout this section, suppose that one has access to a random sample of~$N$ independent continuous-ordinal variable pairs, $(X_i,Y_i), i = 1,\dots, N$, following the unknown sampling distribution~$\Fxy$ in~\eqref{eq:contamdistXY}. Hence, the polyserial model is possibly misspecified for an unknown fraction~$\varepsilon$ of the observed sample.

\subsection{Maximum likelihood revisited}\label{sec:MLErevisited}

The observed sample can be uniquely characterized by a particular empirical density function,
\begin{equation}\label{eq:fhat}
	\fhatxy{x,y} = \frac{1}{N}\sum_{i=1}^N\I{X_i = x, Y_i = y},
\end{equation}
for $x\in\R, y\in\Y$, where the indicator function~$\I{E}$ takes value~1 if an event~$E$ is true, and~0 otherwise. \edit{As such,} \edit{t}he empirical density~$\fhat$ only takes nonzero values when evaluated at points in the observed sample~$\{(X_i,Y_i)\}_{i=1}^N$. 
\edit{T}he maximum likelihood estimator~$\thetahatMLE$ in~\eqref{eq:MLE} can be expressed as the minimizer of the KL divergence between the empirical density~$\fhat$ and model density~$\pxyempty$, which is given by
\begin{equation}\label{eq:KL}
	\KL{\fhat}{\pxyempty}
	=
	\frac{1}{N}\sum_{i=1}^N\Bigg( \log\left(\fhatxy{X_i,Y_i} \right) - \log\left( \pxyfun{X_i,Y_i}{\Btheta}{}\right) \Bigg),
\end{equation}
\edit{and} follows by definition of~$\fhat$ and the KL divergence in~\eqref{eq:KLgeneral} \edit{as well as} the convention $0\log(0)=0$.\footnote{\edit{Alternatively, the ML estimator (and also the proposed robust estimator) can be equivalently defined using statistical divergence functionals, as done in \citet{basu1998}.}}  
We refer to \citet{white1982} for a detailed exposition of the connection between KL divergence and ML.

As alluded to earlier, ML estimation tends to be highly susceptible to contamination in the observed sample. It may therefore be desirable to consider alternatives that are designed to more robust against contamination. It turns out that the idea of minimizing a given divergence between the empirical density~$\fhat$ and the model density~$\pxyempty$ can be exploited to construct contamination-robust estimators by choosing alternative divergences to the KL divergence. This is exactly what \emph{minimum power divergence} estimation \citep{basu1998} does.

\subsection{Minimum power divergence}
\citet{basu1998} propose a class of divergences between two densities that generalizes the KL divergence, but is less affected by data contamination. Suppose~$g_1$ and~$g_2$ are bivariate densities supported on~$\R^2$ or a common subset thereof. For a fixed tuning constant $\alpha > 0$, the divergence of \citet{basu1998} between~$g_1$ and~$g_2$ is defined by 
\begin{equation*}
\divergence{g_1}{g_2}
=
	\int\int
	\Bigg(
		g_2^{1+\alpha}(s,t) - \left(1+\frac{1}{\alpha}\right)g_1(s,t)g_2^\alpha(s,t) + \frac{1}{\alpha}g_1^{1+\alpha}(s,t)
	\Bigg)\d t\d s,
\end{equation*}
where each integral is taken over the densities' domain.  
Note that the divergence~$\divsymb_\alpha$ is not defined at $\alpha = 0$. To overcome this issue, \citet{basu1998} define~$\divsymb_0$ as the limit of $\divsymb_\alpha$ as $\alpha\downarrow 0$, that is,
\begin{equation}\label{eq:D0}
	\divergencenull{g_1}{g_2}
	=
	\lim_{\alpha\downarrow 0}\divergence{g_1}{g_2}
	=
	\int\int 
	g_1(s,t) \log\left( \frac{g_1(s,t)}{g_2(s,t)} \right) \d t \d s
	= \KL{g_1}{g_2},
\end{equation}
where the second equality follows from the identity $z\mapsto \log z = \lim_{\alpha\downarrow 0} \alpha^{-1}(z^\alpha - 1)$ and the third equality from the definition of the KL divergence in~\eqref{eq:KLgeneral}. \citet{basu1998} call the divergence $\divsymb_\alpha, \alpha\geq 0$, the \emph{density power divergence} (DPD). It follows that the DPD generalizes the KL divergence. We explain in a moment the intuition behind the DPD.

As before, if the second dimension of the two densities~$g_1, g_2$ corresponds to a discrete random variable, that is, the first variable in~$g_j$ is continuous, but the second one is discrete, replace the inner integral in the DPD $\divsymb_\alpha, \alpha \geq 0$, by a summation over the second variable's domain.

\subsection{Proposed robust estimator}\label{sec:proposethetahat}
For $\alpha > 0$, the DPD between the empirical density~$\fhat$ in~\eqref{eq:fhat} and the polyserial model's density~$\pxyempty$ reads
\begin{equation}\label{eq:DPD}
\begin{split}
	\divergence{\fhat}{\pxyempty}
	=
	\int_\R \sum_{y\in\Y}\pxyfun{x,y}{\Btheta}{1+\alpha}\d x
	-
	(1+\alpha^{-1}) \frac{1}{N}\sum_{i=1}^N \pxyfun{X_i,Y_i}{\Btheta}{\alpha} + \alpha^{-1},
\end{split}
\end{equation}
which follows from writing out~$\fhat$. For the choice $\alpha = 0$, the DPD $\divergencenull{\fhat}{\pxyempty}$ reduces to the KL divergence in~\eqref{eq:KL} by~\eqref{eq:D0}. 

For a pre-specified tuning constant $\alpha \geq 0$, our proposed estimator minimizes the divergence~$\divsymb_\alpha$  between~$\fhat$ and~$\pxyempty$ with respect to the model parameter~$\Btheta\in\BTheta$. Specifically, our proposed estimator is given by
\begin{equation}\label{eq:estimator}
	\thetahat = \arg\min_{\Btheta\in\BTheta}\divergence{\fhat}{\pxyempty}.
\end{equation}
In particular, if $\alpha = 0$, the estimator~$\thetahat$ coincides with the MLE~$\thetahatMLE$ (see Section~\ref{sec:MLErevisited}). Since it minimizes a DPD, estimator~$\thetahat$ constitutes a \emph{minimum density power divergence estimator} \citep{basu1998}. As such, a limit theory for~$\thetahat$ is readily available from \citet{basu1998}, who derive the theoretical properties of such estimators for general models. Before we turn to the estimator's limit theory, we first provide some intuition as to \emph{why} minimum DPD estimators are more robust than ML whenever $\alpha > 0$.

An estimator defined as the minimum of a (differentiable) loss function can be equivalently characterized as a root of the loss' gradient; a characterization known as \emph{estimating equation}. The estimating equation of the MLE $(\alpha = 0)$ can be written as
\[
	\frac{1}{N} \sum_{i=1}^N \scorefun{\Btheta}{X_i,Y_i} = \vec{0},
\]
which is satisfied for $\Btheta = \thetahatMLE$, and where the $d$-vector
\[
	\scorexy{\Btheta} = \partialderivative{}{\Btheta}\log\pxy{\Btheta}, \qquad x\in\R,y\in\Y,
\]
denotes the log-likelihood \emph{score} function. A closed-form expression of~$\scorexy{\Btheta}$ is provided in Appendix~\ref{app:asymptotics}. Analogously, the estimating equation of the proposed estimator~$\thetahat$ in ~\eqref{eq:estimator} can be shown to read
\begin{equation}\label{eq:estimatingeq}
	\frac{1}{N}\sum_{i=1}^N \pxyfun{X_i, Y_i}{\Btheta}{\alpha} \scorefun{\Btheta}{X_i, Y_i}
	-
	\vec{c}_\alpha(\Btheta)
	=
	\vec{0},
\end{equation}
which is satisfied for $\Btheta = \thetahat$ and where 
$
	\vec{c}_\alpha (\Btheta) = \int_\R\sum_{y\in\Y}\pxyfun{x,y}{\Btheta}{1+\alpha}\scorexy{\Btheta}\d x
$ 
is a correction factor independent of observed data.  
Observe that for $\alpha = 0$, the estimating equation~\eqref{eq:estimatingeq} equals that of the MLE,\footnote{By elementary likelihood theory, $\vec{c}_0 (\Btheta) = \int_\R\sum_{y\in\Y}\pxyfun{x,y}{\Btheta}{}\scorexy{\Btheta}\d x = \Eoperator_{\Pxysymb}\left[\scorefun{\Btheta}{X,Y}\right] = \vec{0}$ for all $\Btheta\in\BTheta$.} in which all observations in the sample $\{(X_i, Y_i)\}_{i=1}^N$ are weighted equally. Conversely, when $\alpha > 0$, the term
\[
	\frac{1}{N}\sum_{i=1}^N \pxyfun{X_i, Y_i}{\Btheta}{\alpha} \scorefun{\Btheta}{X_i, Y_i}
\]
in estimating equation~\eqref{eq:estimatingeq} \emph{``provides a relative-to-the-model downweighting for [contaminated]\footnote{The original quote of \citet[][p.~551]{basu1998} refers to contaminated observations as \emph{outlying} observations. In their terminology, an outlier is any data point that was not generated by the postulated model. However, since one often associates outliers with extreme values rather than discordant model fit, we believe that the term ``outlying'' is potentially misleading, so we prefer ``contaminated''.} observations''} \citep[][p.~551]{basu1998} with individual-specific weights
\begin{equation}\label{eq:weights-raw}
		\tilde{w}_{i,\alpha}(\Btheta) = \pxyfun{X_i, Y_i}{\Btheta}{\alpha}.
\end{equation}
A weight $\tilde{w}_{i,\alpha}(\Btheta)$ is bounded from below by~0 and takes values close to~0 for observations that the polyserial model cannot fit well, such as contaminated data points. As such, \emph{``observations that are widely discrepant to the [polyserial] model will get nearly zero weights''} \citep[][pp.~551--52]{basu1998}. Larger choices of $\alpha > 0$ lead to more stringent downweighting. 
\edit{For instance, a contaminated observation with an ordinal response~$Y_i$ that strongly disagrees with the latent threshold structure implied by the model density~$p_{XY}$ would receive a weight close to~0.}

\edit{I}f contamination is absent $(\varepsilon = 0)$, it can be shown that  the robust estimator~$\thetahat$ and the MLE~$\thetahatMLE$ in~\eqref{eq:MLE} are asymptotically equal to one another for all choices $\alpha \geq 0$. This property is implied by the property of \emph{Fisher consistency} that minimum DPD estimators posses \citep[implied by Theorem~1 in][see p.~551 therein]{basu1998}. 

\edit{Being expressible as a root of a summation over the sample where each summand is a certain function evaluated at a single observation (Eq.~\ref{eq:estimatingeq}), minimum DPD estimators are a special case of $M$-estimation \citep[see][Section~3.1]{basu1998}. However, unlike classic $M$-estimation as proposed in \citet{huber1964}, minimum DPD estimation does not require the data to admit a numerical interpretation due to operationalizing ``outlyingness'' through discrepant model fit rather than extreme values. Consequently, it can be directly applied to models for mixed data, such as the polyserial model.}

\subsection{\edit{Rescaling of weights}}

By construction, for a given tuning constant $\alpha > 0$, the robust estimator's raw weights~$\tilde{w}_{i,\alpha}(\Btheta)$ in~\eqref{eq:weights-raw} are bounded \edit{from below by~0 and} from above by some data-independent finite value~$M_{\alpha}(\Btheta)$ only depending on~$\alpha$ and~$\Btheta$, which is defined by $M_{\alpha}(\Btheta) = \sup\{\pxyfun{x, y}{\Btheta}{\alpha} : x\in\R, y\in\Y\}$. \edit{The fact that the upper bound~$M_{\alpha}(\Btheta)$ varies for different values of~$\Btheta$ and~$\alpha$ makes it difficult to compare raw weights across different estimates and specifications. Since the raw weights are intended as a tool for detecting contaminated observations, it would be beneficial to have such a comparability. In the following, we describe a simple novel procedure to rescale the weights to always be bounded from above by~1. R}escale the raw weights by their upper bound~$M_{\alpha}(\Btheta)$ to obtain
\begin{equation}\label{eq:weights}
		w_{i,\alpha}(\Btheta) = \tilde{w}_{i,\alpha}(\Btheta) \big/ M_{\alpha}(\Btheta), \qquad i = 1,\dots, N.
\end{equation}
\edit{We} henceforth refer to the rescaled weights~$w_{i,\alpha}(\Btheta)$ simply as \emph{weights}. The rescaled weights are contained in the interval~$[0,1]$ and take values close to~0 for poorly fitting observations and value~1 for perfectly fitting observations. This rescaling is without loss of generality because the estimating equation~\eqref{eq:estimatingeq} can equivalently be expressed as
\begin{equation*}
	\frac{1}{N}\sum_{i=1}^N w_{i,\alpha}(\Btheta) \scorefun{\Btheta}{X_i, Y_i}
	-
	\vec{c}_\alpha(\Btheta) / M_{\alpha}(\Btheta)
	=
	\vec{0}.
\end{equation*}
Appendix~\ref{app:algo} describes an algorithm for computing the upper bound~$M_{\alpha}(\Btheta)$. \edit{The usefulness of the rescaled weights as tool for identifying contaminated data points will be demonstrated in an empirical application in Section~\ref{sec:application}, where they successfully detect observations discrepant from the model.}

\section{Statistical and computational properties}\label{sec:properties}
This section is devoted to the properties of minimum DPD estimators for estimating  polyserial models. We first discuss their estimands under contamination, then their asymptotic properties and efficiency properties, and then computational aspects as well as our software implementation.

\subsection{Estimand}\label{sec:estimand}
It is instructive to discuss what the proposed minimum DPD estimator~$\thetahat$ in~\eqref{eq:estimator} actually estimates. Recall that~$\fxy$ denotes the density of the sampling distribution~$\Fxy$ in~\eqref{eq:contamdistXY} that the observed sample follows. Just like~$\Fxy$, the population density~$\fxy$ is completely unspecified and unknown in practice, including contamination fraction~$\varepsilon$. For a fixed tuning constant $\alpha\geq 0$, the estimand~$\Btheta_0$ of estimator~$\thetahat$ is the parameter that minimizes the DPD between the population density and the polyserial model density, that is,
\[
	\Btheta_0 = \arg\min_{\Btheta\in\BTheta}\divergence{\fxy}{\pxyempty}.
\]
Observe that $\divergence{\fxy}{\pxyempty}$ is simply the population analogue of the empirical divergence $\divergence{\fhat}{\pxyempty}$ that is minimized by the estimator~$\thetahat$ in~\eqref{eq:estimator}.

In the absence of contamination ($\varepsilon = 0$), one has that $f_{0,XY}(\cdot,\cdot) = \pxyemptyfun{\Btheta_*}$ by~\eqref{eq:contamdistXY}, and it follows from Theorem~1 in \citet{basu1998} that $\Btheta_0 = \Btheta_*$, so that the population divergence is zero-valued. Subsequently, just like the ML estimator,~$\thetahat$ is unbiased  for the true~$\Btheta_*$ (in population) if the model is correctly specified. This property is known as \emph{Fisher consistency}. We stress that Fisher consistency holds true for \emph{all} choices of $\alpha \geq 0$. In other words, if contamination is absent, then the proposed estimator is unbiased (in population) no matter the choice of tuning constant~$\alpha$.  

On the other hand, if contamination is present $(\varepsilon > 0)$, then the estimand~$\Btheta_0$ will generally differ from the true~$\Btheta_*$. How much they differ depends on the contamination fraction~$\varepsilon$ and contamination type~$\Hxeta$ in the contaminated distribution~\eqref{eq:contamdist}, as well as the choice of tuning constant~$\alpha\geq 0$. Roughly speaking, holding the unknown contaminated distribution~$\Fxeta$ constant, the larger~$\alpha$, the closer~$\Btheta_0$ \edit{tends} to \edit{be to} the true~$\Btheta_*$.\footnote{We stress that this sentence is a heuristic rather than a formal result. Since no assumption is made on the contamination distribution~$\Hxeta$, there may exist contaminated distributions~$\Fxeta$ for which higher values of~$\alpha$ do \emph{not} uniformly reduce bias. We refer to \citet[][Chapter~8.2d]{hampel1986} for a discussion on the fundamental difficulty of deriving bias-reduction results of robust procedures in contamination models such as~\eqref{eq:contamdist}.}
In other words, if the polyserial model is misspecified, larger values of~$\alpha$ asymptotically \edit{tend to} lead to less bias for the true~$\Btheta_*$. 

\subsection{Asymptotic analysis}\label{sec:asymptotics}
We are now ready to study the asymptotic properties of the proposed estimator~$\thetahat$ in~\eqref{eq:estimator}. Under mild standard regularity conditions, it can be shown that~$\thetahat$ is asymptotically consistent for estimand~$\Btheta_0$, i.e.,
\[
	\thetahat\convP\Btheta_0,
\]
as $N\to\infty$, and is furthermore asymptotically Gaussian,
\[
	\sqrt{N} \left( \thetahat - \Btheta_0 \right)
	\convweak
	\gauss_d\Big(\vec{0}, \matfun{\Sigma}{\Btheta_0} \Big),
\]
as $N\to\infty$, where ``$\convP$" and ``$\convweak$" denote convergence in probability and distribution, respectively. These two stochastic convergence results are rigorously established in Theorem~\ref{thm:main} in Appendix~\ref{app:asymptotics}. The theorem follows immediately from a more general result in \citet{basu1998}. 
The asymptotic covariance matrix~$\matfun{\Sigma}{\Btheta_0}$ has a closed-form sandwich-type construction
\begin{equation}\label{eq:sandwich}
	\matfun{\Sigma}{\Btheta_0} = \matinv{J}{\Btheta_0} \matfun{K}{\Btheta_0}\matinv{J}{\Btheta_0},
\end{equation}
where the $d\times d$ full-rank matrices $\Btheta\mapsto \matfun{J}{\Btheta}, \matfun{K}{\Btheta}$ are defined in Appendix~\ref{app:asymptotics}. Neither~$\matfun{J}{\Btheta}$ nor~$\matfun{K}{\Btheta}$ are observed in practice, so neither is~$\matfun{\Sigma}{\Btheta_0}$, but each of these matrices can be consistently estimated, as we will explain momentarily. Owing to this asymptotic normality result, one can construct standard errors and confidence intervals for the estimand~$\Btheta_0$. We stress that the estimator remains asymptotically normal even when the polyserial model is misspecified. Similar results for inference in general misspecified models with sandwich-type covariance constructions have been derived by, e.g., \citet{white1982} and \citet{huber1967} for the MLE, \citet{basu1998} for robust minimum DPD estimators (of which our proposed estimator is a special case), and \citet{welz2024robcat} for a class of robust estimators for general categorical data.

Being a function of the unobserved estimand~$\Btheta_0$ as well as additional unobserved population quantities, the population covariance matrix~$\matfun{\Sigma}{\Btheta_0}$ in~\eqref{eq:sandwich} is also unobserved in practice. Nevertheless, we can consistently estimate it by constructing certain estimators~$\hatmatfun{J}{\Btheta}$ and~$\hatmatfun{K}{\Btheta}$, which are pointwise consistent for~$\matfun{J}{\Btheta}$ and~$\matfun{K}{\Btheta}$, respectively, for a given parameter vector $\Btheta\in\BTheta$. Since both estimators are continuous in~$\Btheta$, it follows from the continuous mapping theorem that the sample analogue of the sandwich-type construction in~\eqref{eq:sandwich},
\[
	\hatmatfun{\Sigma}{\Btheta} = \hatmatinv{J}{\Btheta}  \hatmatfun{K}{\Btheta}\hatmatinv{J}{\Btheta},
\]
is pointwise consistent for~$\matfun{\Sigma}{\Btheta}$. Combining this result with the convergence result $\thetahat\convP\Btheta_0$, the plug-in estimator~$\hatmatfun{\Sigma}{\thetahat}$ is consistent for the asymptotic covariance matrix~$\matfun{\Sigma}{\Btheta_0}$. We refer to Appendix~\ref{app:asymptotics} for the definition of~$\hatmatfun{J}{\Btheta}$ and~$\hatmatfun{K}{\Btheta}$ as well as details.

\subsection{Efficiency}\label{sec:efficiency}

For the tuning constant $\alpha=0$, which corresponds to the MLE~$\thetahatMLE$, it can be shown (see Appendix~\ref{app:covmatcomparison}) that if the polyserial model is correctly specified ($\varepsilon = 0$), the asymptotic covariance matrix in~\eqref{eq:sandwich} reduces to the inverted \emph{Fisher information} of the polychoric model,~$\mat{I}_{\Btheta_0}^{-1}$, where
\[
\fisher{\Btheta}
	=
	\int_{\R} \sum_{y\in\Y} \pxy{\Btheta}\scorexy{\Btheta}\scorexy{\Btheta}^\top\d x,
\]
denotes the Fisher information of the polyserial model at~$\Btheta\in\BTheta$. Hence,~\eqref{eq:sandwich} nests the well-known result that the MLE's asymptotic covariance matrix is equal to the inverted Fisher information matrix. It is furthermore well-known that ML estimation is fully efficient, meaning that no unbiased estimator has a smaller variance \citep[e.g., Theorem~3.10 in][]{lehmann1998point}.

In contrast, for strictly positive tuning constants $\alpha > 0$, the asymptotic covariance matrix of the corresponding estimator~$\thetahat$ in~\eqref{eq:sandwich} is generally \emph{not} equal to the inverse Fisher information matrix at the polyserial model. Thus, recalling the full efficiency property of the MLE, our robust estimator is \emph{less} efficient than the MLE. 
The efficiency loss is due to downweighting of observations with low probability under the polyserial model, which happens even when the model is correctly specified. Consequently, while the robust estimator and MLE are both consistent and unbiased for the true parameter value as long as contamination is absent, the former has a larger estimation variance.

To quantify the efficiency loss, we calculate at the population level the relative efficiency of our robust estimator compared to the MLE when the polyserial model is correctly specified ($\varepsilon = 0$). Recall that in this zero-contamination case, the estimand~$\Btheta_0$ is equal to the true parameter value~$\Btheta_*$ for all $\alpha \geq 0$ due to the property of Fisher consistency. For a given true value~$\Btheta_*$, the \emph{relative efficiency} \citep[e.g.,][Chapter~8.2]{vandervaart1998} of our estimator for the true polyserial correlation~$\rho_*$ with respect to the fully efficient MLE is given by
\[
	\frac{\var{\hatNMLE{\rho}}}{\var{\hatN{\rho}}}
	=
	\left(\mat{I}_{\Btheta_*}^{-1}\right)_{1,1} \Big/ \left(\matfun{\Sigma}{\Btheta_*}\right)_{1,1},
\] 
where the operator $(\cdot)_{1,1}$ picks out the top left element of a matrix. \edit{The efficiency loss of minimum DPD estimators can also be expressed as a function of~$\alpha$ for a given model \citep[see][Section~4.2]{basu1998}.}

\begin{table}
\centering
\begin{tabular}{r | c c c c c c}
$\alpha$ & 0 (MLE) & 0.1 & 0.25 & 0.5 & 0.75 & 1
\\\hline
rel. efficiency & 1.000 & 0.983 & 0.916 & 0.762 & 0.612 & 0.488
\end{tabular}
\caption{Relative efficiency for estimating the polyserial correlation coefficient~$\rho_*$ when~$Y$ has $r=5$ response options, for various choices of the tuning constant~$\alpha$ at a true parameter vector $\Btheta_* = \left(\rho_*, \mu_*, \sigma_*^2, \tau_{*,1}, \tau_{*,2}, \tau_{*,3}, \tau_{*,4} \right)^\top =  \left(0.5, 0, 1, -1.5, -0.5, 0.5, 1.5 \right)^\top$.}
\label{tab:efficiency}
\end{table}

Table~\ref{tab:efficiency} lists the relative efficiencies of various choices of the tuning constant~$\alpha$ for an ordinal~$Y$ with $r=5$ response options and true parameter vector $\Btheta_* = \left(\rho_*, \mu_*, \sigma_*^2, \tau_{*,1}, \tau_{*,2}, \tau_{*,3}, \tau_{*,4} \right)^\top =  \left(0.5, 0, 1, -1.5, -0.5, 0.5, 1.5 \right)^\top$. Up to about $\alpha = 0.25$, the relative efficiency of the robust estimator stays above~90\%, and then decreases roughly linearly to slightly below~50\% for $\alpha = 1$. Choices of $\alpha > 1$, while yielding even more robustness, would result in a comparatively poor relative efficiency of considerably less than~50\% and are therefore not considered in this paper.  The efficiency results for other choices of the true parameter vector~$\Btheta_*$ are very similar and are therefore deferred to Appendix~\ref{app:additional-results} (Figure~\ref{fig:efficiency}). Our software implementation, discussed in more detail in Section~\ref{sec:implementation}, provides functionality to compute \edit{the relative} efficiency at arbitrary user-specified parameter vectors and tuning constants~$\alpha$. 

A loss of efficiency is a common property of many robust estimators involving continuous random variables, and, as demonstrated in this section, our proposed estimator is no exception.\footnote{In contrast, for exclusively categorical variables, it is possible to construct robust estimators that are fully efficient. See \citet{welz2024robcat} for general models of categorical data, and \citet{welz2025polycor} for the special case of the polychoric correlation model.} As such, our estimator is based on a compromise between efficiency and robustness. Borrowing an insurance metaphor from \citet{anscombe1960}, we \emph{``sacrifice some efficiency at the model, in order to insure against accidents caused by deviations from the model"} \citep[][p.~5]{huber2009}. However, it turns out that the efficiency loss is relatively minor for reasonably small positive tuning constants like $\alpha = 0.1$ (Table~\ref{tab:efficiency}), which, as the simulation experiments in Section~\ref{sec:simulation} will reveal, suffices to gain substantial robustness over ML estimation.

\subsection{Implementation}\label{sec:implementation}
We provide a\edit{n} open-source implementation of our robust estimator as part of the package \pkg{robcat} \citep[for ``ROBust CATegorical data analysis";][]{robcat} for the statistical programming environment~\proglang{R} \citep{R}. The package is freely available from CRAN (the Comprehensive \proglang{R} Archive Network) at \url{https://CRAN.R-project.org/package=robcat}. We used this package for obtaining all numerical results in this paper.

In principle, every suitable method for numerical optimization can be used to minimize the robust estimator's minimization problem in~\eqref{eq:estimator}. In our experience, unconstrained optimization via the BFGS algorithm \citep[e.g.,][Section~6.1]{nocedal2006} works well. Yet, additional stability could be gained from making explicit the constraints in the parameter space~$\BTheta$ in~\eqref{eq:Theta}, for which standard algorithms for constrained optimization can be used, such as the simplex method of \citet{nelder1965}. In our implementation, we adopt a similar default behavior as the implementation of robust polychoric correlation estimation \citep{welz2025polycor} in package \pkg{robcat}. Specifically, the implementation first tries unconstrained optimization with the BFGS algorithm. If instability or numerical nonconvergence are encountered or a constraint is violated, it instead uses the algorithm of \citet{nelder1965} for constrained optimization.\footnote{\edit{In our experience, the higher the contamination fraction, the more likely it is that unconstrained optimization fails, so constrained algorithms must be used instead. As an example, we refer to the convergence statistics for the simulation in Section~\ref{sec:simulation}, reported in Appendix~\ref{app:additional-results}.}} Nevertheless, numerous other optimization algorithms are supported, and users can freely choose their preferred option. 

As for the choice of tuning constant~$\alpha \geq 0$ in the DPD function~\eqref{eq:DPD}, higher values lead to greater robustness but less efficiency, whereas values closer to~0 are less robust but more efficient. We shall see in the simulation experiments in the next section that the choice $\alpha = 0.5$ constitutes a good compromise between robustness against contamination up to about~$\varepsilon = 0.3$ while being reasonably efficient with about~76\% of the MLE's efficiency (Table~\ref{tab:efficiency}). Therefore, $\alpha = 0.5$ is the default choice in our implementation. While this is the default choice, we acknowledge that there might be situations in which choices that lead to more (or less) robustness are more appropriate, so we advise users to decide on~$\alpha$ on a case-by-case basis.

Moreover, we do not recommend the two-step estimation procedure in~\eqref{eq:twostep} for robust estimation. Recall from~\eqref{eq:twostep} that the two-step estimators for~$\mu_*$ and~$\sigma_*^2$ are given by the sample mean and sample variance, respectively, of the $X_i, i=1,\dots,N,$ while the estimators for the thresholds are computed from the empirical cumulative frequencies of the $Y_i,i=1,\dots,N$. However, if there is contamination in the sample such as outliers in the data for~$X$ and careless responses in the data for~$Y$, all of these three sample statistics might become heavily biased. This bias is then possibly inherited by the estimate of the correlation coefficient in the second stage. To avoid this problem, our robust estimator estimates all model parameters simultaneously.

\section{Simulation experiments}\label{sec:simulation}

This section conducts a number of simulation experiments to evaluate the robustness of the robust minimum DPD estimator against partial misspecification of the polyserial model. Subsection~\ref{sec:simdesign} describes the simulation design, Subsection~\ref{sec:simresults} the results, and Subsection~\ref{sec:additional-simulations} briefly summarizes additional simulations that are featured in detail in the appendix.

\subsection{Design}\label{sec:simdesign}
We consider a polyserial model under which the random vector $(X, \eta)$ is jointly normally distributed according to~\eqref{eq:normality} with true correlation parameter~$\rho_* = 0.5$ and the true first two moments of the observed~$X$ are $\mu_* = \E{X} = 0$ as well as $\sigma^2_* = \var{X} = 1$. The latent variable~$\eta$ also has zero mean and unit variance and governs the observed ordinal~$Y$ through the discretization process in~\eqref{eq:discretization} with true threshold parameters $\tau_{*,1} = -1.5, \tau_{*,2} = -0.5, \tau_{*,3} = 0.5, \textnormal{ and } \tau_{*,4} = 1.5,$ so that~$Y$ has $r=5$ response categories. Using integer scoring $\Y = \{1,2,\dots,5\}$, the corresponding true point polyserial correlation coefficient in this setting amounts to~$\tilde{\rho}_* = 0.477$ (see Section~\ref{sec:pointpolyserial}).

To simulate contamination, we replace a fraction~$\varepsilon$ of the data for $(X, \eta)$ with draws from a particular contamination distribution~$\Hxeta$, which is not modeled by our robust estimator nor ever assumed to be known.
Specifically, the contamination distribution~$\Hxeta$ in this simulation is set to a shifted bivariate $t$-distribution with noncentrality parameter $(10,-2)^\top$, scale matrix $\textnormal{diag}(0.25, 0.25)$, and~10 degrees of freedom. To obtain contaminated data for the ordinal~$Y$, we discretize this contamination distribution's realizations in the~$\eta$-dimension according to the same thresholds $\tau_{*,1}, \dots, \tau_{*,4}$ as the uncontaminated realizations from the polyserial model. The ensuing contaminated observations are mean-shifted to the right in the continuous~$X$-dimension and primarily inflate the first and second response option in the ordinal~$Y$-dimension. For instance, the contaminated data in Figure~\ref{fig:main-simdesign} were generated by this process (with contamination fraction $\varepsilon = 0.15$) and correspondingly exhibit these features. In the robust statistics literature, such contaminating data are an example of \emph{negative leverage points} because they drag positive correlational estimates towards zero or even negative values, thereby creating \emph{negative} leverage.

We sample $N=500$ observations $(X_i, Y_i), i=1,\dots, N$, from this process with contamination fraction $\varepsilon \in \{0, 0.002, 0.01, 0.05, 0.1, 0.15, 0.2, 0.3, 0.4, 0.49\}$. Note that $\varepsilon = 0.002$ corresponds to only one single contaminated data point for the sample size $N=500$. For each simulated data set, we estimate the true parameter~$\Btheta_*$ by means of a minimum DPD estimator~$\thetahat$ with tuning constants $\alpha \in\{0,0.1,0.25,0.5, 0.75,1\}$. This procedure is repeated~5,000 times.

Recall that the  tuning constant choice $\alpha = 0$ corresponds to the MLE, for which we use the usual Fisher-information based asymptotic covariance matrix to compute standard errors instead of the sandwich-type construction in~\eqref{eq:sandwich}. Further recall that larger values of~$\alpha$ lead to a more robust estimator, at the cost of a loss in efficiency. In fact, the relative efficiencies in Table~\ref{tab:efficiency} were computed for the same parameter value that serves as true value~$\Btheta_*$ in this simulation. We do not consider $\alpha$-values beyond~1 due to their poor efficiency properties of having substantially less than~50\% of the MLE's efficiency (see Table~\ref{tab:efficiency}). In addition to an estimate~$\hatN{\rho}$ of the true polyserial correlation coefficient~$\rho_*$, we also construct an estimate~$\hat{\tilde{\rho}}_N$ of the true point polyserial correlation coefficient~$\tilde{\rho}_*$ from the individual parameter estimates in~$\thetahat$ (see Section~\ref{sec:pointpolyserial} for details). 

Let~$\hatN{\rho}$ be a polyserial correlation estimate computed from a given simulated data set, and~$\SEhat{\hatN{\rho}}$ be its associated standard error estimate constructed from the limit theory in Theorem~\ref{thm:main} in Appendix~\ref{app:asymptotics}. We evaluate performance by means of the following metrics. 

\begin{itemize}
	\item Bias of the polyserial correlation estimate, $\hatN{\rho} - \rho_*$, and the point polyserial correlation estimate, $\hat{\tilde{\rho}}_N - \tilde{\rho}_*$, averaged over the 5,000 repetitions.
	\item Average approximate bias of the standard error estimate, defined as the difference between~$\SEhat{\hatN{\rho}}$ and the sample standard deviation (SD) of the individual correlation estimates~$\hatN{\rho}$ across the 5,000 repetitions, where the latter is a finite-sample approximation of the true population standard error. A rigorous definition is provided in Appendix~\ref{app:perfmeasures}. 
	\item Coverage at the significance level $\gamma = 0.05$, defined as the proportion (across repetitions) of confidence intervals $\left[ \hatN{\rho} \mp q_{1-\gamma/2} \cdot\SEhat{\hatN{\rho}} \right]$ that contain the true~$\rho_*$, where~$q_{1-\gamma/2}$ denotes the $(1-\gamma/2)$ quantile of the standard normal distribution.
	\item Average confidence interval length at significance level $\gamma = 0.05$, where the length of an individual confidence interval is given by $2\cdot q_{1-\gamma/2} \cdot\SEhat{\hatN{\rho}}$.
\end{itemize} 


\subsection{Results}\label{sec:simresults}

\begin{table}
\centering
\notsotiny
\setlength{\tabcolsep}{4.67pt}
\begin{tabular}{l l c r r c c c r c c c}
& & & \multicolumn{3}{c}{Point estimate} & & \multicolumn{2}{c}{Standard error} & & \multicolumn{2}{c}{Confidence interval} \\
\noalign{\smallskip}\cline{4-6}\cline{8-9}\cline{11-12}\noalign{\smallskip}
Contamination & $\alpha$ & & \multicolumn{1}{c}{$\hatN{\rho}$} & \multicolumn{1}{c}{Bias} & SD & & $\SEhat{\hatN{\rho}}$ & \multicolumn{1}{c}{Bias} & & Coverage & Length \\
\noalign{\smallskip}\hline\noalign{\smallskip}
\multirow{6}{*}{$\varepsilon = 0$} 
  & 0 (MLE) 		&& 0.500 	& 0.000    & 0.035 && 0.036 & 0.001 && 0.951 & 0.142 \\ 
  & 0.1 		&& 0.500 	& 0.000    & 0.036 && 0.036 & 0.000 && 0.948 & 0.141 \\ 
  & 0.25 		&& 0.499 	& $-0.001$ & 0.037 && 0.037 & 0.000 && 0.949 & 0.147 \\ 
  & 0.5 		&& 0.499 	& $-0.001$ & 0.041 && 0.041 & 0.000 && 0.950 & 0.162 \\ 
  & 0.75 		&& 0.499 	& $-0.001$ & 0.046 && 0.046 & 0.001 && 0.951 & 0.182 \\ 
  & 1 			&& 0.500 	& 0.000    & 0.051 && 0.052 & 0.001 && 0.953 & 0.205 \\ 
\noalign{\smallskip}
\multirow{6}{*}{$\varepsilon = 0.002$} 
  & 0 (MLE) 		&& 0.416 	& $-0.084$    & 0.035 && 0.039 & 0.003 && 0.395 & 0.152 \\ 
  & 0.1 		&& 0.499 	& $-0.001$    & 0.036 && 0.036 & 0.000 && 0.948 & 0.141 \\ 
  & 0.25 		&& 0.499 	& $-0.001$    & 0.037 && 0.037 & 0.000 && 0.950 & 0.147 \\ 
  & 0.5 		&& 0.499 	& $-0.001$    & 0.041 && 0.041 & 0.000 && 0.949 & 0.162 \\ 
  & 0.75	  	&& 0.499 	& $-0.001$    & 0.046 && 0.046 & 0.001 && 0.951 & 0.182 \\ 
  & 1 			&& 0.499 	& $-0.001$    & 0.051 && 0.052 & 0.001 && 0.953 & 0.205 \\ 
\noalign{\smallskip}
\multirow{6}{*}{$\varepsilon = 0.01$} 
  & 0 (MLE) 		&& 0.215 	& $-0.285$ & 0.033 && 0.057 & 0.024 && 0.000 & 0.224 \\ 
  & 0.1 		&& 0.499 	& $-0.001$ & 0.036 && 0.036 & 0.000 && 0.947 & 0.142 \\ 
  & 0.25 		&& 0.499 	& $-0.001$ & 0.037 && 0.038 & 0.000 && 0.950 & 0.147 \\ 
  & 0.5 		&& 0.499 	& $-0.001$ & 0.041 && 0.041 & 0.001 && 0.951 & 0.162 \\ 
  & 0.75 		&& 0.499 	& $-0.001$ & 0.046 && 0.046 & 0.001 && 0.953 & 0.182 \\ 
  & 1 			&& 0.499 	& $-0.001$ & 0.051 && 0.052 & 0.001 && 0.953 & 0.205 \\ 
\noalign{\smallskip}
\multirow{6}{*}{$\varepsilon = 0.05$} 
  & 0 (MLE) 		&& $-0.190$ 	& $-0.690$ & 0.032 && 0.073 & 0.042 && 0.000 & 0.287 \\ 
  & 0.1 		&& 0.497 	& $-0.003$ & 0.037 && 0.037 & 0.000 && 0.949 & 0.146 \\ 
  & 0.25 		&& 0.498 	& $-0.002$ & 0.038 && 0.038 & 0.000 && 0.948 & 0.150 \\ 
  & 0.5 		&& 0.497 	& $-0.003$ & 0.041 && 0.042 & 0.000 && 0.947 & 0.164 \\ 
  & 0.75 		&& 0.496 	& $-0.004$ & 0.046 && 0.047 & 0.001 && 0.951 & 0.183 \\ 
  & 1 			&& 0.496 	& $-0.004$ & 0.051 && 0.052 & 0.001 && 0.953 & 0.204 \\ 
\noalign{\smallskip}
\multirow{6}{*}{$\varepsilon = 0.1$} 
  & 0 (MLE) 		&& $-0.409$ 	& $-0.909$ & 0.030 && 0.063 & 0.033 && 0.000 & 0.248 \\ 
  & 0.1 		&& 0.494 	& $-0.006$ & 0.039 && 0.039 & 0.000 && 0.948 & 0.151 \\ 
  & 0.25 		&& 0.497 	& $-0.003$ & 0.039 && 0.039 & 0.000 && 0.950 & 0.154 \\ 
  & 0.5 		&& 0.494 	& $-0.006$ & 0.042 && 0.043 & 0.000 && 0.948 & 0.167 \\ 
  & 0.75 		&& 0.492 	& $-0.008$ & 0.046 && 0.047 & 0.001 && 0.950 & 0.183 \\ 
  & 1 			&& 0.491 	& $-0.009$ & 0.051 && 0.052 & 0.001 && 0.952 & 0.202 \\ 
\noalign{\smallskip}
\multirow{6}{*}{$\varepsilon = 0.15$} 
  & 0 (MLE)		&& $-0.529$ 	& $-1.029$ & 0.027 && 0.059 & 0.032    	 && 0.000 & 0.233 \\ 
  & 0.1 		&& 0.369 	& $-0.131$ & 0.327 && 0.042 & $-0.285$   && 0.830 & 0.164 \\ 
  & 0.25 		&& 0.495 	& $-0.005$ & 0.040 && 0.040 & 0.000 	 && 0.948 & 0.157 \\ 
  & 0.5 		&& 0.490 	& $-0.010$ & 0.043 && 0.043 & 0.000 	 && 0.948 & 0.170 \\ 
  & 0.75 		&& 0.487 	& $-0.013$ & 0.046 && 0.047 & 0.001 	 && 0.948 & 0.184 \\ 
  & 1 			&& 0.485 	& $-0.015$ & 0.050 && 0.051 & 0.001 	 && 0.949 & 0.201 \\ 
\noalign{\smallskip}
\multirow{6}{*}{$\varepsilon = 0.2$} 
  & 0 (MLE) 		&& $-0.603$ 	& $-1.103$ & 0.024 && 0.060 & 0.035    && 0.000 & 0.234 \\ 
  & 0.1 		&& $-0.591$	& $-1.091$ & 0.078 && 0.037 & $-0.041$ && 0.004 & 0.145 \\ 
  & 0.25 		&& 0.493 	& $-0.007$ & 0.041 && 0.041 & 0.000    && 0.947 & 0.162 \\ 
  & 0.5 		&& 0.486 	& $-0.014$ & 0.044 && 0.044 & 0.000    && 0.945 & 0.172 \\ 
  & 0.75 		&& 0.481 	& $-0.019$ & 0.047 && 0.047 & 0.000    && 0.941 & 0.185 \\ 
  & 1 			&& 0.478 	& $-0.022$ & 0.050 && 0.051 & 0.001    && 0.939 & 0.199 \\ 
\noalign{\smallskip}
\multirow{6}{*}{$\varepsilon = 0.3$} 
  & 0 (MLE) 		&& $-0.686$ 	& $-1.186$ & 0.020 && 0.063 & 0.043      && 0.000 & 0.247 \\ 
  & 0.1 		&& $-0.694$ 	& $-1.194$ & 0.021 && 0.024 & 0.003 	 && 0.000 & 0.094 \\ 
  & 0.25 		&& 0.464 	& $-0.036$ & 0.171 && 0.043 & $-0.128$   && 0.926 & 0.170 \\ 
  & 0.5 		&& 0.475 	& $-0.025$ & 0.045 && 0.046 & 0.000 	 && 0.923 & 0.179 \\ 
  & 0.75 		&& 0.464 	& $-0.036$ & 0.047 && 0.048 & 0.001 	 && 0.901 & 0.187 \\ 
  & 1 			&& 0.456 	& $-0.044$ & 0.049 && 0.050 & 0.001 	 && 0.882 & 0.196 \\ 
\noalign{\smallskip}
\multirow{6}{*}{$\varepsilon = 0.4$} 
  & 0 (MLE) 		&& $-0.727$ 	& $-1.227$ & 0.018 && 0.065 & 0.047    && 0.000 & 0.257 \\ 
  & 0.1 		&& $-0.732$ 	& $-1.232$ & 0.022 && 0.019 & $-0.003$ && 0.000 & 0.073 \\ 
  & 0.25 		&& $-0.729$ 	& $-1.229$ & 0.024 && 0.019 & $-0.005$ && 0.000 & 0.074 \\ 
  & 0.5 		&& $-0.475$ 	& $-0.975$ & 0.500 && 0.027 & $-0.473$ && 0.191 & 0.106 \\ 
  & 0.75 		&& 0.440 	& $-0.060$ & 0.058 && 0.056 & $-0.002$ && 0.772 & 0.221 \\ 
  & 1 			&& 0.468 	& $-0.032$ & 0.074 && 0.127 & 0.053    && 0.818 & 0.498 \\ 
\noalign{\smallskip}
\multirow{6}{*}{$\varepsilon = 0.49$} 
  & 0 (MLE) 		&& $-0.745$ 	& $-1.245$ & 0.017 && 0.066 & 0.049    && 0.000 & 0.260 \\ 
  & 0.1 		&& $-0.760$ 	& $-1.260$ & 0.016 && 0.016 & 0.000    && 0.000 & 0.064 \\ 
  & 0.25 		&& $-0.782$ 	& $-1.282$ & 0.015 && 0.015 & 0.000    && 0.000 & 0.059 \\ 
  & 0.5 		&& $-0.820$ 	& $-1.320$ & 0.014 && 0.014 & 0.000    && 0.000 & 0.053 \\ 
  & 0.75 		&& $-0.858$ 	& $-1.358$ & 0.038 && 0.031 & $-0.007$ && 0.001 & 0.118 \\ 
  & 1 			&& $-0.379$ 	& $-0.879$ & 0.676 && 0.122 & $-0.554$ && 0.007 & 0.478 \\ 
   \hline
\end{tabular}
\caption{Performance measures for estimating  polyserial correlation coefficients in the simulation in Section~\ref{sec:simdesign} at significance level $\gamma=0.05$ (averaged across 5,000 repetitions).}
\label{tab:main-simresults}
\end{table}

\begin{figure}[!ht]
	\centering
	\includegraphics[width = \textwidth]{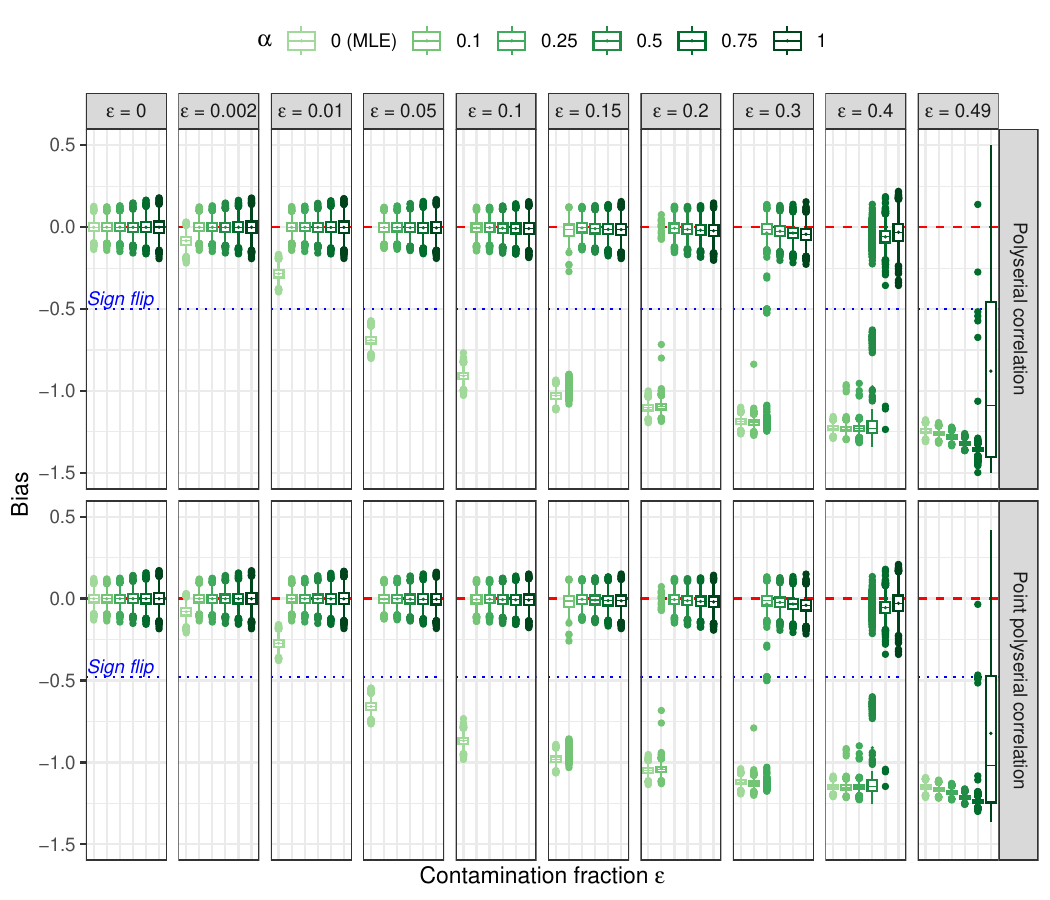}
\caption{Boxplots of the bias of the considered estimators for the polyserial correlation coefficient, $\hatN{\rho} - \rho_*$ (top panel) and the point polyserial correlation coefficient with integer scoring, $\hat{\tilde{\rho}}_N - \tilde{\rho}_*$ (bottom panel), for various contamination fractions in the misspecified polyserial models across 5,000 repetitions. Diamonds represent the respective average bias. The dotted lines at $\rho_* = -0.5$ and $-\tilde{\rho}_* = -0.477$ indicate a sign flip in the respective estimate.}
\label{fig:main-simresults-boxplots}
\end{figure}

We start evaluating the results by visualizing the bias of the polyserial and point polyserial estimates. Figure~\ref{fig:main-simresults-boxplots} illustrates by means of boxplots the biases across the 5,000 repetitions. An analogous plot for the parameter vector~$\Btheta$ is provided in Appendix~\ref{app:additional-results}; the results are similar to those of the correlation estimates. Furthermore, as can be immediately seen from Figure~\ref{fig:main-simresults-boxplots}, the results for polyserial and point polyserial correlation are very much alike, so we restrict the following discussion to polyserial correlation. In particular, Table \ref{tab:main-simresults} contains the performance measures for polyserial correlation.

In the absence of contamination $(\varepsilon = 0)$, all estimators yield accurate results, but the robust estimators (those with $\alpha > 0$) exhibit more variation than the MLE ($\alpha = 0$), reflecting their diminished efficiency (e.g., Table~\ref{tab:efficiency}). Furthermore, all estimators attain the nominal coverage level of~95\%.

However, the MLE's performance rapidly deteriorates once contamination is introduced $(\varepsilon > 0)$. For instance, one single bad data point ($\varepsilon = 0.002$ here) suffices for the MLE to exhibit a notable bias of about $-0.084$, resulting in a deteriorated coverage of less than~40\%. At contamination $\varepsilon = 0.01$,  its coverage even drops to~0\%. At contamination level $\varepsilon = 0.05$, its point estimate experiences a sign flip: instead of yielding a positive estimate of the true value $\rho_* = 0.5$, its average estimate of~$-0.19$ is negative. The ML estimate continues to deteriorate until it roughly stabilizes at about $\varepsilon = 0.2$ with an average bias of about~$-0.7$. 

Conversely, the minimum DPD estimators ($\alpha > 0$) turn out to be substantially more robust to contamination than ML. Up until and including the contamination fraction of $\varepsilon = 0.1$, all minimum DPD estimators are virtually unaffected and maintain the nominal coverage level. At $\varepsilon = 0.15$, the coverage of the estimator with $\alpha = 0.1$ drops slightly to about~83\%. At $\varepsilon = 0.2$, its coverage drops to nearly~0\%, while all choices with $\alpha > 0.1$ remain stable with a coverage above~90\%. Furthermore, all minimum DPD estimators up to this point accurately estimate the standard error of polyserial correlation (except $\alpha = 0.1$ at $\varepsilon > 0.15$, where its coverage is near~0). In the high-contamination setting of $\varepsilon = 0.3$, estimators with $\alpha > 0.1$ start exhibiting a minor but notable bias, while at $\varepsilon = 0.4$, only the choices $\alpha\in\{0.75,1\}$ retain a reasonably good performance, while all other estimators have coverage rates of or near zero. At $\varepsilon = 0.49$, the highest considered contamination level, all estimators break down.\footnote{All point estimators always numerically converged across all configurations. However, in extremely high-contamination settings, the covariance matrix estimator could occasionally not be computed due to singularity issues. Such non-existent standard error estimates were omitted in Figure~\ref{fig:main-simresults-boxplots} and Table~\ref{tab:main-simresults}. Nonconvergence statistics are provided in Appendix~\ref{app:additional-results}.}

This simulation demonstrates that ML estimation of the polyserial correlation coefficient is highly susceptible to already very low levels of contamination. Conversely, using our proposed estimator yields substantial gains in robustness. For instance, at tuning constant $\alpha = 0.1$, which only sacrifices less than~2\% efficiency (Table~\ref{tab:efficiency}), the robust estimator remains accurate up until a contamination level of about $\varepsilon = 0.1$. Robustness against higher contamination levels can be achieved by choosing higher values of~$\alpha$. In addition, the simulation also exposes the boundaries of the robust estimator: In extremely high-contamination settings of beyond~40\% contamination, even the estimator with the highest considered tuning constant value of $\alpha = 1$ breaks down. However, it is questionable whether modeling with data of such extremely poor quality is meaningful to begin with. 

\subsection{Additional simulations}\label{sec:additional-simulations}

We perform three additional simulation studies, all described in detail in Appendix~\ref{app:additional-simulations}. These simulations are intended to explore the benefits and limitations of our proposed methodology.

The first additional simulation, described in Appendix~\ref{app:outlier}, has the same setup as the design in Section~\ref{sec:simdesign}, but the contamination manifests through extreme outliers in both the~$X$- and~$\eta$-dimension. This simulation is primarily intended to study the behavior of the MLE and robust estimator under gross errors in the data. In brief, one single gross error observation in a sample suffices for the MLE to produce sign-flipped estimates that are extremely biased. In contrast, the robust estimator remains nearly unaffected by small to moderate contamination fractions with such gross errors, and only starts exhibiting considerable biases in high-contamination settings.

The second additional simulation, described in Appendix~\ref{app:corshift}, is motivated by the fact that the simulation design in Section~\ref{sec:simdesign} is primarily concerned with mean-shifted contamination. Mean-shifted contamination is characterized by the population mean of the contamination distribution~$\Hxeta$ being markedly different than that of the true normal distribution~$\Pxetasymb$. However, contamination may also manifest through changes in \emph{correlation} rather than means. The second additional simulation therefore focuses on a \emph{correlation-shifted} contamination distribution~$\Hxeta$, which is equal to the true normal distribution~$\Pxetasymb$, except for a sign-flipped correlation coefficient~$-\rho_*$.  In brief, for moderate polyserial correlation~$\rho_*$ (where~$\Hxeta$ and~$\Pxetasymb$ substantially overlap), our robust estimator does not yield an improvement over ML. Conversely, it does provides a notable gain in robustness for larger true correlations~$\rho_*$ (where~$\Hxeta$ and~$\Pxetasymb$ barely overlap). 

Combined with the simulations on mean-shifted contamination in Sections~\ref{sec:simdesign} and~\ref{app:outlier}, the results of the simulation on correlation-shifted contamination suggest that the potential for robustness gain depends on the overlap between the contamination distribution~$\Hxeta$ and true normal distribution~$\Pxetasymb$. If there is much overlap, the robust estimator cannot distinguish well between contamination and regular observations since it does not make any assumptions on the former, so it may not improve upon ML. In contrast, if~$\Hxeta$ and~$\Pxetasymb$ do not overlap much---like with mean-shifted contamination or correlation-shifted contamination with strong true correlation---the robust estimator can identify contaminated observations and subsequently downweigh their influence to achieve considerable robustness gains.

The third additional simulation, described in Appendix~\ref{app:distributional}, is concerned with distributional misspecification, where the polyserial model is misspecified for \emph{all} observations in a sample (cf. Section~\ref{sec:distributional-subsection}). The simulation shows that the potential of robustness gain with our robust estimator depends on the case-specific characteristics of the nonnormal sampling distribution of $(X,\eta)$ for which the polyserial model is distributionally misspecified. In brief, if the sampling distribution can be reasonably well approximated by a mixture of a normal distribution and some other unknown distribution to emulate the contamination model in~\eqref{eq:contamdist}, then the robust estimator can improve upon the MLE. If the sampling distribution does not admit such an approximation, the robust estimator may not yield an improvement \edit{but produce similar estimates as ML, which can be quite poor under distributional misspecification \citep[e.g.,][, and our simulations in Appendix~\ref{app:distributional}]{bedrick1995}}.
\edit{Thus, we advice applied researchers to always test for partially-latent normality when ML and robust estimator yield similar estimates. While such similarity may be due to normality indeed holding true, it may also be due to distributional misspecification from a specific nonnormal distribution for which the robust estimator cannot improve over ML. We provide details and more explicit guidance in Appendix~\ref{app:distributional}.}

\section{Empirical application}\label{sec:application}
To demonstrate our proposed estimator in practice, we apply it to an empirical data set from personality psychology. The data are from an administration of the \emph{Eysenck Personality Inventory} \citep{eysenck1964} to $N=231$ undergraduate students at Northwestern University (collected by William Revelle), and are publicly available in the \proglang{R} package \pkg{psychTools} \citep{psychTools}. We restrict ourselves to the continuous variable \pkg{stateanx} and the ordinal variable \pkg{epilie} in this demonstration.

\begin{figure}
	\centering
	\begin{subfigure}{0.42\textwidth}
	\centering
	\includegraphics[width = \textwidth]{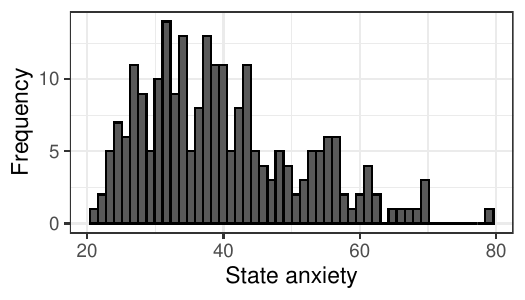}
	\caption{\pkg{stateanx}}
	\label{fig:application-stateanx}
	\end{subfigure}
	\begin{subfigure}{0.42\textwidth}
	\includegraphics[width = \textwidth]{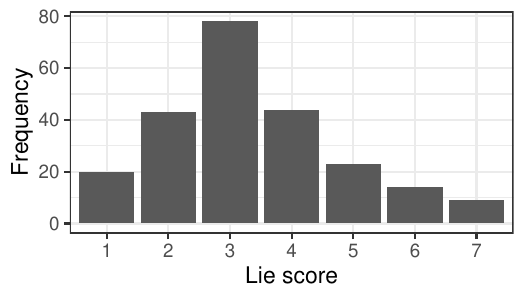}
	\caption{\pkg{epilie}}
	\label{fig:application-epilie}
	\end{subfigure}
\caption{Histogram of the continuous variable \pkg{stateanx} (left) and response frequencies of the ordinal variable \pkg{epilie} (right) in the data of \citet{psychTools}, for $N=231$ observations.}
\label{fig:application-data}
\end{figure}

The continuous variable \pkg{stateanx} is the score of an unspecified scale measuring \emph{state anxiety}, defined as a temporary emotional reaction to adverse events \citep[e.g.,][]{saviola2020}.\footnote{Technically, being an aggregate score of ordinal measurements, the variable \pkg{stateanx} is strictly speaking not continuous. However, due to the granularity of the measurements (see Figure~\ref{fig:application-stateanx}), it may be viewed as being quasi-continuous.} Figure~\ref{fig:application-stateanx} provides a histogram of this variable in the data of \citet{psychTools}. The empirical distribution seems to be skewed to the right. In particular, there is one unusually large observation in its right tail with value~79 that seems to be separated from this the rest of the data, so it may be viewed as an outlier (the sample mean is~39.8). Hence, due to the inflated right tail, the validity of the normality assumption seems questionable. Indeed, the nonparametric Shapiro-Wilk test rejects marginal normality of this variable with a test statistic of $W=0.953$ and a $p$-value of $<0.001$. 

The ordinal variable \pkg{epilie} is the score of a lie scale and measures social desirability of a participant's responses: The higher the score, the less trustworthy their responses. Originally, this variable has~10 unique ordinal response options.\footnote{The Eysenck Personality Inventory constructs the variable~\pkg{epilie} as the number of times that a given participant has responded ``yes'' in nine separate yes-no questions. Since higher scores are associated with higher levels of socially desirable responding, this variable can be interpreted as being ordinal.} However, in the data of \citet{psychTools}, no participant has the highest score of~9 or lowest score of~0, and only two participants have the second-highest score of~8. To avoid numerical instability in the estimation of thresholds,\footnote{It may happen that a robust estimator attempts to effectively eliminate thresholds corresponding to (nearly) empty cells in a certain way that we describe in detail in Appendix~\ref{sec:distributional-simulation}, thereby causing instability. This phenomenon has been noted before by \citet{welz2025polycor}.} we therefore assign the two participants with score~8 the score~7, that is, we merge the seventh and eighth response category, and treat the score~1 as the first response category. Thus, the variable \pkg{epilie} here takes values in $\Y = \{1,2,\dots,7\}$ with a total of $r=7$ response options. Figure~\ref{fig:application-epilie} visualizes the frequency of each response option. 

Suppose we are interested in the polyserial correlation coefficient between state anxiety (\pkg{stateanx}) and the ordinal lie score (\pkg{epilie}) in the data of \citet{psychTools}. We therefore fit the polyserial model to these data using ML and our proposed robust estimator. For the latter, we focus on the results with tuning constant $\alpha = 0.5$, being the default choice in our software, but also report those with other choices of~$\alpha$.

\begin{table}
\centering
\small
\begin{tabular}{c c r r c r r}
& & \multicolumn{2}{c}{Robust} & & \multicolumn{2}{c}{ML}
\\
\noalign{\smallskip}\cline{3-4}\cline{6-7}\noalign{\smallskip}
Parameter & & Estimate & $\widehat{\text{SE}}$ & & Estimate & $\widehat{\text{SE}}$ 
\\
\noalign{\smallskip}\hline\noalign{\smallskip}
$\rho$     & & $-0.107$ &  0.076 & & $-0.142$ &  0.067 \\ 
$\mu$      & &   37.990 &  0.615 & &   39.842 &  \edit{0.900} \\ 
$\sigma^2$ & &  107.730 & 10.280 & &  131.575 & \edit{14.485} \\ 
$\tau_1$   & & $-1.359$ &  0.120 & & $-1.358$ &  0.11\edit{8} \\ 
$\tau_2$   & & $-0.586$ &  0.091 & & $-0.604$ &  0.088 \\ 
$\tau_3$   & &    0.276 &  0.085 & &    0.276 &  0.084 \\ 
$\tau_4$   & &    0.828 &  0.096 & &    0.841 &  0.094 \\ 
$\tau_5$   & &    1.268 &  0.116 & &    1.286 &  0.11\edit{6} \\ 
$\tau_6$   & &    1.849 &  0.152 & &    1.771 &  0.152 
\\
\end{tabular}
\caption{Parameter estimates and standard error estimates ($\widehat{\text{SE}}$) for the correlation between the continuous variable \pkg{stateanx} and the ordinal variable \pkg{epilie} (with $r=7$ response options) in the data of \citet{psychTools}, using the robust estimator with $\alpha=0.5$ and the MLE.}
\label{tab:application-stateanx-lie}
\end{table}

Table~\ref{tab:application-stateanx-lie} summarizes the results of the two estimators. The correlation estimate is weaker for the robust estimator with a value of $-0.107$ than for the MLE with a value of $-0.142$. The largest differences occur in the moment estimates of the continuous variable \pkg{stateanx}: Both~$\mu$ and~$\sigma^2$ are estimated to be notably larger by ML (estimates of about 39.8 and 131.6, respectively) compared to the robust estimator (estimates of about 38 and 107.7, respectively). The threshold estimates are similar. Thus, overall, the robust estimates tend to be weaker in absolute magnitude than those of ML.

\begin{figure}
	\centering
	\includegraphics[width = \textwidth]{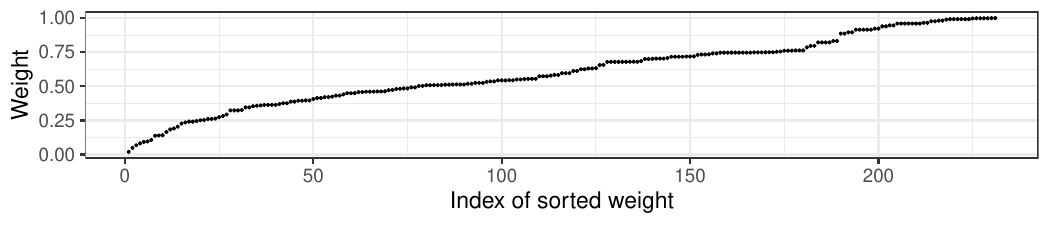}
\caption{Weights $w_{i,\alpha}\left(\thetahat\right), i = 1,\dots, N$, computed with the robust parameter estimates at tuning constants $\alpha = 0.5$, using the data in \citet{psychTools} for the variables \pkg{stateanx} and \pkg{epilie}. For a clearer visualization, the $N=231$ weights are sorted here.}
\label{fig:application-stateanx-lie}
\end{figure}

\begin{table}
\centering
\small
	\begin{tabular}{c| r r r r r r}
 $i$ & 1 & 2 & 3 & 4 & 5 & 6 \\\hline	
$w_{i,\alpha}\left(\thetahat\right)$ & 0.020  &  0.049 &  0.068 &  0.082 &  0.093 &  0.096 \\
\pkg{stateanx} ($X_i$)               & 79     &  69    &  61    &  68    &  70    &  67 \\
\pkg{epilie} ($Y_i$)                 & 3      &  5     &  7     & 1      &   3    &  4 \\
	\end{tabular}
\caption{Weights and values of variables \pkg{stateanx} (continuous) and \pkg{epilie} (ordinal with $r=7$) for observations in the data of \citet{psychTools} whose robustly estimated weights are below~0.1 (using tuning constant $\alpha = 0.5$). The observations are sorted according to the estimated weights.}
\label{tab:application-stateanx-lie-weights}
\end{table}

To obtain insights as to why the robust estimator's results differ from those of the MLE, we calculate the individual-specific weights in~\eqref{eq:weights} using the robust parameter estimates. Figure~\ref{fig:application-stateanx-lie} visualizes the sorted weights. While many observations can be fitted well with weights of nearly~1, there are also observations that receive a weight of close to~0, indicating that strong downweighting has taken place. We therefore investigate in detail all observations whose weight is reasonably close to zero, say, below~0.1. Table~\ref{tab:application-stateanx-lie-weights} lists the weights and values of the two variables for observations whose estimated weights are below~0.1. The smallest weight (valued~0.02) belongs to the observation with the outlying value of~79 in variable \pkg{stateanx} (Figure~\ref{fig:application-stateanx}). Thus, the robust estimator has identified this observation as outlying and subsequently downweighs it in the estimation procedure. In similar fashion, the other observations in Table~\ref{tab:application-stateanx-lie-weights} are those with relatively large \pkg{stateanx} values. It therefore seems that the robust estimator is downweighting observations whose \pkg{stateanx} values are in the heavy right tail; see Figure~\ref{fig:application-stateanx}. A notable exception is the observation that has a somewhat less high \pkg{stateanx} value of only~61. Nevertheless, this observation's lie score amounts to the maximum value of~7, indicating low trustworthiness of the given responses. The fact that this observation was strongly downweigted suggests that it could not sufficiently well modeled by the polyserial model, which might be due to its low trustworthiness. 

\begin{figure}
	\centering
	\includegraphics[width = \textwidth]{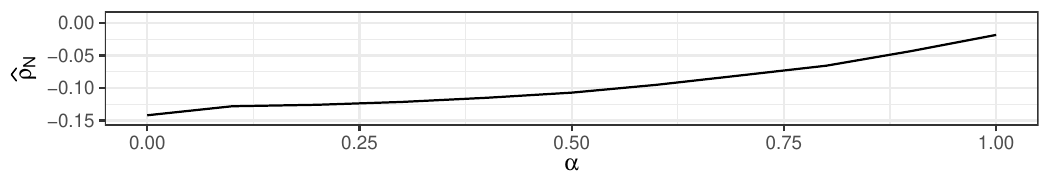}
\caption{Estimates of the polyserial correlation coefficient for different values of tuning constant~$\alpha$ ($\alpha = 0$ is MLE), computed on the variables \pkg{stateanx} and \pkg{epilie} in the data of \citet{psychTools}.}
\label{fig:application-stateanx-lie-alpha}
\end{figure}

As a sensitivity analysis, Figure~\ref{fig:application-stateanx-lie-alpha} plots the polyserial correlation estimates computed for various choices of tuning constant~$\alpha$. All strictly positive choices of~$\alpha$ yield estimates that are weaker in magnitude than the MLE, but stay around~$-0.1$ up until approximately $\alpha = 0.6$. Beyond that tuning constant value, the estimates sharply converge to about~0, primarily reflecting more stringent downweighting of the heavy right tail of the continuous variable \pkg{stateanx}. Overall, though, for tuning constants~$\alpha \leq 0.6$, we obtain the robust finding of a correlation of about~$-0.1$, being weaker than the ML estimate of~$-0.14$. 

This empirical application demonstrated the practical usefulness of our proposed robust estimator by showing how it identifies observations that are likely discrepant to the polyserial model and subsequently downweights their influence. It furthermore demonstrates how the estimated weights are an attractive tool for interpretation and exploratory data analysis. 

\section{Discussion and conclusion} \label{sec:conclusion}
Motivated by recent interest in the psychometric literature in the non-robustness of maximum likelihood (ML) estimation of latent variable models under model misspecification, we study estimation of the polyserial correlation model, which models the association between a continuous variable and an ordinal variable. We consider a \emph{partial} misspecification framework stemming from the robust statistics literature where the polyserial model is misspecified for an unknown, possibly zero-valued fraction of an observed sample. Crucially, this framework does not impose any assumptions on \emph{how} the model is possibly misspecified, which might be due to, for instance but not limited to, outliers or careless responses. We show that one single observation not generated by the polyserial model suffices for the commonly used ML estimator to produce arbitrarily poor results. 

As a remedy, we propose a novel estimator designed to be robust against partial misspecification of the polyserial model. The robust estimator leverages the theory of minimum \emph{density power divergence} \citep[DPD;][]{basu1998} estimation, which, as we show, is able to cope with mixed-type variables, unlike standard approaches for robust estimation with exclusively continuous or exclusively categorical variables. The proposed minimum DPD estimator achieves robustness by implicitly downweighting observations that cannot be sufficiently well fitted by the polyserial model. The ensuing weights are a useful analytical tool for identifying observations discrepant to the polyserial model, such as (but not limited to) outliers and careless responses. Our proposed methodology is implemented as part of the free open-source \proglang{R} package \pkg{robcat} \citep{robcat}, which is publicly available at \url{https://CRAN.R-project.org/package=robcat}.

We show that the proposed robust estimator is consistent as well as asymptotically normally distributed, thereby enabling inference. The price to pay for enhanced robustness is diminished statistical efficiency of our estimator compared to ML. However, we show that substantial robustness can be gained with our estimator while maintaining more than~98\% of ML efficiency.

We verify the estimator's robustness and theoretical properties by means of numerous simulation studies. An empirical application on a data set from personality psychology furthermore demonstrates its practical usefulness, where it identifies outlying data points. 

\edit{A central practical consideration is the choice of tuning constant~$\alpha$,  governing the robustness-efficiency tradeoff. While simulations suggest that $\alpha = 0.5$ is a good compromise with about~76\% of ML efficiency, we acknowledge that a detailed investigation is an important avenue of further research. We recommend to always compare various choices of~$\alpha$, like in Figure~\ref{fig:application-stateanx-lie-alpha}, to assess estimation stability and severity of (partial) model misspecification. Alternatively, one may fix the efficiency loss and select the corresponding value of~$\alpha$ as part of an ex-ante power analysis \citep[cf.][Section~5]{basu1998}.}

This paper suggests a number of extensions. For instance, while our software implementation is reasonably fast with about \edit{two} seconds for computing a robust estimate, substantial speedups could be achieved by rewriting the source code in \proglang{C++} via \pkg{Rcpp} \citep{eddelbuettel2013}. Moreover, the ability to robustly estimate polyserial correlation could be particularly useful in structural equation modeling with mixed data to robustify such analyses against partial model misspecification. We leave these avenues to further research. Overall, we believe that our robust estimator can contribute to the growing literature on making latent variable analyses less dependent on easily violated modeling  assumptions.

\bibliography{bib_polyserial}
\newpage
\appendix
\renewcommand\thefigure{\thesection.\arabic{figure}}   
\renewcommand\thetable{\thesection.\arabic{table}}   
\renewcommand{\theequation}{\thesection.\arabic{equation}}
\renewcommand{\thefootnote}{\roman{footnote}}
\setcounter{figure}{0} 
\setcounter{table}{0}   
\setcounter{equation}{0}  
\setcounter{footnote}{0}   


\section{Technical details}\label{app:asymptotics}

\edit{
This appendix section contains technical details on the polyserial model and robust estimation thereof. Concerning the polyserial model, Subsection~\ref{app:pointpolyserial} covers identification of point polyserial correlation and Subsection~\ref{app:olsson1982notation} provides alternative but equivalent definitions of the polyserial model density. Concerning estimation, Subsection~\ref{app:notation} introduces compact notation for the asymptotic analysis, Subsection~\ref{app:main-theorem} rigorously establishes the main theorem, Subsection~\ref{app:covmatcomparison} compares the asymptotic covariance matrices of ML to the robust estimator, Subsection~\ref{app:covmat-estimation} discusses covariance matrix estimation, and Subsection~\ref{app:expressions} provides closed-form gradient and Hessian expressions.
}

\subsection{\edit{Identification of point polyserial correlation}}\label{app:pointpolyserial}

Point polyserial correlation is the correlation between the observed variable~$X$ and the observed ordinal variable~$Y$ under the partially-latent normality assumption. To identify point polyserial correlation, one needs to introduce a scoring system for the response options of the ordinal~$Y$.  
Recall that we assume that~$Y$ takes values in the set $\Y = \{1,2,\dots, r\}$, where the elements in~$\Y$ need not admit a numerical interpretation. Assigning a scoring system introduces a numeric meaning to each response category. We refer to \citet{fernandez2020}, \citet{ivanova2001}, and \citet[][pp.~340--341]{olsson1982} for discussions of different scoring systems.

We temporarily drop the assumption that the response set of~$Y$ is given by adjacent integers $1,2,\dots,r$, and instead assume that $\Y = \{y_1,y_2,\dots,y_r\}$, where the response options $-\infty < y_1 < y_2 < \cdots < y_r <+\infty$ are real-valued \emph{and} admit a numerical interpretation. As such, there is a particular scoring system associated with the response options. Then, under the polyserial model at a parameter vector~$\Btheta$, the marginal probability of the~$j$-th response option,~$y_j$, is given by
\[
	\pymarg{y_j}{\Btheta} = \CDF{\tau_j} - \CDF{\tau_{j-1}}, \qquad j = 1,\dots, r,
\]
so the population mean~$\mu_Y$ and variance~$\ssigma_Y$ of~$Y$ are respectively identified through
\begin{equation}\label{eq:momentsY}
\begin{split}
	\mu_Y &= \E{Y} = \sum_{j=1}^r  y_j\cdot\pymarg{y_j}{\Btheta}, \qquad \textnormal{and}
	\\
	\ssigma_Y &= \var{Y} = \sum_{j=1}^r y_j^2\cdot \pymarg{y_j}{\Btheta} - \mu_Y^2.
\end{split}
\end{equation}
To clearly distinguish between the population moments of~$X$ and~$Y$, we write $\mu_X = \mu$ and $\ssigma_X = \ssigma$ whenever there is a risk of ambiguity. It can be shown that the product moment of the two observed variables is equal to
\begin{equation}\label{eq:EXY}
	\mu_{XY} = \E{XY} 
	=
	\mu_X \left( y_r - \sum_{j=1}^{r-1} \CDF{\tau_j} \big(y_{j+1} - y_j \big) \right) + \rho\sigma_X  \sum_{j=1}^{r-1} \PDF{\tau_j} \big(y_{j+1} - y_j \big),
\end{equation}
where $\rho$ denotes the polyserial correlation coefficient from the normality model in~\eqref{eq:normality} and~$\PDF{\cdot}$ denotes the univariate standard normal density function. 
It follows from the identity $\cov{X}{Y} = \mu_{XY} - \mu_X\mu_Y$  
that the population correlation between~$X$ and~$Y$ can be written as
\begin{equation}\label{eq:pointpolyserial}
	\tilde{\rho} = \cor{X}{Y} = \frac{\cov{X}{Y}}{\sigma_X\sigma_Y}
	=
	\frac{1}{\sigma_X\sigma_Y} \left(\mu_{XY} - \mu_X\mu_Y\right),
\end{equation}
where explicit expressions for~$(\mu_Y, \ssigma_Y)$ and~$\mu_{XY}$ are provided in Equations~\eqref{eq:momentsY} and~\eqref{eq:EXY}, respectively. The correlation coefficient~$\tilde{\rho}$ between the continuous~$X$ and the ordinal~$Y$ is referred to as \emph{point polyserial coefficient}. For a given scoring system,~$\tilde{\rho}$ can be computed from the parameter vector~$\Btheta = \left(\rho, \mu, \ssigma, \Btau^\top\right)^\top$ of the polyserial model, a property established by \citet{olsson1982}. In particular, an estimator of~$\tilde{\rho}$ can be constructed by substituting~$\Btheta$ for an estimate thereof in Equations~\eqref{eq:momentsY}--\eqref{eq:pointpolyserial}.

For the remainder of this appendix, we resort back to the original assumption that the support of the ordinal~$Y$ is given by a set of adjacent integers without numeric interpretation, $\Y = \{1,2,\dots,r\}$, which, we reiterate, is without loss of generality for identifying the polyserial correlation coefficient.

\subsection{Alternative expressions of the polyserial model density}\label{app:olsson1982notation}

The expressions of the polyserial model's density given in this paper in~\eqref{eq:jointpdf} are different but equivalent to the expression used in \citet{olsson1982}. For completeness, we provide the expressions of \citet{olsson1982} in this subsection. 

Under the polyserial model, the joint density of observing a realization~$x\in\R$ of~$X$ and a response $y\in\Y = \{1,\dots,r\}$ of the ordinal~$Y$ at a parameter vector~$\Btheta\in\BTheta$ can decomposed as
\begin{equation}\label{eq:jointpdffactor}
	\pxy{\Btheta} = \px{\Btheta}\py{\Btheta},
\end{equation}
where the right-hand side follows by Bayes' theorem. The marginal density~$\px{\Btheta}$ and the conditional density~$\py{\Btheta}$ are implied by the bivariate normality model~\eqref{eq:normality} and are defined as follows. 

For the marginal density of~$X$ in~\eqref{eq:jointpdffactor}, we have 
\[
	\px{\Btheta} = \frac{1}{\sqrt{2\pi\ssigma}} \exp\left(-\frac{1}{2\ssigma}(x-\mu)^2\right), \qquad x\in\R,
\]
which is the univariate normal density with mean~$\mu$ and variance~$\ssigma$. This density only depends on these two parameters, so it holds true that $\px{\Btheta} = \px{\mu,\ssigma}$.

For the conditional density $Y|X$ in~\eqref{eq:jointpdffactor}, we have 
\[
	\py{\Btheta} = \CDF{\taustar{y}} - \CDF{\taustar{y-1}},\qquad y \in\Y ,\ x\in\R,
\]
where $\CDF{\cdot}$ denotes the univariate standard normal cumulative distribution function, and
\[
	\taustar{y} = \frac{\tau_y - \rho(x-\mu)/\sigma}{\sqrt{1-\rho^2}}.
\]

Finally, it follows from the density~$\pxy{\Btheta}$ in~\eqref{eq:jointpdffactor} that the polyserial model distribution function in~\eqref{eq:jointcdf} can also be expressed as
\begin{equation*}
	\Pxy{\Btheta}  = \int_{-\infty}^x \sum_{\{w\in\Y : w\leq y\}} \pxfun{u}{\Btheta}{} \pyfun{w}{u}{\Btheta}{} \d u,
\end{equation*}
for $x\in\R, y\in\Y$, and parameter vector $\Btheta\in\BTheta$. 

\subsection{Notation}\label{app:notation}
To simplify notation, we introduce the following compact notation. First, write $\z = (x,y)^\top$ to collect continuous-ordinal pairs $(x,y)\in\R\times\Y$, and, analogously, write $\Z = (X,Y)^\top$ for the corresponding random variables. Thus, the polyserial model density~$\pxy{\Btheta}$ now reads $\pzfun{\z}{\Btheta}{}$. Throughout this section, assume that one has access to a random sample of observations $\Z_i = (X_i,Y_i)^\top, i=1,\dots, N$, that are distributed according to the unknown sampling distribution~$\Fxy$ in~\eqref{eq:contamdistXY}.

Furthermore, in a slight abuse of integral notation, for some possibly vector-valued function~$\psi(\z;\Btheta)$ of $\z\in\R\times\Y = \Zset$ and $\Btheta\in\Btheta$, we define the shorthand
\[
	\int_{\Zset} \psi(\z;\Btheta) \pzfun{\z}{\Btheta}{} \d\z
	=
	\int_\R\sum_{y\in\Y} \psi(x,y; \Btheta) \pxyfun{x,y}{\Btheta}{}\d x.
\]
In addition, let
\[
	\gradient\psi(\z; \Btheta) = \partialderivative{\psi(\z; \Btheta)}{\Btheta}
	\qquad\textnormal{and}\qquad
	\hessian\psi(\z; \Btheta) = \partialderivativetwice{\psi(\z; \Btheta)}{\Btheta}
\]
respectively denote the gradient and Hessian with respect to~$\Btheta$ of~$\psi(\z; \Btheta)$.

The Fisher information matrix of the polyserial model at a parameter vector~$\Btheta\in\BTheta$ is defined as
\begin{equation}\label{eq:fisher}
	\fisher{\Btheta}
	=
	\int_{\Zset} \pz{\Btheta}\scorez{\Btheta}\scorez{\Btheta}^\top\d \z,
\end{equation}
where the log-likelihood score function is given by
\begin{align*}
	\scorez{\Btheta} 
	&= \gradient\log\pz{\Btheta} \\
	&= \frac{1}{\pz{\Btheta}}\gradient\pz{\Btheta} \\
	&= \frac{1}{\py{\Btheta}} \gradient\py{\Btheta}
	+
	\frac{1}{\px{\Btheta}} \gradient\px{\Btheta},
\end{align*}
where the third equality follows from~\eqref{eq:jointpdffactor} and the chain rule. Alternatively, under the polyserial model, the Fisher information can be equivalently defined as
\[
	\fisher{\Btheta}
	=
	\int_{\Zset} \pz{\Btheta}\Qfun{\z}{\Btheta}\d \z,
\]
where
\begin{align*}
	\Qfun{\z}{\Btheta}
	&=
	- \hessian\log \pzfun{\z}{\Btheta}{}
	\\
	&=-\Bigg\{
		\frac{1}{\py{\Btheta}} \hessian\py{\Btheta}
		-
		\frac{1}{\pyfun{y}{x}{\Btheta}{2}}
		\gradient\py{\Btheta}\gradient^\top\py{\Btheta} +
	\\
	&\qquad\qquad \frac{1}{\px{\Btheta}} \hessian
	\px{\Btheta} - \frac{1}{\pxfun{x}{\Btheta}{2}} \gradient\px{\Btheta}\gradient^\top\px{\Btheta}
	\Bigg\}
\end{align*}
denotes the negative Hessian matrix of the log-likelihood, where the second equality follows by~\eqref{eq:jointpdffactor} in conjunction with the chain rule and product rule. We provide closed-form expressions of the gradient and Hessian of the densities~$\py{\Btheta}$ and~$\px{\Btheta}$ in Appendix~\ref{app:expressions}. 

\subsection{Main theorem}\label{app:main-theorem}

We start by introducing certain regularity conditions under which \citet{basu1998} establish consistency for the estimand~$\Btheta_0$ as well as asymptotic normality of minimum DPD estimators. These regularity assumptions are stated in Assumption~\ref{assumption} and are adapted to the polyserial model. We omit assumptions that are automatically satisfied for the polyserial model, such as certain differentiability conditions on the postulated model. We refer to \citet{basu1998} for a complete presentation of the assumptions for minimum DPD estimation.

\begin{assumption}\label{assumption}
For any given tuning constant $\alpha \geq 0$, impose the following regularity conditions.
\begin{enumerate}[label={\textnormal{A.\arabic*}}]
	\item The distribution $\Hxy$ in~\eqref{eq:contamdistXY} has the same support as the polyserial model distribution~$\Pxysymb$, that is, $\R\times\Y$.  \label{ass:commonsupport}
	\item The estimand~$\Btheta_0$, being defined as the minimizer of the population DPD $\divergence{\feps}{\pzempty}$, is a global minimum, unique, and an interior point of the parameter space~$\BTheta$. \label{ass:estimand}
	\item The $d\times d$ population matrix $\matfun{J}{\Btheta},\Btheta\in\BTheta$, defined by
\begin{equation}\label{eq:Jmat}
\begin{split}
	\matfun{J}{\Btheta}
	&=
	\int_{\Zset} \pzfun{\z}{\Btheta}{1+\alpha}\scorez{\Btheta}\scorez{\Btheta}^\top \d\z\ +
	\\
	&\qquad 
	\int_{\Zset} \pzfun{\z}{\Btheta}{\alpha} \left( \Qz{\Btheta} - \alpha\scorez{\Btheta}\scorez{\Btheta}^\top \right)\Big(\fepsz - \pz{\Btheta}\Big)\d\z,
\end{split}
\end{equation}
is positive definite in an open neighborhood of the estimand~$\Btheta_0$. \label{ass:Jmat}
	\item In an open neighborhood of~$\Btheta_0$, all third-order partial derivatives (with respect to the parameters) of the population DPD $\divergence{\feps}{\pzempty}$ exist and are finite. \label{ass:finite}
\end{enumerate}
\end{assumption}

Assumption~\ref{ass:commonsupport} requires that the second argument of the distribution~$\Hxy$ takes values in~$\Y$ and that its first argument is real-valued. Such an assumption is natural for compatibility of the two distributions, and ensures well-definedness of the sampling distribution~$\Fxy$ in~\eqref{eq:contamdistXY}. Assumption~\ref{ass:estimand} ensures that the estimand~$\Btheta_0$ is point-identified by the sampling distribution~$\Fxy$. Assumption~\ref{ass:Jmat} is necessary for well-definedness of the estimator's asymptotic covariance matrix because the latter requires invertibility of~$\matfun{J}{\Btheta_0}$, as we shall see. In similar fashion, Assumption~\ref{ass:finite} is also required for existence and finiteness of the asymptotic covariance matrix.

The following theorem establishes the asymptotic properties of our proposed estimator. It follows immediately by Theorem~2 in \citet{basu1998}.

\begin{theorem}\label{thm:main}
	Grant Assumption~\ref{assumption}. For fixed $\alpha \geq 0$, it holds true that
	\[
		\thetahat \convP \Btheta_0
	\]
	as well as
	\[
	\sqrt{N} \left( \thetahat - \Btheta_0 \right)
	\convweak
	\gauss_d\Big(\vec{0}, \matfun{\Sigma}{\Btheta_0} \Big),
	\]
	as $N\to\infty$, where
	\[
		\matfun{\Sigma}{\Btheta_0} = \matinv{J}{\Btheta_0} \matfun{K}{\Btheta_0}\matinv{J}{\Btheta_0}.
	\]
	The $d\times d$ matrices $\matfun{J}{\Btheta_0}$ and $\matfun{K}{\Btheta_0}$ are respectively defined in~\eqref{eq:Jmat} and 
\[
	\matfun{K}{\Btheta}
	=
	\int_{\Zset} \fepsz \pzfun{\z}{\Btheta}{2\alpha} \scorez{\Btheta}\scorez{\Btheta}^\top\d\z
	-
	\vecfun{\xi}{\Btheta}\vecfun{\xi}{\Btheta}^\top,
	\qquad \Btheta\in\BTheta,
\]
where
\[
	\vecfun{\xi}{\Btheta} 
	=
	\int_{\Zset} \fepsz \pzfun{\z}{\Btheta}{\alpha} \scorez{\Btheta}\d\z
\]
is a $d$-vector, and both matrices are positive definite at $\Btheta_0$.
\end{theorem}

It is worth mentioning that the proof of \citet{basu1998} follows a classic argument for establishing asymptotic normality, namely the proof of Theorem~3.10 in \citet{lehmann1998point}, but adapts it to the more general case of minimum DPD estimation because the proof of \citet{lehmann1998point} is restricted to ML estimation.

Being a population object, the asymptotic covariance matrix~$\matfun{\Sigma}{\Btheta_0}$ is unobserved in practice. We describe in Appendix~\ref{app:covmat-estimation} how it can be estimated in practice. Before that, though, we describe in the following section how it relates to the asymptotic covariance matrix of the MLE.

\subsection{Uncertainty quantification: Comparison with ML}\label{app:covmatcomparison}

The matrix $\matfun{J}{\Btheta}$ in~\eqref{eq:Jmat} can be compactly expressed as
\[
	\matfun{J}{\Btheta} = 
	\matfun{A}{\Btheta} + \matfun{B}{\Btheta}, 
\]
where
\[
	\matfun{A}{\Btheta} =
	\int_{\Zset}\pzfun{\z}{\Btheta}{1+\alpha}\left( (1+\alpha) \scorez{\Btheta}\scorez{\Btheta}^\top - \Qz{\Btheta} \right)\d\z 
\]
only depends on functions of the polyserial model,
and
\[
	\matfun{B}{\Btheta} = \int_{\Zset} \fepsz\pzfun{\z}{\Btheta}{\alpha}
	\left( \Qz{\Btheta} - \alpha\scorez{\Btheta}\scorez{\Btheta}^\top \right)\d\z,
\]
depends on the possibly contaminated sampling density~$\feps$. 

Recall from Section~\ref{sec:estimand} that if the polyserial model is correctly specified ($\varepsilon = 0$), we have that
the estimand corresponds to the true parameter, that is, $\Btheta_0 = \Btheta_*$, and, furthermore, $f_{0,\vec{Z}}(\z) = \pz{\Btheta_*}$ for all $\z\in\Z$. In this zero-contamination case (and this case only), one can easily verify that we have for the MLE (tuning constant choice $\alpha = 0$) that
\[
	\int_{\Zset} \fepsz \pzfun{\z}{\Btheta_*}{2\alpha} \scorez{\Btheta_*}\scorez{\Btheta_*}^\top\d\z = \fisher{\Btheta_*}
\]
as well as
\begin{align*}
	\matfun{A}{\Btheta_*} 
	&=
	\fisher{\Btheta_*}-\fisher{\Btheta_*} = \mat{0},
	\\
	\matfun{B}{\Btheta_*} &= \fisher{\Btheta_*},
	\\
	\vecfun{\xi}{\Btheta_*}
	&=
	\vec{0},
\end{align*}
so it follows from the expressions derived in Theorem~\ref{thm:main} that
\[
	\matfun{\Sigma}{\Btheta_*} = \fisher{\Btheta_*}^{-1}.
\] 
The fact that under correct model specification, the asymptotic covariance matrix of $\sqrt{N}\left(\thetahatMLE - \Btheta_*\right)$ equals the inverse of the Fisher information matrix at the true parameter~$\Btheta_*$ is a familiar property of the MLE, which is nested by Theorem~\ref{thm:main}. Conversely, if the polyserial model is misspecified ($\varepsilon > 0$), one can show that the MLE's asymptotic covariance matrix, being $\matfun{\Sigma}{\Btheta_0}$ at $\alpha = 0$, is equal to well-known sandwich-type covariance matrix expressions derived by \citet{white1982} and \citet{huber1967}. 

For strictly positive choices of~$\alpha$, we obtain a more robust estimator than the MLE. However, when $\alpha > 0$ and the model is correctly specified ($\varepsilon = 0$), one can immediately see that
\begin{align*}
\int_{\Zset} \fepsz \pzfun{\z}{\Btheta_*}{2\alpha} \scorez{\Btheta_*}\scorez{\Btheta_*}^\top\d\z 
&\neq \fisher{\Btheta_*},
\\
	\matfun{A}{\Btheta_*} 
	&\neq\mat{0},
	\\
	\matfun{B}{\Btheta_*} &\neq \fisher{\Btheta_*},
	\\
	\vecfun{\xi}{\Btheta_*}
	&\neq
	\vec{0},
\end{align*}
which is due to the presence of terms involving~$\alpha$ that do \emph{not} vanish when $\alpha > 0$. Thus, in the absence of contamination, our estimator's asymptotic covariance matrix is different than that of the MLE. It now follows from the Cramér-Rao lower bound that our estimator is \emph{not} efficient and, consequently, has a larger estimation variance than the MLE. As discussed in Section~\ref{sec:asymptotics}, a loss of efficiency at the postulated model is a common property of many robust estimators, which can be seen as the price of robustness.

\subsection{Covariance matrix estimation}\label{app:covmat-estimation}
The asymptotic covariance matrix $\matfun{\Sigma_0}{\Btheta_0}$ is unobserved in practice because it depends on the unknown sampling density~$\feps$ as well as the unknown estimand~$\Btheta_0$. This section explains how a consistent estimator of~$\matfun{\Sigma}{\Btheta_0}$ can be constructed. 

First, in the definitions of the population objects $\big(\matfun{B}{\Btheta}, \matfun{J}{\Btheta}, \vecfun{\xi}{\Btheta},  \matfun{K}{\Btheta}\big)$, replace the unknown sampling density~$\fepsz$ by its empirical counterpart in~\eqref{eq:fhat}, that is,
\[
	\fhat(\z) = \frac{1}{N}\sum_{i=1}^N\I{\Z_i = \z},\qquad \z\in\Zset,
\]
resulting in the sample objects
\begin{align*}
	\hatmatfun{B}{\Btheta}
	&=
	\frac{1}{N}\sum_{i=1}^N \pxyfun{X_i,Y_i}{\Btheta}{\alpha} 
	\left(
		\Qfun{X_i, Y_i}{\Btheta} -\alpha\scorefun{\Btheta}{X_i, Y_i}\scorefun{\Btheta}{X_i, Y_i}^\top
	\right),
	\\
	\hatmatfun{J}{\Btheta} 
	&=
	 \matfun{A}{\Btheta} + \hatmatfun{B}{\Btheta},
	\\
	\hatmatfun{\xi}{\Btheta}
	&=
	\frac{1}{N}\sum_{i=1}^N\pxyfun{X_i,Y_i}{\Btheta}{\alpha} \scorefun{\Btheta}{X_i, Y_i},
	\\
	\hatmatfun{K}{\Btheta}
	&=
	\frac{1}{N} \sum_{i=1}^N \pxyfun{X_i, Y_i}{\Btheta}{2\alpha} \scorefun{\Btheta}{X_i, Y_i}\scorefun{\Btheta}{X_i, Y_i}^\top - \hatmatfun{\xi}{\Btheta}\hatmatfun{\xi}{\Btheta}^\top.
\end{align*}
Observe that the matrix $\matfun{A}{\Btheta}$ in $\hatmatfun{J}{\Btheta}$ does not require a sample counterpart because it does not depend on the unknown~$\feps$, so it can be computed for a given parameter~$\Btheta$.

\citet{basu1998} show that for a given $\Btheta\in\BTheta$, the sample matrices $\big(\hatmatfun{B}{\Btheta}, \hatmatfun{J}{\Btheta}, \hatmatfun{\xi}{\Btheta},  \hatmatfun{K}{\Btheta}\big)$ are pointwise consistent estimators of their corresponding population counterparts $\big(\matfun{B}{\Btheta}, \matfun{J}{\Btheta}, \vecfun{\xi}{\Btheta},  \matfun{K}{\Btheta}\big)$, as $N\to\infty$. Thus, it follows from the continuous mapping theorem that 
\[
	\hatmatfun{\Sigma}{\Btheta} = \hatmatinv{J}{\Btheta}\hatmatfun{K}{\Btheta}\hatmatinv{J}{\Btheta}
\]
is pointwise consistent for $\matfun{\Sigma}{\Btheta} = \matinv{J}{\Btheta} \matfun{K}{\Btheta}\matinv{J}{\Btheta}$. In particular, by $\thetahat\convP\Btheta_0$ in conjunction with the continuous mapping theorem, the matrix $\hatmatfun{\Sigma}{\thetahat}$ is a consistent estimator of the asymptotic covariance matrix~$\matfun{\Sigma}{\Btheta_0}$ in Theorem~\ref{thm:main}.

Regarding computation, for a given $\Btheta\in\BTheta$, computation of the objects $\big(\edit{\hatmatfun{B}{\Btheta}}, \edit{\hatmatfun{\xi}{\Btheta}},  \edit{\hatmatfun{K}{\Btheta}}\big)$ is cheap because they only require computation of finite summations of terms with closed-form expressions. The only computational bottleneck is the matrix~$\matfun{A}{\Btheta}$, which is a component of~$\hatmatfun{J}{\Btheta}$ and is defined as
\begin{align*}
		\matfun{A}{\Btheta} 
		&=
	\int_{\Zset}\pzfun{\z}{\Btheta}{1+\alpha}\left( (1+\alpha) \scorez{\Btheta}\scorez{\Btheta}^\top - \Qz{\Btheta} \right)\d\z 
	\\
	&=
	\int_\R
	\pxfun{x}{\Btheta}{1+\alpha} \sum_{y\in\Y} \pyfun{y}{x}{\Btheta}{1+\alpha}
	\left( (1+\alpha) \scorexy{\Btheta}\scorexy{\Btheta}^\top - \Qxy{\Btheta} \right)\d x ,
\end{align*}
where the second equality follows by~\eqref{eq:jointpdffactor}. As such, due to the presence of the one-dimensional integral over the real numbers, the matrix~$\matfun{A}{\Btheta}$ does not posses a closed-form expression. Hence, in practice, the integral needs to be numerically approximated. Since~$\matfun{A}{\Btheta}$ is a $d\times d$ matrix,~$d^2$ one-dimensional integrals need to be computed in elementwise manner. However, since~$\matfun{A}{\Btheta}$ is symmetric---owing to the symmetry of the Hessian~$\Qxy{\Btheta}$ and the outer matrix~$\scorexy{\Btheta}\scorexy{\Btheta}^\top$---it suffices to only compute the $d(d+1)/2$ unique one-dimensional integrals. 

As for directly estimating the Fisher information matrix, a commonly used ML-based estimator is the sample average
\[
	\frac{1}{N} \sum_{i=1}^N  \scorefun{\thetahatMLE}{\vec{Z}_i}\scorefun{\thetahatMLE}{\vec{Z}_i}^\top,
\] 
which is consistent for the true Fisher information matrix~$\fisher{\Btheta_*}$ in~\eqref{eq:fisher} if the polyserial model is correctly specified for all observations $(\varepsilon = 0)$. This estimator is, for instance, the default way ML standard errors for the polyserial model are computed in the package \pkg{lavaan} \citep{lavaan}. Likewise, we use this estimator for computing covariance matrix estimates for the MLE.

\subsection{Gradient and Hessian expressions}\label{app:expressions}
This section derives closed-form expressions of the gradient and Hessian of the model density~$\pxy{\Btheta}$ for $(x,y)\in\R\times\Y$ and 
\[
	\Btheta = \big(\rho, \mu, \sigma^2, \vec{\tau}^\top\big)^\top\in\BTheta,
\] 
where $\vec{\tau} = (\tau_1, \dots, \tau_{-1})^\top$, and we have the restrictions that $\rho\in(-1,1)$, $\sigma > 0$, and $-\infty < \tau_1 < \cdots < \tau_{r-1} < \infty$. We use the density expressions used in \citet{olsson1982} that are provided in Appendix~\ref{app:olsson1982notation}.

For further reference, the density of the univariate standard normal distribution is defined as
\[
	\PDF{v} = \frac{1}{\sqrt{2\pi}} \exp\big( -v^2/2 \big)
\]
with first derivative
\[
	\PDFprime{v} = -v\PDF{v}
\]
and associated cumulative distribution function
\[
	\CDF{v} = \int_{-\infty}^v \PDF{s}\d s,
\]
for $v\in\R$.

\subsubsection{First order expressions}\label{app:firstorder}
By the chain rule and definition of~$\pxy{\Btheta}$ in~\eqref{eq:jointpdffactor}, it holds true that 
\[
	\partialderivative{}{\Btheta}\pxy{\Btheta}
	=
	\px{\Btheta}\partialderivative{\py{\Btheta}}{\Btheta}
	+
	\py{\Btheta}\partialderivative{\px{\Btheta}}{\Btheta},
\]
where
\[
	\partialderivative{\px{\Btheta}}{\Btheta}
	=
	\left(
		\partialderivative{\px{\Btheta}}{\rho}, 
		\partialderivative{\px{\Btheta}}{\mu},
		\partialderivative{\px{\Btheta}}{(\sigma^2)},
		\partialderivative{\px{\Btheta}}{\tau_1},
		\dots,
		\partialderivative{\px{\Btheta}}{\tau_{r-1}}
	\right)^\top
\]
is the gradient of the marginal density, and
\begin{align*}
	&\partialderivative{\py{\Btheta}}{\Btheta}
	=
	\\
	&\quad\left(
		\partialderivative{\py{\Btheta}}{\rho}, 
		\partialderivative{\py{\Btheta}}{\mu},
		\partialderivative{\py{\Btheta}}{(\sigma^2)},
		\partialderivative{\py{\Btheta}}{\tau_1},
		\dots,
		\partialderivative{\py{\Btheta}}{\tau_{r-1}}
	\right)^\top
\end{align*}
is the gradient of the conditional density. In the following, we provide closed-form expressions for each component in these two gradients.

Starting with $\partialderivative{\px{\Btheta}}{\Btheta}$, \citet{olsson1982} show that 
\[
	\partialderivative{\px{\Btheta}}{\mu}
	=
	\px{\Btheta}\frac{x - \mu}{\sigma^2}
\]
and
\[
	\partialderivative{\px{\Btheta}}{(\sigma^2)}
	=
	\frac{\px{\Btheta}}{2\sigma^2} \left( \left( \frac{x-\mu}{\sigma} \right)^2 - 1  \right).
\]
Since the marginal density only depends on~$\mu$ and~$\sigma^2$, all of its remaining derivatives are zero-valued, that is,
\[
	\partialderivative{\px{\Btheta}}{\rho}
	=
	\partialderivative{\px{\Btheta}}{\tau_1}
	=
	\cdots
	=
	\partialderivative{\px{\Btheta}}{\tau_{r-1}}
	=
	0.
\]

For the conditional density's gradient  $\partialderivative{\py{\Btheta}}{\Btheta}$, \citet{olsson1982} show that
\begin{align*}
	&\partialderivative{\py{\Btheta}}{\rho}
	=
	\\
	&\qquad
	(1-\rho^2)^{-3/2}
	\left(
	\PDF{\taustar{y}} \left( \rho\tau_y - \frac{x-\mu}{\sigma}\right)
	-
	\PDF{\taustar{y-1}} \left( \rho\tau_{y-1} - \frac{x-\mu}{\sigma}\right)
	\right)
\end{align*}
and
\[
	\partialderivative{\py{\Btheta}}{\mu}
	=
	\frac{\rho}{\sigma\sqrt{1-\rho^2}}
	\Big(
		\PDF{\taustar{y}} - \PDF{\taustar{y-1}}
	\Big)
\]
and
\[
	\partialderivative{\py{\Btheta}}{(\sigma^2)}
	=
	\frac{\rho(x-\mu)}{2\sigma^3\sqrt{1-\rho^2}}
	\Big(
		\PDF{\taustar{y}} - \PDF{\taustar{y-1}}
	\Big)
\]
and, for $k=1, \dots, r-1$,
\[
	\partialderivative{\py{\Btheta}}{\tau_k}
	=
	\begin{cases}
		\PDF{\taustar{y}} \Big/ \sqrt{1-\rho^2} & \textnormal{ if } k = y,
		\\
		-\PDF{\taustar{y-1}} \Big/ \sqrt{1-\rho^2} & \textnormal{ if } k = y-1,
		\\
		0 & \textnormal{ otherwise.}
	\end{cases}
\]

\subsubsection{Ancillary derivatives}
Before we turn to second order derivatives, it is useful to derive the gradient of~$\taustar{y}$ with respect to~$\Btheta$. It can be shown that
\[
	\partialderivative{\taustar{y}}{\rho}
	=
	\frac{1}{\sqrt{1-\rho^2}}
	\left(
		\frac{\taustar{y}}{\sqrt{1-\rho^2}} - \frac{x-\mu}{\sigma}
	\right)
\]
and
\[
	\partialderivative{\taustar{y}}{\mu}
	=
	\frac{\rho}{\sigma\sqrt{1-\rho^2}}
\]
and
\[
	\partialderivative{\taustar{y}}{(\sigma^2)}
	=
	\frac{\rho(x-\mu)}{2\sigma^3\sqrt{1-\rho^2}}
\]
and, for $k=1,\dots,r-1$,
\[
	\partialderivative{\taustar{y}}{\tau_k}
	=
	\begin{cases}
	1 \Big/\sqrt{1-\rho^2} & \textnormal{ if } k = y,
	\\
	0 & \textnormal{ otherwise}.
	\end{cases}
\]

\subsubsection{Second order expressions}
The (symmetric) Hessian  matrix of the marginal density is given by
\begin{equation}\label{eq:hessian}
\partialderivativetwice{\px{\Btheta}}{\Btheta}
=
\begin{pmatrix}
	\partialderivativetwiced{\px{\Btheta}}{\rho}{\rho}
	&
	\partialderivativetwiced{\px{\Btheta}}{\rho}{\mu}
	&
	\partialderivativetwiced{\px{\Btheta}}{\rho}{(\sigma^2)}
	&
	\partialderivativetwiced{\px{\Btheta}}{\rho}{\tau_1}
	&
	\cdots
	&
	\partialderivativetwiced{\px{\Btheta}}{\rho}{\tau_{r-1}}
	\\
	\partialderivativetwiced{\px{\Btheta}}{\mu}{\rho}
	&
	\partialderivativetwiced{\px{\Btheta}}{\mu}{\mu}
	&
	\partialderivativetwiced{\px{\Btheta}}{\mu}{(\sigma^2)}
	&
	\partialderivativetwiced{\px{\Btheta}}{\mu}{\tau_1}
	&
	\cdots
	&
	\partialderivativetwiced{\px{\Btheta}}{\mu}{\tau_{r-1}}
	\\
	\partialderivativetwiced{\px{\Btheta}}{(\sigma^2)}{\rho}
	&
	\partialderivativetwiced{\px{\Btheta}}{(\sigma^2)}{\mu}
	&
	\partialderivativetwiced{\px{\Btheta}}{(\sigma^2)}{(\sigma^2)}
	&
	\partialderivativetwiced{\px{\Btheta}}{(\sigma^2)}{\tau_1}
	&
	\cdots
	&
	\partialderivativetwiced{\px{\Btheta}}{(\sigma^2)}{\tau_{r-1}}
	\\
	\partialderivativetwiced{\px{\Btheta}}{\tau_1}{\rho}
	&
	\partialderivativetwiced{\px{\Btheta}}{\tau_1}{\mu}
	&
	\partialderivativetwiced{\px{\Btheta}}{\tau_1}{(\sigma^2)}
	&
	\partialderivativetwiced{\px{\Btheta}}{\tau_1}{\tau_1}
	&
	\cdots
	&
	\partialderivativetwiced{\px{\Btheta}}{\tau_1}{\tau_{r-1}}
	\\
	\vdots
	&
	\vdots
	&
	\vdots
	&
	\vdots
	&
	\ddots
	&
	\vdots
	\\
	\partialderivativetwiced{\px{\Btheta}}{\tau_{r-1}}{\rho}
	&
	\partialderivativetwiced{\px{\Btheta}}{\tau_{r-1}}{\mu}
	&
	\partialderivativetwiced{\px{\Btheta}}{\tau_{r-1}}{(\sigma^2)}
	&
	\partialderivativetwiced{\px{\Btheta}}{\tau_{r-1}}{\tau_1}
	&
	\cdots
	&
	\partialderivativetwiced{\px{\Btheta}}{\tau_{r-1}}{\tau_{r-1}}
\end{pmatrix},
\end{equation}
and the Hessian of the conditional density~$\partialderivativetwice{\py{\Btheta}}{\Btheta}$ is constructed analogously by replacing~$\px{\Btheta}$ by~$\py{\Btheta}$ in~\eqref{eq:hessian}. We proceed by deriving the second order derivatives of both the marginal and the conditional density.

For the second order derivatives of the marginal density~$\px{\Btheta}$, tedious but straightforward applications of the chain and product rule yield
\[
	 \partialderivativetwiced{\px{\Btheta}}{\mu}{\mu}
	 =
	 \frac{1}{\sigma^2} \left( (x-\mu) \partialderivative{\px{\Btheta}}{\mu} - \px{\Btheta} \right)
\]
and 
\[
	 \partialderivativetwiced{\px{\Btheta}}{\mu}{\sigma^2}
	 =
	 \frac{x-\mu}{\sigma^2}\left( \partialderivative{\px{\Btheta}}{(\sigma^2)} - \frac{\px{\Btheta}}{\sigma^2} \right)
\]
and
\[
 \partialderivativetwiced{\px{\Btheta}}{(\sigma^2)}{(\sigma^2)}
 =
 \frac{1}{2\sigma^2}
 \left(
 	\partialderivative{\px{\Btheta}}{(\sigma^2)} \left( \left(\frac{x-\mu}{\sigma}\right)^2 - 1 \right)
 +
 \frac{\px{\Btheta}}{\sigma^2}
 \left(1 - 2\left(\frac{x-\mu}{\sigma}\right)^2\right)	
 \right).
\]
Because the marginal density only depends on~$\mu$ and~$\sigma^2$, its remaining second order derivatives are all zero. It follows that among all unique elements in the Hessian in~\eqref{eq:hessian}, only $\partialderivativetwiced{\px{\Btheta}}{\mu}{\mu}$, $\partialderivativetwiced{\px{\Btheta}}{\mu}{(\sigma^2)}$, and $\partialderivativetwiced{\px{\Btheta}}{(\sigma^2)}{(\sigma^2)}$ are nonzero. 

For the second order derivatives of the conditional density~$\py{\Btheta}$, tedious but straightforward applications of the chain and product rule yield
\begin{align*}
	&\partialderivativetwiced{\py{\Btheta}}{\rho}{\rho}
	=
	\\
	&\quad\frac{3\rho}{(1-\rho^2)^{5/2}}
	\left( 
		\PDF{\taustar{y}}\left( \rho\tau_y - \frac{x-\mu}{\sigma} \right)
		-
		\PDF{\taustar{y-1}}\left( \rho\tau_{y-1} - \frac{x-\mu}{\sigma} \right)
	\right) 
	+
	\\
	&\qquad
	\frac{1}{(1-\rho^2)^{3/2}}
	\Bigg[
		\PDFprime{\taustar{y}}\left( \rho\tau_y - \frac{x-\mu}{\sigma} \right) \partialderivative{\taustar{y}}{\rho} + \tau_y\PDF{\taustar{y}} \edit{-} 
		\\
		&\qquad\qquad\qquad\qquad
		\PDFprime{\taustar{y-1}}\left( \rho\tau_{y-1} - \frac{x-\mu}{\sigma} \right) \partialderivative{\taustar{y-1}}{\rho} \edit{-} \tau_{y-1}\PDF{\taustar{y-1}}
	\Bigg]
\end{align*}
and 
\[
	\partialderivativetwiced{\py{\Btheta}}{\mu}{\mu}
	=
	\frac{\rho}{\sigma\sqrt{1-\rho^2}}
	\left(
		\PDFprime{\taustar{y}}\partialderivative{\taustar{y}}{\mu}
		-
		\PDFprime{\taustar{y-1}}\partialderivative{\taustar{y-1}}{\mu}
	\right)
\]
and
\begin{align*}
	\partialderivativetwiced{\py{\Btheta}}{(\sigma^2)}{(\sigma^2)}
	&=
	\frac{\rho(x-\mu)}{2\sigma^3\sqrt{1-\rho^2}}
	\Bigg[
		-\frac{3}{2\sigma^2}\Big( 
		\PDF{\taustar{y}} - \PDF{\taustar{y-1}} \Big)	 + 
	\\
	&\qquad 
		\PDFprime{\taustar{y}}	\partialderivative{\taustar{y}}{(\sigma^2)}
		-
		\PDFprime{\taustar{y-1}}	\partialderivative{\taustar{y-1}}{(\sigma^2)}
	\Bigg]
\end{align*}
and, for $k=1,\dots, r-1$,
\[
	\partialderivativetwiced{\py{\Btheta}}{\tau_k}{\tau_k}
	=
	\begin{cases}
		\frac{1}{\sqrt{1-\rho^2}}
		\PDFprime{\taustar{y}} \partialderivative{\taustar{y}}{\tau_y}
		& \textnormal{ if } k = y,
		\\
		-\frac{1}{\sqrt{1-\rho^2}}
		\PDFprime{\taustar{y-1}} \partialderivative{\taustar{y-1}}{\tau_{y-1}}
		& \textnormal{ if } k = y-1,
		\\
		0 &\textnormal{ otherwise}.
	\end{cases}
\]
For the cross-derivatives, it can be shown that
\begin{align*}
	\partialderivativetwiced{\py{\Btheta}}{\rho}{\mu}
	&=
	\frac{1}{\sigma\sqrt{1-\rho^2}}
	\Bigg[
		\bigg( 1+\frac{\rho^{\edit{2}}}{1-\rho^2} \bigg)
		\bigg( \PDF{\taustar{y}} - \PDF{\taustar{y-1}} \bigg) +
	\\
	&\qquad \rho\bigg(
		\PDFprime{\taustar{y}}\partialderivative{\taustar{y}}{\rho}
		-
		\PDFprime{\taustar{y-1}}\partialderivative{\taustar{y-1}}{\rho}
	\bigg)
	\Bigg]
\end{align*}
and
\begin{align*}
	\partialderivativetwiced{\py{\Btheta}}{\rho}{(\sigma^2)}
	&=
	\frac{x-\mu}{\edit{2}\sigma^3\sqrt{1-\rho^2}}
	\Bigg[
		\bigg( 1+\frac{\rho^{\edit{2}}}{1-\rho^2} \bigg)
		\bigg( \PDF{\taustar{y}} - \PDF{\taustar{y-1}} \bigg) +
	\\
	&\qquad \rho\bigg(
		\PDFprime{\taustar{y}}\partialderivative{\taustar{y}}{\rho}
		-
		\PDFprime{\taustar{y-1}}\partialderivative{\taustar{y-1}}{\rho}
	\bigg)
	\Bigg]
\end{align*}
and
\begin{align*}
	\partialderivativetwiced{\py{\Btheta}}{(\sigma^2)}{\mu}
	&=
	\frac{\rho}{\sigma\sqrt{1-\rho^2}}
	\Bigg[
		\PDFprime{\taustar{y}}\partialderivative{\taustar{y}}{(\sigma^2)}
		-
		\PDFprime{\taustar{y-1}}\partialderivative{\taustar{y-1}}{(\sigma^2)} - 
	\\
	&\qquad 
	\frac{1}{2\sigma^2}\bigg(
		\PDF{\taustar{y}} - \PDF{\taustar{y-1}}	
	\bigg)
	\Bigg]
\end{align*}
and, for $k=1,\dots,r-1$,
\[
	\partialderivativetwiced{\py{\Btheta}}{\mu}{\tau_k}
	=
	\begin{cases}
		\frac{\rho}{\sigma(1-\rho^2)} \PDFprime{\taustar{y}} & \textnormal{ if } k = y,
		\\
		-\frac{\rho}{\sigma(1-\rho^2)} \PDFprime{\taustar{y-1}} & \textnormal{ if } k = y-1,
		\\
		0 & \textnormal{ otherwise,}
	\end{cases}
\]
and
\[
	\partialderivativetwiced{\py{\Btheta}}{(\sigma^2)}{\tau_k}
	=
	\begin{cases}
		\frac{\rho(x-\mu)}{2\sigma^3(1-\rho^2)} \PDFprime{\taustar{y}} & \textnormal{ if } k = y,
		\\
		-\frac{\rho(x-\mu)}{2\sigma^3(1-\rho^2)} \PDFprime{\taustar{y-1}} & \textnormal{ if } k = y-1,
		\\
		0 & \textnormal{ otherwise,}
	\end{cases}
\]
and 
\[
	\partialderivativetwiced{\py{\Btheta}}{\rho}{\tau_k}
	=
	\begin{cases}
	\partialderivative{}{\rho}\left[ \frac{\PDF{\taustar{y}}}{\sqrt{1-\rho^2}}\right] & \textnormal{ if } k = y,
	\\
	-\partialderivative{}{\rho}\left[ \frac{\PDF{\taustar{y-1}}}{\sqrt{1-\rho^2}}\right] & \textnormal{ if } k = y-1,
	\\
	0 & \textnormal{ otherwise,}
	\end{cases}
\]
where
\[
	\partialderivative{}{\rho}\left[ \frac{\PDF{\taustar{k}}}{\sqrt{1-\rho^2}}\right] 
	=
	\frac{\rho}{(1-\rho^2)^{3/2}} \PDF{\taustar{k}} + \frac{1}{\sqrt{1-\rho^2}}\PDFprime{\taustar{k}} \partialderivative{\taustar{k}}{\rho},
\]
for $k\in\{1,\dots, r-1\}$\edit{, and, finally,}
\edit{
\[
	\partialderivativetwiced{\py{\Btheta}}{\tau_i}{\tau_j} = 0
\]
}
\edit{for $i\neq j$.} 

\edit{We verified the correctness of all analytical expressions of derivatives provided in this Appendix~\ref{app:expressions} using numerical approximations. The corresponding code is provided in the online replication material.}

\newpage
\section{Algorithm for upper bound of raw weights}\label{app:algo}
Recall from Section~\ref{sec:proposethetahat} that the nonnegative individual-specific raw weights in estimating equation~\eqref{eq:estimatingeq} are defined as
\[
	\tilde{w}_{i,\alpha}(\Btheta) = \pxyfun{X_i, Y_i}{\Btheta}{\alpha}, \qquad i = 1,\dots, N,
\]
which are bounded from below by~0 and bounded from above by
\begin{equation}\label{eq:upperbound}
	M_{\alpha}(\Btheta) = \sup\big\{\pxyfun{x, y}{\Btheta}{\alpha} : x\in\R, y\in\Y\big\}.
\end{equation}
The upper bound $M_{\alpha}(\Btheta)$ does not depend on observed data and is finite because it is assumed in the polyserial model that $\sigma^2 > 0$ and $\rho\in (-1,1)$. We require this upper bound to construct the rescaled weights
\[
	w_{i,\alpha}(\Btheta) = \tilde{w}_{i,\alpha}(\Btheta) \big/ M_{\alpha}(\Btheta)
\] 
in Equation~\eqref{eq:weights} so that the (rescaled) weights take values in $[0,1]$. In the following, we describe an algorithm to compute the upper bound~$M_{\alpha}(\Btheta)$. The algorithm exploits the fact that the domain of the second argument of~$\pxyfun{\cdot, \cdot}{\Btheta}{\alpha}$ is finite, namely~$\Y$.

\begin{enumerate}
  \setcounter{enumi}{-1}
	\item Fix $\Btheta = (\rho,\mu,\sigma^2,\tau_1,\dots,\tau_{r-1})^\top\in\BTheta$ and $\alpha > 0$. Usually, these are respectively an estimate~$\thetahat$ and the tuning constant used to obtain that estimate in problem~\eqref{eq:estimator}.
	\item Fix a response option~$y\in\Y$ and maximize the objective function
	\begin{equation}\label{eq:weights-objective}
		S_y(x) = \pxyfun{x, y}{\Btheta}{\alpha}
	\end{equation}
	with respect to $x\in\R$, keeping~$y$ fixed. Denote by~$S_y^*$ the value of the objective at the argmax. 
	\item Repeat Step~1 for all $y\in\Y$ and return the upper bound $\edit{M}_{\alpha}(\Btheta) = \max\left\{S^*_y : y \in\Y\right\}$. 
\end{enumerate}
Note that the scalar gradient of the objective~$S_y(x)$ in~\eqref{eq:weights-objective} can be shown to be equal to
	\[
		\alpha \pxyfun{x, y}{\Btheta}{\alpha-1} \int_{\tau_{y-1}}^{\tau_y} \left( \partialderivative{}{x} \pxetafun{x,v}{\Btheta}{}  \right) \d v,
	\]
	which follows by the chain rule in conjunction with the Leibniz integral rule, and where
	\[
		\partialderivative{}{x} \pxetafun{x,v}{\Btheta}{}
		=
		-\frac{1}{\sigma (1-\rho^2)} \left(\frac{x-\mu}{\sigma} - \rho v\right)\pxetafun{x,v}{\Btheta}{}, \qquad x,v\in\R,
	\]
	which follows by the definition of the bivariate normal density corresponding to the distribution in~\eqref{eq:normality}. This gradient can be used for maximizing the objective~$S_y(x)$ by using standard methods for numerical optimization. In our implementation, we use the BFGS algorithm \citep[e.g.,][Section~6.1]{nocedal2006} and initialize at~$x=\mu$.

\newpage
\section{Additional results from the main text}\label{app:additional-results}
\subsection{Additional results for efficiency}
\begin{figure}
	\centering
	\includegraphics[width = \textwidth]{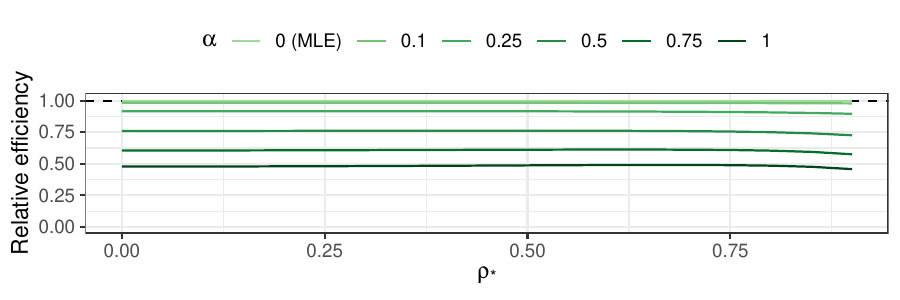}
\caption{Relative efficiency of the robust estimator (with respect to the MLE) for various choices of the tuning constant~$\alpha$ (lines), expressed as a function of the true polyserial correlation coefficient~$\rho_*$ ($x$-axis) where the remaining true parameters are fixed at $\left(\mu_*, \sigma_*^2, \tau_{*,1}, \tau_{*,2}, \tau_{*,3}, \tau_{*,4} \right)^\top =  \left(0, 1, -1.5, -0.5, 0.5, 1.5 \right)^\top$.}
\label{fig:efficiency}
\end{figure}

Table~\ref{tab:efficiency} in Section~\ref{sec:efficiency} lists the relative efficiency (with respect to the MLE) of the robust estimator at the true parameter vector $\Btheta_* = \left(\rho_*, \mu_*, \sigma_*^2, \tau_{*,1}, \tau_{*,2}, \tau_{*,3}, \tau_{*,4} \right)^\top =  \left(0.5, 0, 1, -1.5, -0.5, 0.5, 1.5 \right)^\top$, for various choices of tuning constant~$\alpha$. Figure~\ref{fig:efficiency} visualizes the relative efficiencies as a function of the true polyserial correlation coefficient~$\rho_*$ with the remaining parameters in~$\Btheta_*$ fixed. The relative efficiencies of the robust estimators stay nearly constant, except for a small dip towards large correlations of about~$\rho_* = 0.9$. Our package \pkg{robcat} provides functionality to compute statistical efficiency at custom values of~$\Btheta_*$.

\subsection{Definition of performance measures}\label{app:perfmeasures}
A performance measure used in Section~\ref{sec:simulation} is the approximate bias of the standard error estimation, which is defined as the difference between the average estimated standard error (averages over the simulation repetitions) and the standard deviation of the simulated sampling distribution of the correlation estimates. We rigorously define this performance measure in the following.

Denote by~$\rhohatt$ an estimate of the polyserial correlation coefficient in a $t$-th simulation repetition, $t=1,\dots,T$, where $T=5,000$ for all simulation studies in this paper. The sample standard deviation of the~$T$ individual correlation estimates is given by 
\[
		\SEapprox{\rhohat} = \sqrt{\frac{1}{T-1}\sum_{t=1}^T \left(\rhohatt - \rhohatmean \right)^2},
\]
where
\[
		\rhohatmean = \frac{1}{T}\sum_{t=1}^T\hatN{\rho}^{(t)},
\]
is the sample mean of the estimates. The estimates' sample standard deviation~$\SEapprox{\rhohat}$ is an approximation to the unknown finite-sample standard error of the correlation estimator. While an asymptotic expression of the standard error exists---being the square root of the top left element of~$\matfun{\Sigma}{\Btheta_0}$ in Theorem~\ref{thm:main}---the true standard error~$\SE{\rhohat}$ in finite samples is unknown. We therefore use the sample standard deviation~$\SEapprox{\rhohat}$ as an approximation.

The performance measure of the approximate bias of standard error estimation is now defined as
\[
	\SEbarhat{\rhohat} - \SEapprox{\rhohat},
\]
where
\[
		\SEbarhat{\rhohat} = \frac{1}{T}\sum_{t=1}^T \SEhat{\rhohatt},
\]
is the sample mean of the individual correlation standard error estimates. If the asymptotic theory in Theorem~\ref{thm:main} is correct, then, for a sufficiently large sample size,~$\SEbarhat{\rhohat}$ should be close to the finite sample approximation for the true standard error,~$\SEapprox{\rhohat}$.

\subsection{Additional simulation results}
Figure~\ref{fig:main-simresults-boxplots} in Section~\ref{sec:simulation} visualizes the bias of the estimates for the polyserial correlation coefficient in the employed simulation. Since this coefficient is only one of the~$d$ parameters of the polyserial model (alongside the population mean and variance of~$X$ as well as the thresholds underlying~$Y$), it is of interest to also assess the estimation accuracy for the remaining parameters in the parameter vector $\Btheta = (\rho, \mu, \sigma^2, \vec{\tau}^\top)^\top$. To compare an estimate~$\thetahat$ to the true value~$\Btheta_*$, we use a geometric approach, namely the angle between~$\thetahat$ and~$\Btheta_*$, which is defined as
\[
	\textnormal{angle}\left( \thetahat, \Btheta_* \right)
	=
	\frac{180}{\pi}\cdot
	\cos^{-1}\left( \frac{\left|\Btheta_*^\top\thetahat\right|}{\|\Btheta_*\|\|\thetahat\|}  \right),
\]
where $\|\cdot\|$ denotes the Euclidean norm. Multiplying by the constant $180/\pi$ transforms the measurement unit from radians to degrees. Hence, the closer the angle to zero degrees, the better the estimation accuracy. Conversely, an angle of ninety degrees indicates a poor performance because the two vectors are perpendicular to one another. As performance measure over~$T$ repetitions, we use the root mean squared error (RMSE) of the squared angles, that is,
\[
	\textnormal{RMSE}
	=
	\sqrt{\frac{1}{T}\sum_{t=1}^T \textnormal{angle}^2\left( \hat{\Btheta}_N^{(t)}, \Btheta_* \right) },
\]
where $\hat{\Btheta}_N^{(t)}$ is the estimate of the $t$-th repetition. This performance measure has been used before in the robust statistics literature, like in, e.g., \citet{alfons2017}. 

\begin{figure}
	\centering
	\includegraphics[width = \textwidth]{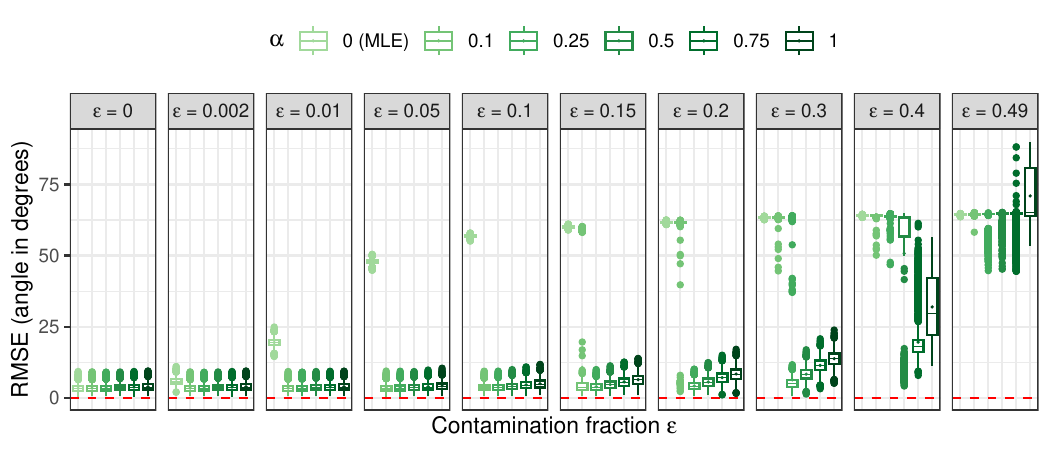}
\caption{Boxplots of the RMSE (in degrees) of the considered estimators, for various contamination fractions in the misspecified polyserial models across 5,000 repetitions. Diamonds represent the respective average RMSE.}
\label{fig:appendix-simresults-boxplots}
\end{figure}

Figure~\ref{fig:appendix-simresults-boxplots} visualizes the RMSE (in degrees) of the different estimators across the considered contamination fractions. The results are very similar in quality as those for estimating the polyserial correlation coefficient in Figure~\ref{fig:main-simresults-boxplots}: As soon as contamination is present, the MLE starts to exhibit a notable bias, and stabilizes at about $\varepsilon = 0.1$ with an angle of about~$60^\circ$. Conversely, the robust estimators ($\alpha > 0$) are more accurate in the presence of contamination until they break gradually start to down as well as the contamination fraction is increased. Beyond $\varepsilon = 0.4$, all estimators have broken down and perform similarly poorly. 

\begin{figure}
	\centering
	\includegraphics[width = \textwidth]{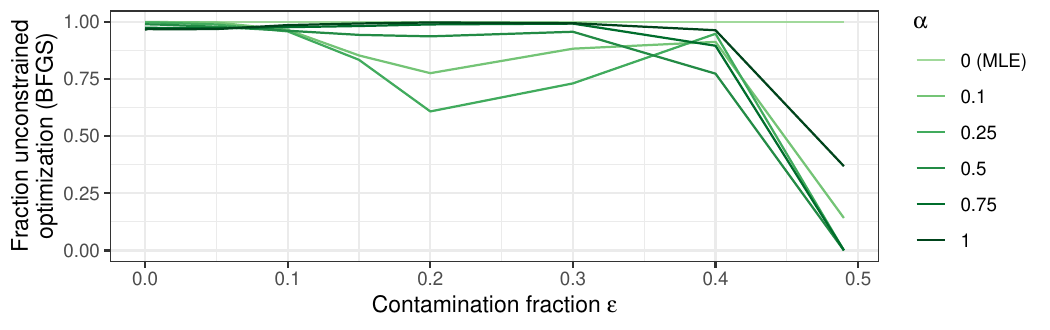}
\caption{\edit{Fraction of the 5,000 simulation repetitions in which unconstrained optimization via BFGS did not converge, so constrained Nelder-Mead optimization was subsequently used instead.}}
\label{fig:appendix-simresults-unconstrained}
\end{figure}

\begin{figure}
	\centering
	\includegraphics[width = \textwidth]{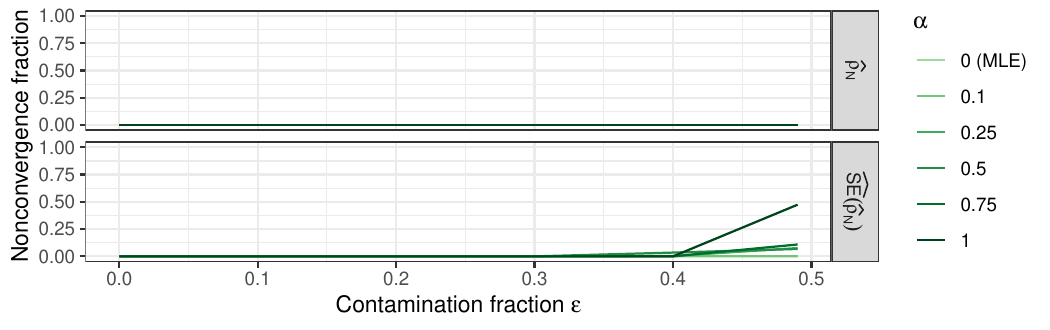}
\caption{Fraction of the 5,000 simulation repetitions in which a point estimator did not numerically converge\edit{---neither with the unconstrained BFGS nor the constrained Nelder-Mead algorithm---}(top), and its associated standard error could not be computed due to singularity of~$\hatmatfun{J}{\thetahat}$ in the covariance matrix estimator (bottom). The point estimators always converged, resulting in overplotted lines.}
\label{fig:appendix-simresults-nonconvergence}
\end{figure}

\edit{Recall from Section~\ref{sec:implementation} that by default, our implementation first tries unconstrained optimization via the BFGS algorithm, and, in case of nonconvergence or an error, subsequently tries constrained optimization via the Nelder-Mead algorithm. Figure~\ref{fig:appendix-simresults-unconstrained} visualizes the frequency with which unconstrained optimization converged. In case of low contamination fractions, unconstrained optimization nearly always succeeded. Conversely, as the contamination fraction increases to higher values, unconstrained optimization fails more often, so a constrained algorithm must be used instead. Similar behavior was observed in all other simulations in this paper.}

Furthermore, Figure~\ref{fig:appendix-simresults-nonconvergence} illustrates the fraction of the 5,000 repetitions in which the correlation estimate~$\hatN{\rho}$ of given estimator~$\thetahat$ did not numerically converge to a solution, \edit{neither with the unconstrained BFGS nor the constrained Nelder-Mead algorithm (see Section~\ref{sec:implementation})}. In similar fashion, it visualizes the fraction in which its associated estimated standard error~$\SEhat{\hatN{\rho}}$ could not be computed due to singularity of matrix~$\hatmatfun{J}{\thetahat}$ in the covariance matrix estimator in Section~\ref{sec:asymptotics}. The point estimates~$\hatN{\rho}$ always numerically converged to a solution across all configurations and repetitions. Up to (and including) contamination fraction~$\varepsilon = 0.3$, the standard errors could virtually always be computed. Only in extremely high-contamination settings with $\varepsilon > 0.3$, as the estimators gradually break down (Figure~\ref{fig:appendix-simresults-boxplots}), standard errors could occasionally also not be computed. In particular, at the highest considered contamination fraction $\varepsilon = 0.49$, the estimator with tuning constant $\alpha = 1$ fails to compute standard errors in about half of the repetitions. However, this estimator  itself seems to be unstable in this setting to begin with (see Figure~\ref{fig:main-simresults-boxplots}), which may cause frequent singularity of matrix~$\hatmatfun{J}{\thetahat}$. 

\subsection{\edit{Discussion on robustifying the two-step approach}} \label{app:twostep}

As described in Section~\ref{sec:MLE}, polyserial correlation is often estimated through the two-step approach in~\eqref{eq:twostep}, which is computationally faster but less efficient than joint ML estimation. It turns out that the two-step approach is also a maximum likelihood (ML) estimator in which model parameters are estimated sequentially rather than jointly. In particular, it can be shown \citep{muthen1984} that 
\begin{itemize}
	\item the first-step estimators $\left(\hat{\mu}_{\textrm{TS}}, \hat{\sigma}^2_{\textrm{TS}}\right)$ of $(\mu, \sigma^2)$ are ML estimates of a marginal location-scale normality model for~$X$;
	\item the first-step threshold estimators $\hat{\tau}_{k,\textrm{TS}}$ are ML estimates of thresholds $\tau_k, k = 1,\dots,r-1,$ in marginal latent normality model for~$Y$,
	\item the second-step correlation estimator of~$\rho$ is a conditional ML estimator in a bivariate normality model, conditional on the first-step estimates.
\end{itemize}

It is therefore natural to ask whether the two-step approach can also be robustified by using DPD estimators. Indeed, it is possible to apply separate DPD estimators to separately estimate each of the three models above in a robust manner, owing to the generality of the results in \citet{basu1998}. However, there are substantial theoretical and computational downsides of such a robustified two-stage approach, as we shall explain in the following. 

First, using DPD for jointly estimating all model parameters already incurs an efficiency loss, as discussed in detail in Section~\ref{sec:efficiency}. Since using DPD in a two-stage approach requires fitting three separate models, the efficiency loss of such a robustified two-step approach would potentially be substantial due to using three estimation procedures with diminished efficiency. One of the attractive features of joint DPD estimation (as proposed in Section~\ref{sec:estimator}) is that robustness can be gained with only a relatively minor efficiency loss of less than~2\% (see Section~\ref{sec:efficiency}). 

Second, every DPD estimator requires numerically solving an integral over the postulated model's density to the power of~$(1+\alpha)$ in every iteration of the employed optimization routine \citep[][Eq.~2.2]{basu1998}. This numerical integral is the main computational bottleneck that causes the robust estimator in Section~\ref{sec:estimator} to be computationally slower than joint ML. Recall that in our case, the integral is one-dimensional because the $Y$-dimension is a computationally cheap sum of~$r$ summands; see Eq.~\eqref{eq:DPD}. 
In a robustified two-step approach, however, there are three separate DPD estimators, each of which requires numerically solving a one-dimensional integral in every iteration. Hence, we expect the robustified two-stage approach to take even \emph{longer} to compute than simultaneous estimation because the former requires separately solving three univariate integrals, while the latter only needs to solve one. While the main advantage of the two-step approach as in \citet{olsson1982} is reduced computing time, such an advantage would therefore be absent (even reversed) in a DPD-based robustification thereof.

Thus, overall, robustifying the two-step approach via DPD would result in a procedure that is slower to compute and less efficient than joint estimation. Consequently, we believe that the latter approach is strictly preferable to the former.

\newpage
\section{Additional simulations}\label{app:additional-simulations}

\subsection{Contamination through gross errors}\label{app:outlier}
This simulation is concerned with \emph{gross errors}, that is, extreme outliers. We are interested in how ML and our robust estimator react to such observations. 

For the true polyserial model parameters, we choose the same configurations as in Section~\ref{sec:simulation}, namely $\rho_* = 0.5, \mu_* = 0, \sigma_*^2 = 1, \tau_{*,1} = -1.5, \tau_{*,2} = -0.5, \tau_{*,3} = 0.5, \tau_{*,4} = 1.5$, so that the ordinal variable again has five response categories. As contaminating distribution of $(X,\eta)$, we set a bivariate normal distribution with covariance matrix equal to identity and population mean $(a, -a)^\top$, where $a = 1,000,000$, and discretize the ensuing realizations in the~$Y$-dimension with thresholds $\tau_{*,j}, j=1,\dots,4$. Thus, contamination manifests through extreme outliers in the~$X$-dimension and inflation of the first response category in the~$Y$-dimension. We consider the same contamination fractions as in Section~\ref{sec:simulation}, namely $\varepsilon \in \{0, 0.002, 0.01, 0.05, 0.1, 0.15, 0.2, 0.3, 0.4, 0.49\}$, as well as the same sample size, $N=500$, number of repetitions, $T=5,000$, and performance measures. 

\begin{figure}
	\centering
	\begin{subfigure}{\textwidth}
	\centering
	\includegraphics[width = \textwidth]{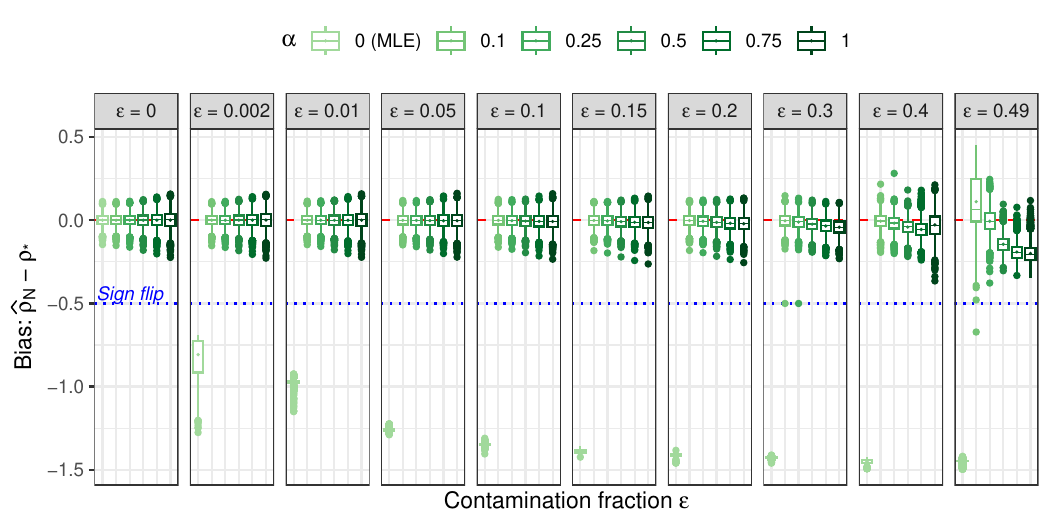}
	\caption{Boxplot visualization of the bias. Diamonds represent the respective average bias.}
	\end{subfigure}
	\begin{subfigure}{\textwidth}
	\includegraphics[width = \textwidth]{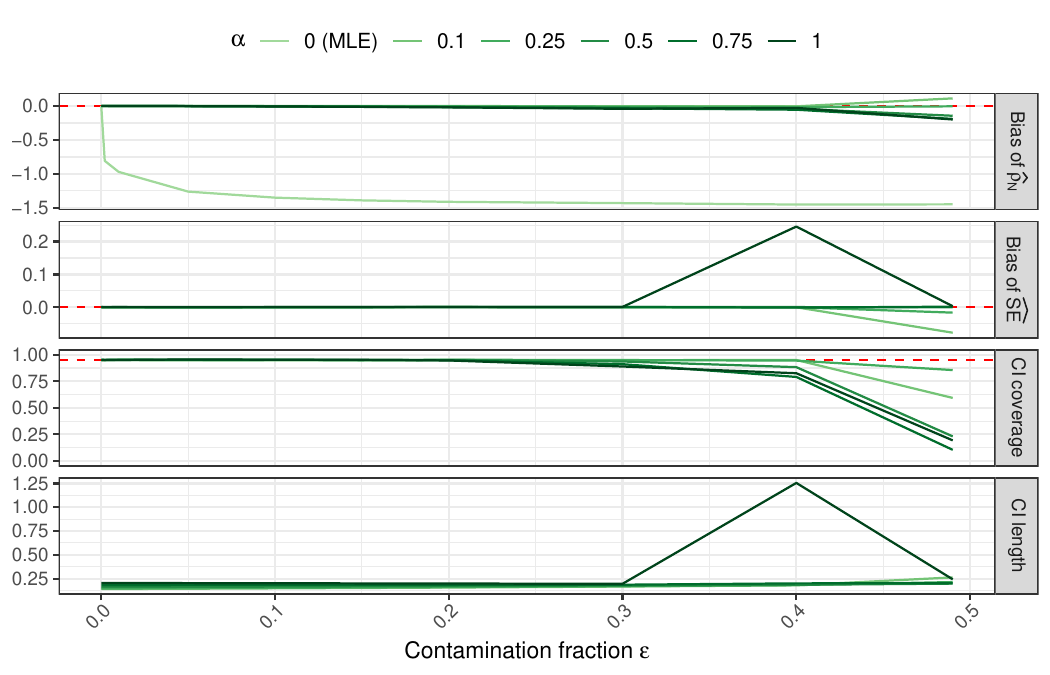}
	\caption{Performance measures of $\hatN{\rho}$. For $\varepsilon > 0$, no standard errors (SE) could be computed for ML, so no SE estimation bias, coverage, or confidence interval length could be computed (cf. Figure~\ref{fig:outlier-nonconvergence}).}
	\end{subfigure}
\caption{Visualizations of the results of the ``gross error'' simulation design in Section~\ref{app:outlier} for t\edit{he} considered estimators across 5,000 repetitions, at significance level $\gamma = 0.05$.}
\label{fig:outlier-simresults}
\end{figure}

Figure~\ref{fig:outlier-simresults} visualize the simulation results. Already one single gross error observation ($\varepsilon = 0.002$ with the considered sample size) suffices for the MLE to suffer from a sign flip: While the true value is moderately positive ($\rho_* = 0.5$), the ML correlation estimates are always negative.  Furthermore, the extreme values of the outliers cause the Fisher information matrix to become (computationally) singular, so no inferential statistics could be computed for the MLE for any positive contamination fraction in any repetition. Conversely, the robust estimators stay nearly unaffected by the presence of such gross errors up until about $\varepsilon = 0.2$ with accurate point estimates and coverage at the nominal level of~95\%. Beyond $\varepsilon = 0.2$, the coverage of the robust estimators slightly decreases, but the point estimates remain fairly accurate. Only at the extreme contamination level of $\varepsilon = 0.49$ considerable bias occurs also for the robust estimators. 

Interestingly, the robust estimator is more accurate under gross error contamination than under the less extreme contamination in Section~\ref{sec:simulation} (see Figure~\ref{fig:main-simresults-boxplots}). It seems plausible that gross error contamination is easier to distinguish from regular observations from the polyserial model than less extreme forms of contamination. Consequently, it might be easier for the robust estimator to downweight the correct observations compared to a situation when contamination and regular observations are more alike.

\begin{figure}
	\centering
	\includegraphics[width = \textwidth]{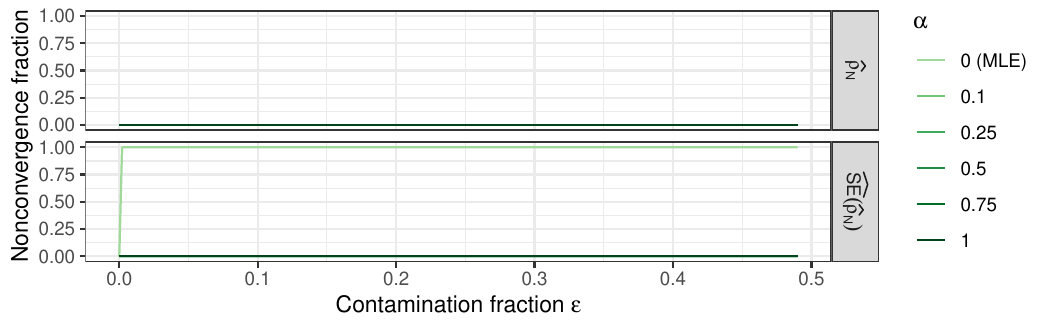}
\caption{For the ``gross error'' design: Fraction of the 5,000 simulation repetitions in which a point estimator did not numerically converge\edit{---neither with the unconstrained BFGS nor the constrained Nelder-Mead algorithm---}(top), and its associated standard error could not be computed due to singularity of~$\hatmatfun{J}{\thetahat}$ in the covariance matrix estimator (bottom). The point estimators always converged, resulting in overplotted lines. \edit{For $\varepsilon > 0$, no standard errors could ever be computed for ML, reflecting the instability of ML estimation in the presence of gross errors.}}
\label{fig:outlier-nonconvergence}
\end{figure}

For completeness, Figure~\ref{fig:outlier-nonconvergence} summarizes numerical nonconvergence. The point estimators always converged. With a positive contamination fraction $\varepsilon > 0$, ML standard errors could never be computed due to the normality-based Fisher information matrix becoming singular in this setting, further reflecting the instability of ML in the presence of gross errors.

\subsection{Contamination through correlation shift}\label{app:corshift}
The simulations in Sections~\ref{sec:simulation} and~\ref{app:outlier} have been concerned with mean-shifted contamination, that is, the contamination has been generated by a contamination distribution of~$(X,\eta)$ whose population mean differs from the population mean of the true normal distribution~$\Pxetafun{\cdot,\cdot}{\Btheta_*}$. However, contamination may manifest in an unlimited variety of ways, with some not characterized by mean shifts but \emph{correlation shifts}. A correlation shift occurs if the contamination distribution has the same population means and variances as the true distribution, but a different correlation structure. In our context of polyserial correlation, the contamination distribution~$\Hxeta$ is equal to the 
true normal distribution~$\Pxetafun{\cdot,\cdot}{\Btheta_*}$, except for a sign-flipped correlation parameter,~$-\rho_*$. 

\begin{figure}
	\centering
	\includegraphics[width = 0.93\textwidth]{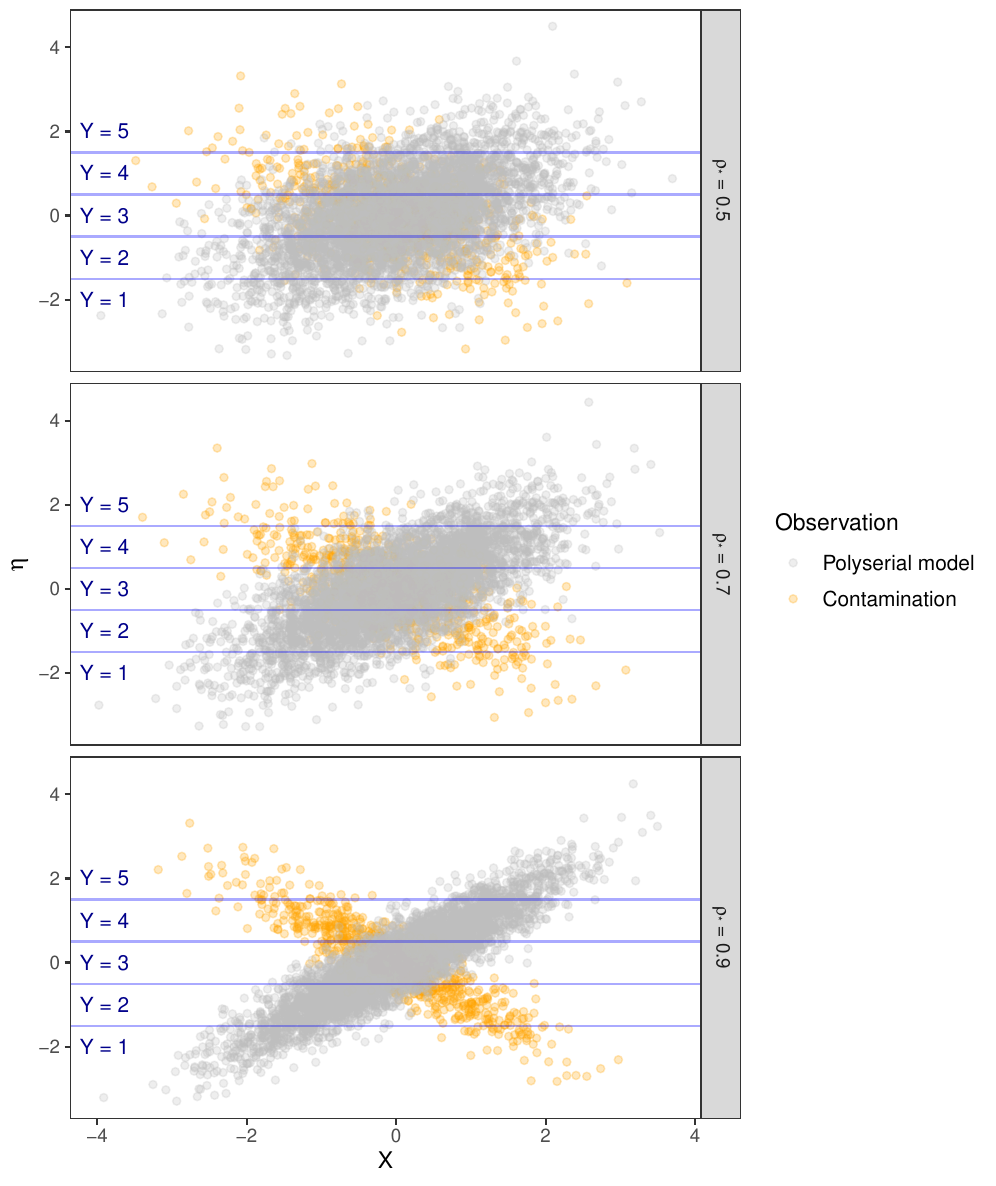}
\caption{Example data sets generated by the process with correlation-shifted combination described in Section~\ref{app:corshift}, for three true correlations~$\rho_*\in\{0.5, 0.7, 0.9\}$ (rows) with contamination fraction $\varepsilon = 0.15$. The remaining true parameters are set to $\mu_* = 0, \sigma_*^2 = 1, \tau_{*,1} = -1.5, \tau_{*,2} = -0.5, \tau_{*,3} = 0.5, \tau_{*,4} = 1.5$. Gray dots indicate random draws from the true normal distribution~$\Pxetafun{\cdot,\cdot}{\Btheta_*}$, while orange dots indicate draws from the correlation-shifted contamination distribution~$\Hxeta$. Horizontal lines denote the discretization thresholds. For a clearer visualization, the sample size is set to $N=5,000$ in this example data set.}
\label{fig:corshift-simdesign}
\end{figure}

In this simulation, we consider partial misspecification of the polyserial model through correlation-shifted contamination. As true parameter values, we set $\mu = 0, \sigma^2 = 1$, and $\vec{\tau}_*^\top = (-1.5,-0.5,0.5,1.5)^\top$ so that the ordinal variable has five response options. For the true polyserial correlation coefficient, we consider $\rho_*\in\{0.5,0.7,0.9\}$. Figure~\ref{fig:corshift-simdesign} visualizes an example data set generated by this process. For the weakest considered correlation, $\rho_* = 0.5$, contamination (orange dots) and polyserial model points (gray points) substantially overlap and are therefore hard to distinguish. Conversely, the stronger the correlation~$\rho_*$, the more distinct the contamination becomes from polyserial model points. We therefore expect that the robustness gain of the robust estimator should increase with stronger true correlations. 

We generate $N=500$ observations from the described process and repeat this procedure 5,000 times. The same performance measures and estimators as in Section~\ref{sec:simulation} are used. We stress that this simulation design has been used before by \citet{welz2025polycor} in the context of the polychoric correlation model. 

\begin{figure}
	\centering
	\includegraphics[width = \textwidth]{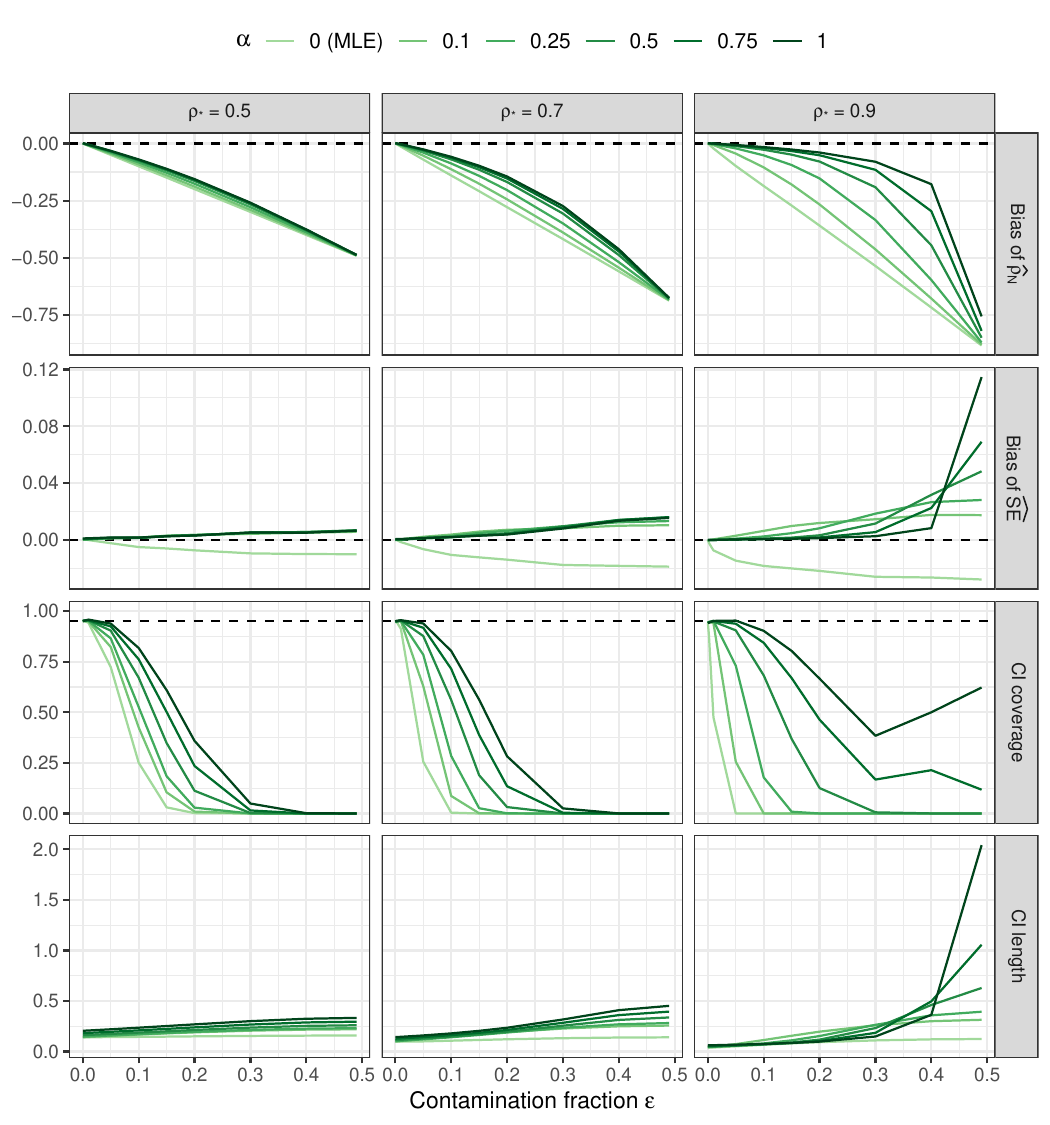}
\caption{Performance measures for the simulation in Section~\ref{app:corshift} with correlation-shifted contamination, averaged across the~5,000 repetitions, for true correlation coefficients $\rho_*\in\{0.5,0.7,0.9\}$ (rows) and different estimators (solid colored lines). The performance measures are the bias of the correlational estimates ($\rhohatmean - \rho_*$), the bias of their estimated standard errors $\left(\SEbarhat{\rhohat} - \SEapprox{\rhohat}\right)$, coverage of the~95\% confidence intervals (with the nominal level as dashed horizontal line), and the average length of the confidence intervals. Cases for which no standard errors could be computed due to singularity were excluded from the inferential performance measures; see Figure~\ref{fig:appendix-corshift-nonconvergence} for a summary.}
\label{fig:corshift-simresults}
\end{figure}

Figure~\ref{fig:corshift-simresults} illustrates the simulation results. In the weakest considered correlation size $(\rho_* = 0.5)$, the magnitude of bias is similar for all considered estimators and the robust methods only provides a minor improvement compared to the MLE. For the stronger true correlations, the robust estimators yields a notable gain in robustness, particularly for the strongest considered correlation of $\rho_* = 0.9$. It seems plausible that this gain in robustness is driven by contamination being better distinguishable from regular data points when the true correlation that is being shifted has larger magnitude (Figure~\ref{fig:corshift-simdesign}). Furthermore, at contamination fraction $\varepsilon = 0.49$, all considered estimators converge to similar biases across all considered correlations. Specifically, they converge to the value $-\rho_*$, which indicates that a correlation of nearly~0 was estimated. It is not surprising that for two approximately equally-sized groups of data points with the same mean but whose respective population correlations are mutually sign-flipped, the best fit can be achieved at a correlation estimate of about~0. 

\begin{figure}
	\centering
	\includegraphics[width = \textwidth]{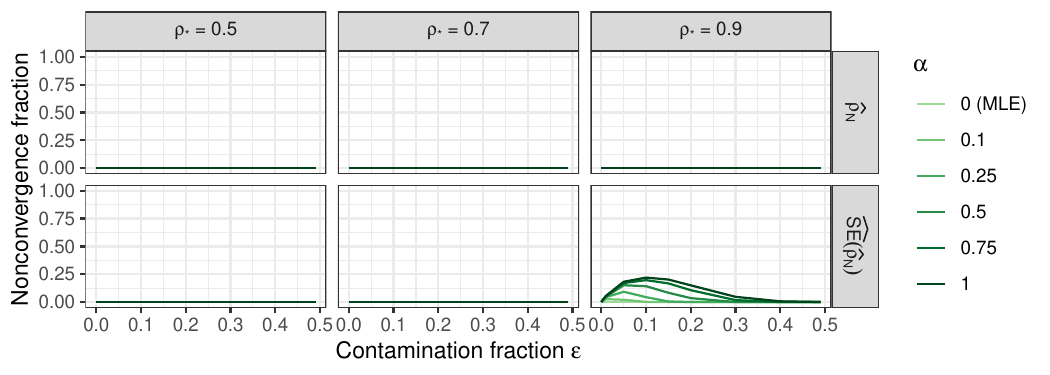}
\caption{For the ``correlation shift'' design: Fraction of the 5,000 simulation repetitions in which a point estimator did not numerically converge\edit{---neither with the unconstrained BFGS nor the constrained Nelder-Mead algorithm---}(top), and its associated standard error could not be computed due to singularity of~$\hatmatfun{J}{\thetahat}$ in the covariance matrix estimator (bottom). The point estimators always converged, resulting in overplotted lines.}
\label{fig:appendix-corshift-nonconvergence}
\end{figure}

Figure~\ref{fig:appendix-corshift-nonconvergence} summarizes numerical nonconvergence. Occasional non-computability of standard errors only occurs in the high correlation setting of $\rho_* = 0.9$. Most nonconvergent cases occur with $\alpha = 1$ at $\varepsilon = 0.1$, where about~25\% of the standard errors cannot be computed.

Overall, this simulation study demonstrates that the magnitude of our proposed estimator's robustness gain depends on the overlap between the contamination distribution~$\Hxeta$ and true normal distribution~$\Pxetafun{\cdot,\cdot}{\Btheta_*}$. If these distributions substantially overlap---like in correlation-shifted contamination with weak to moderate true correlation---then the robust estimator may not be able to clearly distinguish between contamination and regular observations since it makes no assumptions on the former, and is subsequently unable to provide a tangible robustness gain. On the other hand, when~$\Hxeta$ and~$\Pxetafun{\cdot,\cdot}{\Btheta_*}$ are sufficiently distinct---like in correlation-shifted contamination with strong true correlation or mean-shifted correlation (cf. Sections~\ref{sec:simulation} and~\ref{app:outlier})---then the robust estimator can identify the contaminated observations and subsequently downweigh them to achieve considerable robustness gains. In either scenario, we observe that the robust estimator is always at least as accurate as the ML estimator, and therefore constitutes an overall improvement in terms of robustness. Similar findings for a different robust estimator in the context of estimating polychoric correlation have been reported by \citet{welz2025polycor}.

\subsection{Distributional misspecification}\label{app:distributional}
\subsubsection{Preliminaries}
The estimator proposed in this paper is designed to be robust against \emph{partial} misspecification, which is characterized by the polyserial model being misspecified for only a (possibly empty) subset of the observed sample (Section~\ref{sec:misspecification}). In contrast, in the framework of \emph{distributional} misspecification, the model is misspecified for \emph{all} observations in a sample, not just a subset thereof. Distributional misspecification of the polyserial model manifests in the observed-latent variable pair~$(X,\eta)$ being jointly distributed according to an unknown and unspecified nonnormal distribution~$G = G_{X,\eta}$. In this framework, the object of interest is the population correlation between~$X$ and~$\eta$ under the distribution~$G$, that is, $\rho_G = \corF{G}{X}{\eta}$, instead of the polyserial correlation coefficient. Estimators for situations where~$G$ is nonnormal have been proposed by \citet{bedrick1996}, \citet{lord1963}, and \citet{brogden1949}. All of these estimators are based on normality-based maximum likelihood, but are less prone to distributional misspecification when the sampling distribution is not normal. 

Although conceptually different, robustness to partial and distributional misspecification frameworks are \textit{``practically synonymous notions"} \citep[][p.~4]{huber2009}. Thus, despite our robust estimator being designed for a fundamentally different type of model misspecification---partial rather than distributional misspecification---it is a relevant question whether or not our estimator can also offer enhanced robustness to distributional misspecification. 
\citet[][p.~32]{welz2025polycor} argue that the potential for robustness gains of partial-misspecification-robust estimators in distributional misspecification depends on the properties of the unknown sampling distribution~$G$. 
If the nonnormal~$G$ can be decently approximated by mixture between a normal distribution and some other distribution---thereby approximating the partial misspecification framework in~\eqref{eq:contamdist}---then our robust estimator should perform reasonably well. If~$G$ cannot be approximated by such a mixture, neither ML nor our estimator may be expected to perform well.

To investigate the performance of our robust estimator when the polyserial model is distributionally misspecified, we perform a simulation study in the following subsection.

\subsubsection{Simulation study} \label{sec:distributional-simulation}

To simulate data for which the polyserial model is distributionally misspecified, we aim to sample from a nonnormal bivariate distribution~$G$ with a prespecified value of the population correlation $\rho_G = \corF{G}{X}{\eta}$. This is exactly what the VITA simulation method of \citet{gronneberg2017vita} does. For a prespecified value of~$\rho_G$, the VITA method models the random vector~$(X,\eta)$ so that the individual variables~$X$ and~$\eta$ have correlation~$\rho_G$, are both marginally normally distributed, but are \emph{not} jointly normally distributed. Instead, the joint distribution~$G$ is set to a prespecified nonnormal copula distribution. \citet{gronneberg2017vita} show that the VITA method is particularly useful when VITA-modeled variables are discretized because the ensuing ordinal variables could not have been generated by a latent bivariate normal distribution, as assumed by e.g., polychoric correlation.

\begin{figure}
	\centering
	\includegraphics[width = \textwidth]{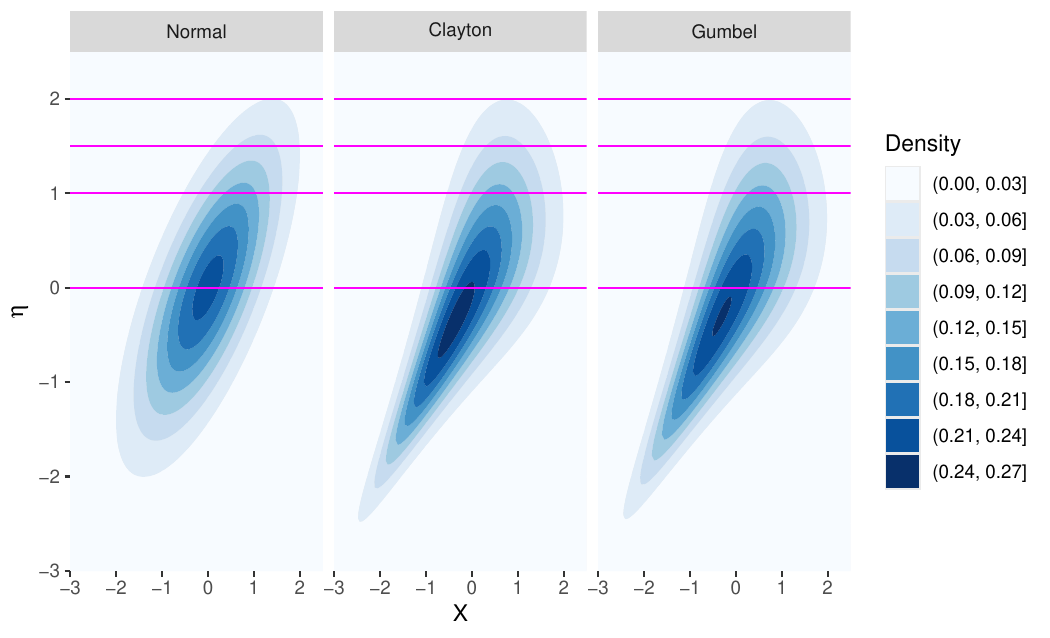}
\caption{Bivariate densities of the standard bivariate normal distribution, Gumbel copula, and Clayton copula with correlation $\rho_G = 0.7$ and standard normal marginal distributions. The horizontal lines indicate the thresholds with which the latent~$\eta$ is discretized to the ordinal~$Y$.}
\label{fig:distributional-simdesign}
\end{figure}

\begin{figure}
	\centering
	\includegraphics[width = \textwidth]{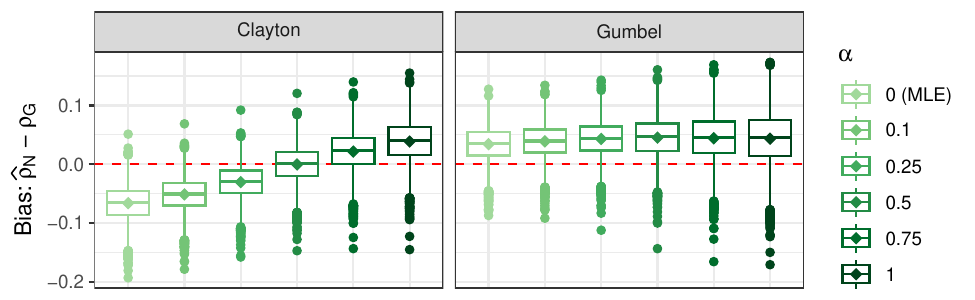}
\caption{Boxplot visualization of the bias of the considered bias, $\hatN{\rho} - \rho_G$, for the polyserial correlation coefficient under distributional misspecification through Clayton and Gumbel copulas, for 5,000 repetitions and true correlation $\rho_G = 0.7$. Diamonds represent average biases.}
\label{fig:distributional-simresults}
\end{figure}

In our simulation, we use the VITA method to generate data for the variables $(X,\eta)$ so that their joint distribution is a Clayton or Gumbel copula with population correlation $\rho_G = 0.7$, and~$X$ and~$\eta$ are both marginally standard normal. Figure~\ref{fig:distributional-simdesign} visualizes these copulas. We then discretize the latent variable~$\eta$ according to~\eqref{eq:discretization} with discretization thresholds $\tau_1 = 0,  \tau_2 = 1,  \tau_3 = 1.5,  \tau_4 =  2$, 
so that the ensuing observed ordinal variable~$Y$ has five response options. We generate $N=500$ observations for~$(X,Y)$ according to this process and repeat this procedure 5,000 times. As implementation of the VITA method, we use the package \pkg{covsim} \citep{covsim}. The same performance measures as in Section~\ref{sec:simulation} are used.

To summarize, this simulation emulates a situation where misspecification of the polyserial model would not be detectable by testing for marginal nonnormality of the observed~$X$ by, for instance, a Kolmogorov-Smirnov test, because~$X$ is marginally normal. Hence, the simulated situation might be seen as a challenging one for any normality-based estimator because the distributional misspecification only manifests in the unobserved joint distribution.

\begin{table}[t]
\centering
\small
\setlength{\tabcolsep}{4.67pt}
\begin{tabular}{l l c r r c c c r c c c}
& & & \multicolumn{3}{c}{Point estimate} & & \multicolumn{2}{c}{Standard error} & & \multicolumn{2}{c}{Confidence interval} \\
\noalign{\smallskip}\cline{4-6}\cline{8-9}\cline{11-12}\noalign{\smallskip}
Distribution~$G$ & $\alpha$ & & \multicolumn{1}{c}{$\hatN{\rho}$} & \multicolumn{1}{c}{Bias} & SE & & $\widehat{\text{SE}}$ & Bias & & Coverage & Length \\
\noalign{\smallskip}\hline\noalign{\smallskip}
\multirow{6}{*}{Clayton} 
  & 0 (MLE) 		&& 0.634 & $-0.066$ & 0.003 && 0.037 &  0.007     &&  0.613  &  0.145\\
  & 0.1 		&& 0.649 & $-0.051$ & 0.029 && 0.029 &  0.000  	  &&  0.597  &  0.113\\
  & 0.25 		&& 0.670 & $-0.030$ & 0.029 && 0.029 &  0.000     &&  0.851  &  0.113\\
  & 0.5 		&& 0.700 &  $0.000$ & 0.031 && 0.031 &  0.000     &&  0.947  &  0.120\\
  & 0.75 		&& 0.722 &    0.022 & 0.033 && 0.033 &  0.000	  &&  0.871  &  0.130\\
  & 1 			&& 0.739 &    0.039 & 0.036 && 0.036 &  0.000     &&  0.774  &  0.142\\
\noalign{\medskip}
\multirow{6}{*}{Gumbel} 
  & 0 (MLE) 		&& 0.734 & 0.034 & 0.029 && 0.022 & $-0.007$ && 0.600 & 0.087 \\ 
  & 0.1 		&& 0.739 & 0.039 & 0.029 && 0.029 & 0.000    && 0.690 & 0.113 \\ 
  & 0.25 		&& 0.743 & 0.043 & 0.030 && 0.030 & 0.000    && 0.665 & 0.118 \\ 
  & 0.5 		&& 0.745 & 0.045 & 0.035 && 0.034 & 0.000    && 0.690 & 0.135 \\ 
  & 0.75	  	&& 0.744 & 0.044 & 0.040 && 0.004 & 0.000    && 0.743 & 0.157 \\ 
  & 1 			&& 0.744 & 0.044 & 0.046 && 0.046 & 0.001    && 0.788 & 0.182 \\ 
  \hline
\end{tabular}
\caption{Performance measures for estimating  polyserial correlation coefficients under distributional misspecification, at significance level $\gamma=0.05$ (averaged across 5,000 repetitions).}
\label{tab:distributional-simresults}
\end{table}

\begin{figure}
	\centering
	\includegraphics[width = \textwidth]{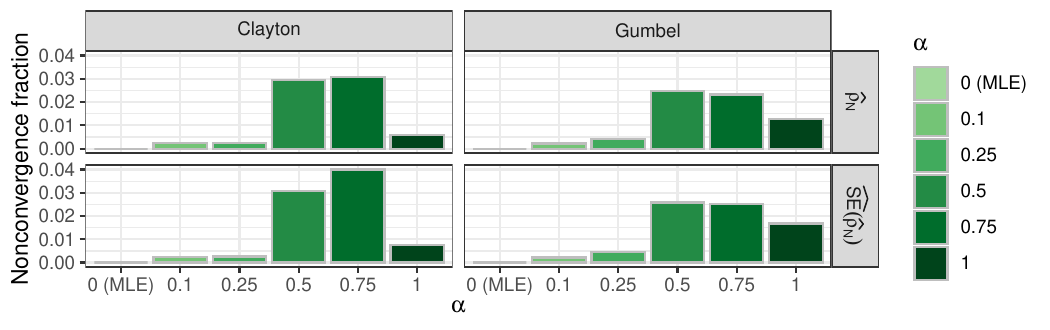}
\caption{For the ``distributional misspecification'' design: Fraction of the 5,000 simulation repetitions in which a point estimator did not numerically converge\edit{---neither with the unconstrained BFGS nor the constrained Nelder-Mead algorithm---}(top), and its associated standard error could not be computed due to singularity of~$\hatmatfun{J}{\thetahat}$ in the covariance matrix estimator (bottom).}
\label{fig:distributional-nonconvergence}
\end{figure}

Figure~\ref{fig:distributional-simresults} and Table~\ref{tab:distributional-simresults} summarize the simulation results. When the joint distribution~$G$ is a Clayton copula, the MLE exhibits a bias of about~$-0.066$, whereas all robust estimators are less biased (in absolute magnitude). The MLE's coverage is about~60\%, but it also has the widest confidence intervals (wider then the robust estimators). In contrast, despite having narrower confidence intervals, the estimators with $\alpha > 0.1$ achieve a coverage of about~60\% or higher. Especially the choices $\alpha = 0.5, 0.75$ are remarkably accurate with coverages of about~95\% and~8\edit{7}\%, respectively. Conversely, when distribution~$G$ is a Gumbel copula, all estimators exhibit a notable bias of similar magnitude. The slightly higher coverage of the robust estimators is primarily due to wider confidence intervals.  Thus, in this setting, the robust estimators do not improve upon the MLE.

We noticed that in a small number of repetitions, the robust estimator did not numerically converge. This situation occurs when multiple response options of~$Y$ are almost never chosen (e.g., only once or twice in the sample). In such a situation, the robust estimator may attempt to eliminate a threshold by either pushing the first or last threshold to~$\pm\infty$ or pulling adjacent thresholds together as closely as numerically possible. Similar behavior has been observed for a robust estimator of the polychoric correlation model \citep{welz2025polycor}. We follow \citet{welz2025polycor} by identifying numerical instability through adjacent thresholds being unreasonably far from each other, namely by a minimum distance of~3.92. Under the polyserial model, this distance covers as much as~95\% of all probability mass of the standard normal marginal distribution of~$\eta$. We subsequently omitted the estimates where such nonconvergence occurs from the analyses in this section. Figure~\ref{fig:distributional-nonconvergence} summarizes the numerical nonconvergence statistics. The highest fraction of nonconvergent cases occurs for the choice $\alpha = 0.\edit{7}5$ with about~\edit{4}\%, which is a reasonably small value considering that our proposed estimator is not designed for distributional misspecification.

Overall, the results suggest that for population correlation~0.7, the Clayton copula might be reasonably well-approximable by a mixture between a normal distribution and some other distribution, but not the Gumbel copula. Figure~\ref{fig:distributional-simdesign} indicates that while the two copulas are of comparable shape, the Clayton copula has substantially more probability mass in its center than the Gumbel distribution, which may enable it to be somewhat approximable by a mixture involving a normal distribution. A similar result for the Clayton copula has been found in \citet{welz2025polycor} for distributional misspecification of the polychoric correlation model.

To conclude, this simulation study demonstrates that in certain situations with distributional misspecification of the polyserial model, our estimator can offer enhanced robustness compared to the MLE. On the other hand, it also demonstrates that there are situations where no robustness can be gained because MLE and robust estimators perform very similarly. Thus, while distributional misspecification is not covered by the partial misspecification framework for which our robust estimator is intended, in certain situations our estimator can yield a robustness gain under distributional misspecification, thereby constituting an overall gain in robustness compared to ML estimation.

\edit{As a practical consequence, we recommend applied researchers to stay wary of potential partially-latent nonnormality even when ML and our robust estimator produce very similar point estimates. While such similarity may be due to normality holding true (in which case both estimators yield asymptotically equivalent estimates), it may also be due to distributional misspecification from a specific nonnormal distribution for which the robust estimator cannot improve over ML (such as the Gumbel copula considered in this simulation).\footnote{A third  possibility is that partially-latent normality is violated in the $\eta$-dimension, but the misspecification is inconsequential because the ensuing density of the observed $(X,Y)$ is equal to the polyserial model distribution in~\eqref{eq:jointcdf} (see Section~\ref{sec:misspecification}). In this case, both ML and robust estimators will yield accurate point estimates despite the model being technically misspecified. Hence, in this theoretical case, there are no adverse effects for point estimation with either estimator. We refer to \citet{gronneberg2020} and \citet{foldnes2020polycor} for detailed discussions on testing for underlying normality from discretized variables.} To investigate the possibility of distributional misspecification whenever ML and robust estimator yield similar estimates, a~$\chi^2$-test for partially-latent normality may be considered, which is, for example, implemented in the \proglang{R} package \pkg{polycor} \citep{R:polycor}.}\footnote{\edit{Recently, \citet{foldnes2020polycor} have proposed a bootstrap test for testing underlying multivariate normality when \emph{all} observed variables are ordinal. Therefore, the bootstrap test as described in \citet{foldnes2020polycor} is not directly applicable to mixed data. However, we believe that it should be possible to extend their methodology to mixed data. We leave this to future research.}}

\end{document}